\documentclass[12pt]{article}
\pdfoutput=1
\usepackage[colorlinks,linkcolor=Blue,citecolor=Blue,bookmarks,bookmarksnumbered]{hyperref}
\usepackage[scaled=0.85]{helvet}
\usepackage{amsmath,amssymb,accents,mathrsfs,XoohmE}
\usepackage{svg}
\usepackage{graphicx,color}
\graphicspath{{Figures/}}
\usepackage{booktabs}
\usepackage{multirow}
\usepackage{placeins}
\usepackage{amsmath}
\usepackage{subfigure}
\usepackage{multirow}
\usepackage{lipsum}  
\usepackage{float}
\usepackage{algorithm}
\usepackage{algpseudocode}

\usepackage{fancyvrb,newverbs,xcolor}
\usepackage{colortbl}
\definecolor{cverbbg}{gray}{0.93}

\newenvironment{lcverbatim}
 {\SaveVerbatim{cverb}}
 {\endSaveVerbatim
  \flushleft\fboxrule=0pt\fboxsep=.5em
  \colorbox{cverbbg}{%
    \makebox[\dimexpr\linewidth-2\fboxsep][l]{\BUseVerbatim{cverb}}%
  }
  \endflushleft
}

\newverbcommand{\cverb}
  {\setbox\verbbox\hbox\bgroup}
  {\egroup\colorbox{cverbbg}{\box\verbbox}}

\usepackage{amsfonts}

\DeclareMathSymbol{\shortminus}{\mathbin}{AMSa}{"39}
\definecolor{Green}  {rgb}{0.10,0.70,0.10} 
\definecolor{Orange} {rgb}{1.00,0.50,0.15} 
\definecolor{Red}    {rgb}{0.90,0.00,0.12} 
\definecolor{Purple} {rgb}{0.50,0.25,0.55} 
\definecolor{Turque} {rgb}{0.00,0.65,0.85} 
\definecolor{Blue}   {rgb}{0.00,0.00,1.00} 
\definecolor{Magenta}{rgb}{1.00,0.00,1.00} 
\definecolor{Gold}   {rgb}{1.00,0.75,0.25} 
\definecolor{Seaweed}{rgb}{0.01,0.24,0.09} 
\definecolor{Brown}  {rgb}{0.43,0.26,0.32} 
\definecolor{grey1}  {rgb}{0.20,0.20,0.20} 
\definecolor{grey2}  {rgb}{0.40,0.40,0.40} 
\definecolor{grey3}  {rgb}{0.60,0.60,0.60} 
\definecolor{grey4}  {rgb}{0.80,0.80,0.80} 
\definecolor{grey5}  {rgb}{0.90,0.90,0.90} 
\def\C#1#2{{\ifcase#1\or
             \color{Green}\or \color{Orange}\or \color{Red}\or
              \color{Purple}\or \color{Turque}\or \color{Blue}\or
               \color{Magenta}\or \color{Gold}\or \color{Seaweed}\or
                \color{Brown}\or\color{grey1}\or\color{grey2}\or
                 \color{grey3}\else\color{grey4}\fi#2}}

\definecolor{Slate} {rgb}{0.00,0.45,0.55}

\usepackage{color} 
\usepackage{listings} 
\usepackage{setspace} 
 
\definecolor{Code}{rgb}{0,0,0} 
\definecolor{Decorators}{rgb}{0.5,0.5,0.5} 
\definecolor{Numbers}{rgb}{0.5,0,0} 
\definecolor{MatchingBrackets}{rgb}{0.25,0.5,0.5} 
\definecolor{Keywords}{rgb}{0,0,1} 
\definecolor{self}{rgb}{0,0,0} 
\definecolor{Strings}{rgb}{0,0.63,0} 
\definecolor{Comments}{rgb}{0,0.63,1} 
\definecolor{Backquotes}{rgb}{0,0,0} 
\definecolor{Classname}{rgb}{0,0,0} 
\definecolor{FunctionName}{rgb}{0,0,0} 
\definecolor{Operators}{rgb}{0,0,0} 
\definecolor{Background}{rgb}{0.98,0.98,0.98}

\lstdefinelanguage{Python}{ 
numbers=left, 
numberstyle=\footnotesize, 
numbersep=1em, 
xleftmargin=1em, 
framextopmargin=2em, 
framexbottommargin=2em, 
showspaces=false, 
showtabs=false, 
showstringspaces=false, 
frame=l, 
tabsize=4, 
basicstyle=\ttfamily\small\setstretch{1}, 
backgroundcolor=\color{Background}, 
commentstyle=\color{Comments}\slshape, 
stringstyle=\color{Strings}, 
morecomment=[s][\color{Strings}]{"""}{"""}, 
morecomment=[s][\color{Strings}]{'''}{'''}, 
morekeywords={import,from,class,def,for,while,if,is,in,elif,else,not,and,or,print,break,continue,return,True,False,None,access,as,,del,except,exec,finally,global,import,lambda,pass,print,raise,try,assert}, 
keywordstyle={\color{Keywords}\bfseries}, 
morekeywords={[2]@invariant,pylab,numpy,np,scipy}, 
keywordstyle={[2]\color{Decorators}\slshape}, 
emph={self}, 
emphstyle={\color{self}\slshape}, 
} 

\def\rD{{\rm D}}
\def\rI{{\rm I}}
\def\rJ{{\rm J}}
\def\rK{{\rm K}}
\def\rL{{\rm L}}
\def\rR{{\rm R}}

\def\fracm#1#2{\hbox{\large{${\frac{{#1}}{{#2}}}$}}}

\def\be{\begin{equation}}
\def\ee{\end{equation}}
\newcommand{\bea}{\begin{eqnarray}}
\newcommand{\eea}{\end{eqnarray}}
\newcommand{\ena}{\end{eqnarray}}

\def\pp{{\mathchoice
          {
              \kern 1pt%
              \raise 1pt
              \vbox{\hrule width5pt height0.4pt depth0pt
                    \kern -2pt
                    \hbox{\kern 2.3pt
                          \vrule width0.4pt height6pt depth0pt
                          }
                    \kern -2pt
                    \hrule width5pt height0.4pt depth0pt}%
                    \kern 1pt
           }
            {
              \kern 1pt%
              \raise 1pt
              \vbox{\hrule width4.3pt height0.4pt depth0pt
                    \kern -1.8pt
                    \hbox{\kern 1.95pt
                          \vrule width0.4pt height5.4pt depth0pt
                          }
                    \kern -1.8pt
                    \hrule width4.3pt height0.4pt depth0pt}%
                    \kern 1pt
            }
            {
              \kern 0.5pt%
              \raise 1pt
              \vbox{\hrule width4.0pt height0.3pt depth0pt
                    \kern -1.9pt  
                    \hbox{\kern 1.85pt
                          \vrule width0.3pt height5.7pt depth0pt
                          }
                    \kern -1.9pt
                    \hrule width4.0pt height0.3pt depth0pt}%
                    \kern 0.5pt
            }
            {
              \kern 0.5pt%
              \raise 1pt
              \vbox{\hrule width3.6pt height0.3pt depth0pt
                    \kern -1.5pt
                    \hbox{\kern 1.65pt
                          \vrule width0.3pt height4.5pt depth0pt
                          }
                    \kern -1.5pt
                    \hrule width3.6pt height0.3pt depth0pt}%
                    \kern 0.5pt
            }
        }}

\def\mm{{\mathchoice
                       {
                             \kern 1pt
               \raise 1pt    \vbox{\hrule width5pt height0.4pt depth0pt
                                  \kern 2pt
                                  \hrule width5pt height0.4pt depth0pt}
                             \kern 1pt}
                       {
                            \kern 1pt
               \raise 1pt \vbox{\hrule width4.3pt height0.4pt depth0pt
                                  \kern 1.8pt
                                  \hrule width4.3pt height0.4pt depth0pt}
                             \kern 1pt}
                       {
                            \kern 0.5pt
               \raise 1pt
                            \vbox{\hrule width4.0pt height0.3pt depth0pt
                                  \kern 1.9pt
                                  \hrule width4.0pt height0.3pt depth0pt}
                            \kern 1pt}
                       {
                           \kern 0.5pt
             \raise 1pt  \vbox{\hrule width3.6pt height0.3pt depth0pt
                                  \kern 1.5pt
                                  \hrule width3.6pt height0.3pt depth0pt}
                           \kern 0.5pt}
                       }}

\def\ad{{\kern0.5pt
                   \alpha \kern-5.05pt \raise5.8pt\hbox{$\textstyle.$}\kern
0.5pt}}

\def\bd{{\kern0.5pt
                   \beta \kern-5.05pt \raise5.8pt\hbox{$\textstyle.$}\kern
0.5pt}}

\def\qd{{\kern0.5pt
                   q \kern-5.05pt \raise5.8pt\hbox{$\textstyle.$}\kern
0.5pt}}
\def\Dot#1{{\kern0.5pt
     {#1} \kern-5.05pt \raise5.8pt\hbox{$\textstyle.$}\kern
0.5pt}}

\catcode`@=11
\def\un#1{\relax\ifmmode\@@underline#1\else
        $\@@underline{\hbox{#1}}$\relax\fi}
\catcode`@=12

\def\a{\alpha}
\def\b{\beta}
\def\c{\chi}
\def\d{\delta}
\def\e{\epsilon}

\def\i{\iota}
\def\j{\psi}
\def\k{\kappa}
\def\l{\lambda}

\def\o{\omega}

\def\r{\rho}
\def\s{\sigma}
\def\t{\tau}

\def\z{\zeta}

\def\L{\Lambda}
\def\O{\Omega}
\def\P{\Pi}

\def\S{\Sigma}

\def\ca{{\cal A}}

\def\cf{{\cal F}}

\def\dslash{\not{\hbox{\kern-2pt $\partial$}}}
\def\Dslash{\not{\hbox{\kern-4pt $D$}}}
\def\pslash{\not{\hbox{\kern-2.3pt $p$}}}
 \newtoks\slashfraction
 \slashfraction={.13}
 \def\slash#1{\setbox0\hbox{$ #1 $}
 \setbox0\hbox to \the\slashfraction\wd0{\hss \box0}/\box0 }
 
\def\kcr{{\hbox{\ro \char'170}}}                
\def\ktl{{\hbox{\ro \char'170}}}        
\def\ktr{{\hbox{\ro \char'170}}}        
\def\kbl{{\hbox{\ro \char'170}}}        
\def\kbr{{\hbox{\ro \char'170}}}        

\def\plpl{\raise-2pt\hbox{$\raise3pt\hbox{$_+$}\hskip-6.67pt\raise0.0pt
\hbox{$^+$}\hskip 0.01pt$}}
\def\mimi{\raise-2pt\hbox{$\raise3pt\hbox{$_-$}\hskip-6.67pt\raise0.0pt
\hbox{$^-$}\hskip 0.01pt$}} 

\def\bo{{\raise.15ex\hbox{\large$\Box$}}}               
\def\TH{{\raise.2ex\hbox{$\displaystyle \bigodot$}\mskip-4.7mu \llap H \;}}
\def\face{{\raise.2ex\hbox{$\displaystyle \bigodot$}\mskip-2.2mu \llap {$\ddot
        \smile$}}}                                      

\def\dt#1{\on{\hbox{\bf .}}{#1}}                
\def\Dot#1{\dt{#1}}

\def\Tilde#1{\widetilde{#1}}                    
\def\Hat#1{\widehat{#1}}                        
\def\leftrightarrowfill{$\mathsurround=0pt \mathord\leftarrow \mkern-6mu
        \cleaders\hbox{$\mkern-2mu \mathord- \mkern-2mu$}\hfill
        \mkern-6mu \mathord\rightarrow$}
\def\dvec#1{\vbox{\ialign{##\crcr
        \leftrightarrowfill\crcr\noalign{\kern-1pt\nointerlineskip}
        $\hfil\displaystyle{#1}\hfil$\crcr}}}           
\def\dt#1{{\buildrel {\hbox{\LARGE .}} \over {#1}}}     

\def\fracm#1#2{\hbox{\large{${\frac{{#1}}{{#2}}}$}}}
\def\sfrac#1#2{{\vphantom1\smash{\lower.5ex\hbox{\small$#1$}}\over
        \vphantom1\smash{\raise.4ex\hbox{\small$#2$}}}} 
\def\bfrac#1#2{{\vphantom1\smash{\lower.5ex\hbox{$#1$}}\over
        \vphantom1\smash{\raise.3ex\hbox{$#2$}}}}       
\def\afrac#1#2{{\vphantom1\smash{\lower.5ex\hbox{$#1$}}\over#2}}    

\let\bm\relax
\newcommand{\bm}[1]{{\boldsymbol{#1}}}
\usepackage{float}
\def\ad{{\dot{\alpha}}}
\def\bd{{\dot{\beta}}}

 \font\rOpe=cmsy10                        
 \def\ktl{{\hbox{\rOpe\char'170}}}        
 \def\kbl{{\hbox{\rOpe\char'170}}}        
 \def\kcr{{\reflectbox{\rOpe\char'170}}}        
 \def\ktr{{\reflectbox{\rOpe\char'170}}}        
 \def\kbr{{\reflectbox{\rOpe\char'170}}}        
 \def\Border{\vbox{\hsize0pt
        \setlength{\unitlength}{1mm}
        \newcount\xco
        \newcount\yco
        \xco=-21
        \yco=12
        \begin{picture}(0,0)(-7.5,0)
        \put(\xco,\yco){$\ktl$}
        \advance\yco by-1
        {\loop
        \put(\xco,\yco){$\kcr$}
        \advance\yco by-2
        \ifnum\yco>-240
        \repeat
        \put(\xco,\yco){$\kbl$}}
        \xco=170
        \yco=12
        \put(\xco,\yco){$\ktr$}
        \advance\yco by-1
        {\loop
        \put(\xco,\yco){$\kcr$}
        \advance\yco by-2
        \ifnum\yco>-240
        \repeat
        \put(\xco,\yco){$\kbr$}}
        \put(-19.5,13){\scalebox{.6065}{%
         University of Maryland Center for String and Particle  Theory \&\ Physics Department%
        |University of Maryland Center for String and Particle  Theory \&\ Physics Department}}
        \put(-19.5,-241.5){\scalebox{.5835}{%
         ****University of Maryland * Center for String and
         Particle  Theory* Physics Department****University of Maryland *Center
        for String and Particle  Theory* Physics Department}}
        \end{picture}
        \par\vskip-8mm}}
\definecolor{UMred}{rgb}{.9,.05,.2}
\definecolor{HUblue}{rgb}{.0,.3,.7}

\definecolor{Red}    {rgb}{0.90,0.00,0.12} 
\definecolor{Blue}   {rgb}{0.00,0.00,1.00} 
\definecolor{Green}  {rgb}{0.10,0.70,0.10} 
\definecolor{Turque} {rgb}{0.00,0.65,0.85} 
\definecolor{Orange} {rgb}{1.00,0.50,0.15} 
\definecolor{Magenta}{rgb}{1.00,0.00,1.00} 
\definecolor{Gold}   {rgb}{1.00,0.75,0.25} 
\definecolor{Seaweed}{rgb}{0.01,0.24,0.09} 
\definecolor{Purple} {rgb}{0.50,0.25,0.55} 
\definecolor{Brown}  {rgb}{0.43,0.26,0.32} 
\definecolor{grey1}  {rgb}{0.20,0.20,0.20} 
\definecolor{grey2}  {rgb}{0.40,0.40,0.40} 
\definecolor{grey3}  {rgb}{0.60,0.60,0.60} 
\definecolor{grey4}  {rgb}{0.80,0.80,0.80} 
\definecolor{grey5}  {rgb}{0.90,0.90,0.90} 
\def\C#1#2{{\ifcase#1\or
             \color{Red}\or \color{Green}\or \color{Blue}\or\
              \color{Turque}\or \color{Orange}\or \color{Magenta}\or 
               \color{Gold}\or \color{Seaweed}\or \color{Purple}\or
                \color{Brown}\or\color{grey1}\or\color{grey2}\or
                 \color{grey3}\else\color{grey4}\fi#2}}

\definecolor{Slate} {rgb}{0.00,0.45,0.55}

\newdimen\parshift\parshift=\parindent
\catcode`@=11
 \long\def\@footnotetext#1{\insert\footins{\reset@font\footnotesize
           \interlinepenalty\interfootnotelinepenalty\splittopskip%
            \footnotesep\splitmaxdepth\dp\strutbox\floatingpenalty\@MM%
             \hsize\columnwidth\addtolength{\hsize}{-2\parindent}
              \@parboxrestore\protected@edef\@currentlabel%
              {\csname p@footnote\endcsname\@thefnmark}%
                \color@begingroup%
                 \@makefntext{\rule\z@\footnotesep\ignorespaces#1%
                  \@finalstrut\strutbox}%
                \color@endgroup}}
 \long\def\@makefntext#1{\hglue\parshift%
           \vbox{\noindent\baselineskip=11pt plus.5pt minus.5pt\hb@xt@0em{\hss\@makefnmark\kern1pt}#1}}
\catcode`@=12

\newskip\humongous \humongous=0pt plus 1000pt minus 1000pt
\def\caja{\mathsurround=0pt}
\def\eqalign#1{\,\vcenter{\openup2\jot \caja
        \ialign{\strut \hfil$\displaystyle{##}$&$
        \displaystyle{{}##}$\hfil\crcr#1\crcr}}\,}
\newif\ifdtup

\makeatletter
\def\section{\@startsection{section}{1}{\z@}
        {3ex plus-1ex minus-.2ex}{1pt plus1pt}{\large\sf\bfseries\boldmath}}
\def\subsection{\@startsection{subsection}{2}{\z@}
         {1.5ex plus-1ex minus-.2ex}{0.01pt plus1pt}{\sf\slshape}}
\def\subsubsection{\@startsection{subsubsection}{3}{\z@}
          {1.5ex plus-1ex minus-.2ex}{0.01pt plus0.2pt}{\sf\boldmath}}
\def\paragraph{\@startsection{paragraph}{4}{\z@}
           {.75ex \@plus.5ex \@minus.2ex}{-2mm}{\sf\bfseries\boldmath}}
\makeatother

 \allowdisplaybreaks
 \seceq

\usepackage{lipsum}
\usepackage{listings}
\definecolor{MyDarkGreen}{rgb}{0.0,0.4,0.0} 
\usepackage[enableskew,vcentermath]{youngtab}

\definecolor{Hey}{rgb}{.9,.05,.4}
\definecolor{orange}{rgb}{1,.5,0}
\definecolor{plum}{rgb}{.4,0,.6}
\definecolor{R}{rgb}{1,0,0}
\definecolor{G}{rgb}{0.1,0.7,0}
\definecolor{B}{rgb}{0,0,1}
\begin{document}
\newcommand{\numberthis}{\addtocounter{equation}{1}\tag{\theequation}}

\thispagestyle{empty}
\noindent{\small
\hfill{  \\ 
$~~~~~~~~~~~~~~~~~~~~~~~~~~~~~~~~~~~~~~~~~~~~~~~~~~~~~~~~~~~~~~~~~$
$~~~~~~~~~~~~~~~~~~~~~~~~~~~~~~~~~~~~~~~~~~~~~~~~~~~~~~~~~~~~~~~~~$
{}
}}
\vspace*{0mm}
\begin{center}
{\large \bf
Unfolded Adinkra Properties of 
 Supermultiplets (I)\\[2pt]
}   \vskip0.3in
{\large {
$~~~~~~~~~~~~~$
Aleksander J.\ Cianciara\footnote{acianciara@princeton.edu}$^{,a,b,c,d}$,
S.\ James Gates, Jr.\footnote{gatess@umd.edu}$^{,e}$, 
\newline
$~~~~~~~~~~~~~$
Youngik (Tom) Lee\footnote{youngik${}_-$lee@alumni.brown.edu}${}^{,c,d}$,
Ethan T. Levy\footnote{elevy127@umd.edu}$^{,e}$,
\newline
$~~~~~~~~~~~~~~~~$
Tarek O. Razzaz\footnote{tarek${}_-$razzaz@brown.edu}${}^{,c,d}$, and
Jacob Richardson\footnote{jacob${}_-$richardson@brown.edu, ${}^*$Affiliation began September 6, 2023}$^{,c^*,d^*,e}$ 
$~~~~~~$
\newline
}}
\\*[4mm]
\emph{
\centering
$^{a}$Joseph Henry Laboratories, Princeton University, 
Princeton, NJ 08544, USA, \\
$^{b}$Institute for Advanced Study,
Princeton, NJ 08540, USA, \\[12pt]
$^{c}$Brown University, Department of Physics,
\\[1pt]
Box 1843, 182 Hope Street, Barus \& Holley 545,
Providence, RI 02912, USA,
\\[12pt]
$^{d}$Brown Center for Theoretical Physics, 
\\[1pt]
340 Brook Street, Barus Hall,
Providence, RI 02912, USA,
\\[12pt] and \\[4pt]
$^{e} $Department of Physics, University of Maryland,
\\[1pt]
College Park, MD 20742-4111, USA
}
 \\*[10mm]
{ ABSTRACT}\\[4mm]
\parbox{142mm}{\parindent=2pc\indent\baselineskip=14pt plus1pt
Adinkra networks arise in the Carroll limit of supersymmetric QFT.  Extensions of adinkras 
that are infinite dimensional graphs have never previously been discussed in the 
literature.  We call these ``infinite unfolded'' adinkras and study the properties of 
their realization on familiar 4D, $\cal N$ = 1 supermultiplets.  A new feature in 
``unfolded'' adinkras is the appearance of quantities whose actions resemble BRST operators 
within Verma-like modules. New ``net-centric" quantities ${\Tilde \chi}_{(1)}$ and 
${\Tilde \chi}_{(2)}$ are introduced, which along with quantity $\chi_{\rm o}$, 
describe distinctions between familiar supermultiplets in 
4D, $\cal N $ = 1 theories.  A previously unobserved property in all adinkras that we call ``adinkra vorticity" is noted.
}

 \end{center}
\vfill
\vfill
\clearpage
%

%
\tableofcontents
\newpage
\section{Introduction}

In a 1965 publication by J.-M. L\' evy-Leblond, \cite{CRT} considerations of novel non-relativistic limits of the Poincar\' e Group were introduced into the literature.
One of these limits ($c$ $\to$ 0) is often referred to by the
appellation, the ``Carrollian limit."  In a work by K.\ Koutrolikos and M.\
Najafizadeh \cite{K&N}, the observation appears that the Carrollian limit of supersymmetric multiplets leads to adinkra graphs \cite{Adnk1} where
network theory and graph theory are powerful tools for the discovery and study of ``hidden'' mathematical properties such as symmetries and other unsuspected mathematical structures. For example, it was discovered 
\cite{codes1,codes2,codes3} that irreducible representations of 4D, ${\cal N} = 1$ supersymmetry which propagates physical degrees of freedom are {\em {all}} 
related to the presence of classical error-correcting codes.

Adinkras constitute a class of bipartite graphs whose properties as networks
were proposed to contain information about the gauge-invariant component
fields of supermultiplets along with information about the orbits of these fields
under the actions of supercharges.  It has been proposed that this information
is the key to the discovery of structures that can play the roles of familiar
structures from the representation theory of compact Lie Algebras. For the latter, the Jordan-Chevalley decomposition provides access to maximal torus, roots, weights, and raising (lowering) operators.  The study of adinkras has given rise to two additional types of graphs that appear important in this effort.
 
The first of these is related to a well-known set of structures in mathematics.   
Associated with the Permutation Groups of arbitrary order, there are graphs known 
as ``permutahedra'' or ``permutahedrons \cite{perms1,perms2,perms3,perms4}.''  All of the nodes in permutahedrons may be 
regarded as being members of a single class of objects.  Hence these are {\em {not}} 
bipartite graphs.  From our efforts studying these objects, it is clear that it is 
possible to define quantities (purely associated with intrinsic properties of the 
permutation operators) that play the role of weights.  This clears the way to a 
definition of raising and lowering operators that appear identical to the 
representation theory of compact Lie Algebras.
 
A second type of bipartite graph that extends adinkras are graphs where the links and 
the nodes carry representations of the A-, B-, and D- type compact Lie algebras in the 
classification scheme of Cartan.  This class of graphs has been given the name of 
``adynkras'' due to the fact that the information carried by the nodes is most 
conveniently represented as Dynkin Labels (or equivalently as Young Tableaux). 
A recent example of the potency of this approach was provided by the first physics 
literature demonstration of a ``supercurrent'' for eleven-dimensional supergravity 
theory.  In this construction, a conformal graviton representation was found at the 
sixteenth level of an 11D, $\cal N$ = 1 scalar superfield. The arguments
that lead to this result echo the original discovery of the 4D, $\cal N$ = 1 supergravity supercurrent superfield... in either case there is no a priori
need to consider geometry in superspace.
Thus, the graph-based approach leverages techniques from algebraic geometry, graph theory, and IT coding to provide new insights into the structure of supersymmetric theories of particles, fields, and strings. 

In this work, we discuss a third type of extension to the concept of the original proposed structure of adinkras.  We will call these newest types of extensions ``unfolded adinkras.''  Unfolded adinkras are, like the original adinkras, bipartite graphs. {But unlike the original adinkras, they can be
infinite and thus have structures that resemble unidirectional Verma Modules. Additionally, unfolded adinkras strictly ascend by level. They must be read upwards starting with the bottom-most fields and increasing in level, adding a temporal derivative after each supersymmetric transformation.} In a sense these are not new as within the DFGHIL collaboration, that first studied adinkras, members were well aware of the existence of unfolded adinkras in the
internal deliberations \cite{KIP} of its research.

\section{Unfolded Adinkras}

There is a hitherto {\em {largely unexplored}} way to construct adinkras, which we shall call the ``unfolded adinkra''. The unfolded adinkra shows the field relations directly as a connection between bosonic and fermionic fields with emphasis on engineering dimensions \cite{3}.
In this section, let us use the 4D, ${\cal N}$ = 1 chiral, vector, and tensor multiplets as toy models to draw the unfolded adinkra for each supermultiplet.

\subsection{Valise Adinkras}
By the ``Hanging Garden Theorem'' \cite{HGT}, any arbitrary adinkra can be redrawn such that all bosons appear at a common altitude (or height) and all fermions appear at a common height that differs from that of the bosons.
Such an adinkra is called a ``valise adinkra'' \cite{adnkVaL}. In valise adinkras, all bosons have the same engineering dimension, while all fermions share the same engineering dimension. Through a series of operations, any adinkra can be put into valise form. 
However, in the event that the L matrices are non-orthogonal, then the operations required to change the adinkra to valise form will involve redefining fields (linear combinations of fields as can be seen in Chapter 4). That means the valise adinkra won’t always contain the original fields.
Any adinkra that is {\em {not}} a valise adinkra may be referred to as an ``unfolded adinkra''.

\begin{figure}[H]
	\centering	\includegraphics[scale=0.5]{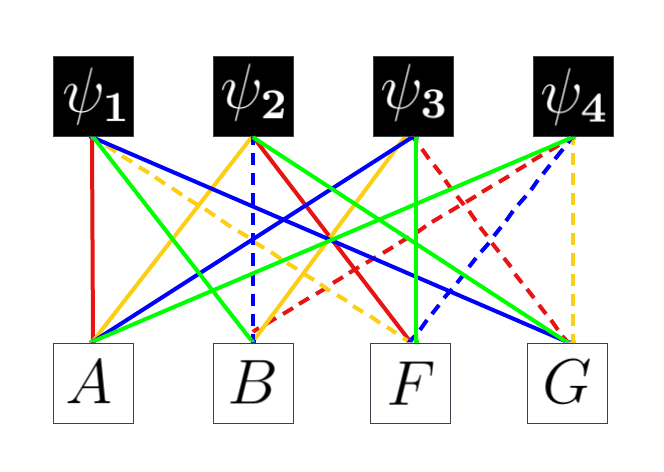}
	\caption{A valise adinkra for the 4D, $\mathcal{N}=1$ chiral multiplet.}
	\label{fig001}
\end{figure}
For example, let us consider the valise adinkra for the chiral supermultiplet (Fig. \ref{fig001}). 
Here, the black nodes are fermionic fields, and white nodes are bosonic fields. The dotted edges between the nodes represent the negative coefficients that appear after applying the supercovariant derivative on lower-height positioned fields, and the solid edges represent the positive coefficients.
Here, red, yellow, blue, and green indicate the $\rD_1, \rD_2, \rD_3, \rD_4$ cases respectively.

\begin{figure}[H]
	\centering	\includegraphics[scale=0.5]{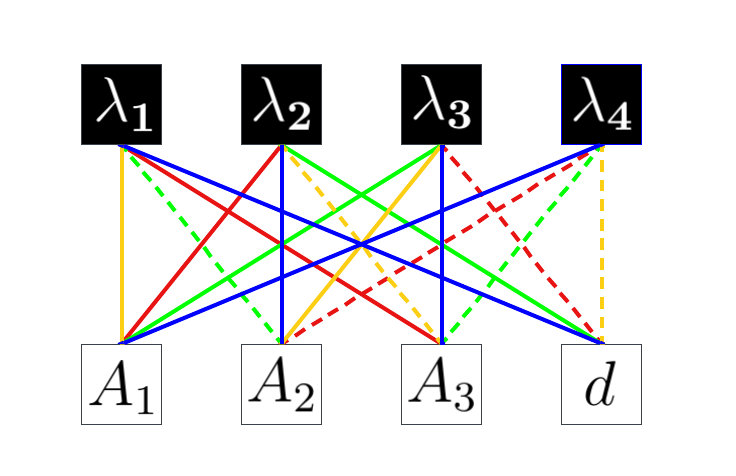}
	\caption{The valise adinkra for the 4D, $\mathcal{N}=1$ vector multiplet.}
	\label{fig003}
\end{figure}

\begin{figure}[H]
	\centering	\includegraphics[scale=0.5]{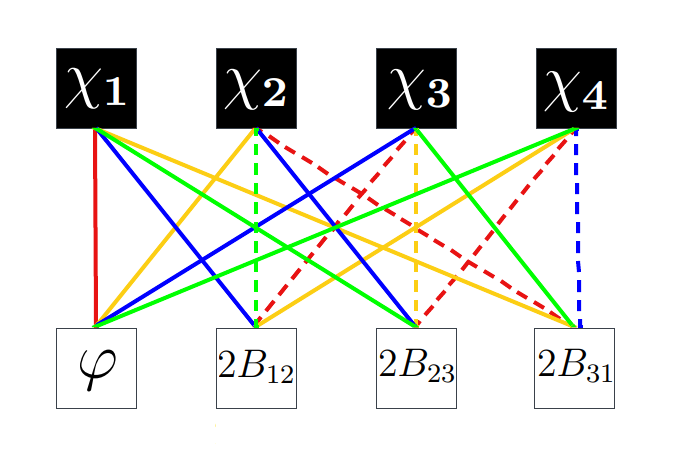}
	\caption{The valise adinkra for the 4D, $\mathcal{N}=1$ tensor multiplet.}
	\label{fig004}
\end{figure}

In the same way, Fig. \ref{fig003} and \ref{fig004} show the valise adinkra for 4D, $\mathcal{N}=1$ vector and tensor supermultiplets.
 
\subsection{Unfolded Adinkras}
The engineering dimensions of each field are determined by their appearances in an action. By comparing the engineering dimensions of each field, we can ``unfold" the valise adinkra into three layers as shown below in Fig. \ref{fig002}, where the auxiliary bosonic fields $F$ and $G$ have the highest engineering dimension and therefore the highest height in the unfolded adinkra.

\begin{figure}[H]
	\centering	\includegraphics[scale=0.45]{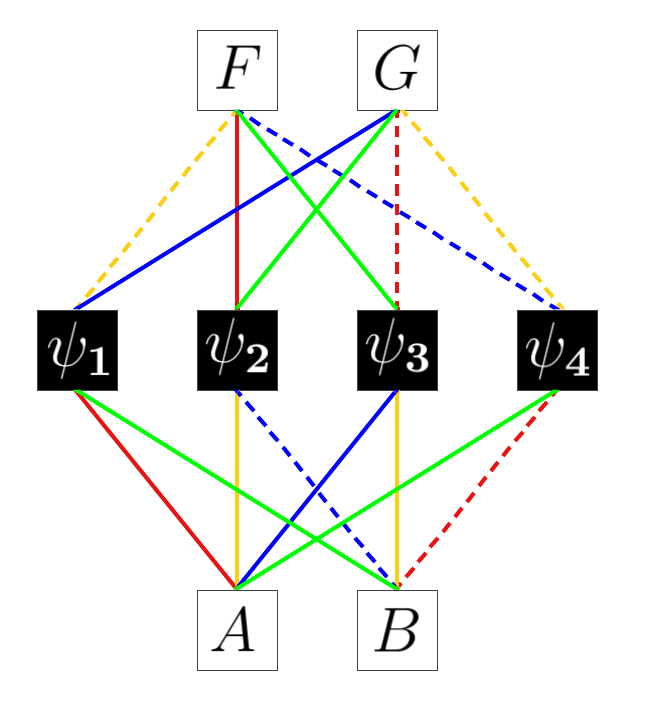}
	\caption{An unfolded adinkra for the 4D, $\mathcal{N}=1$ chiral multiplet.}
	\label{fig002}
\end{figure}

\begin{figure}[H]
	\centering	\includegraphics[scale=0.45]{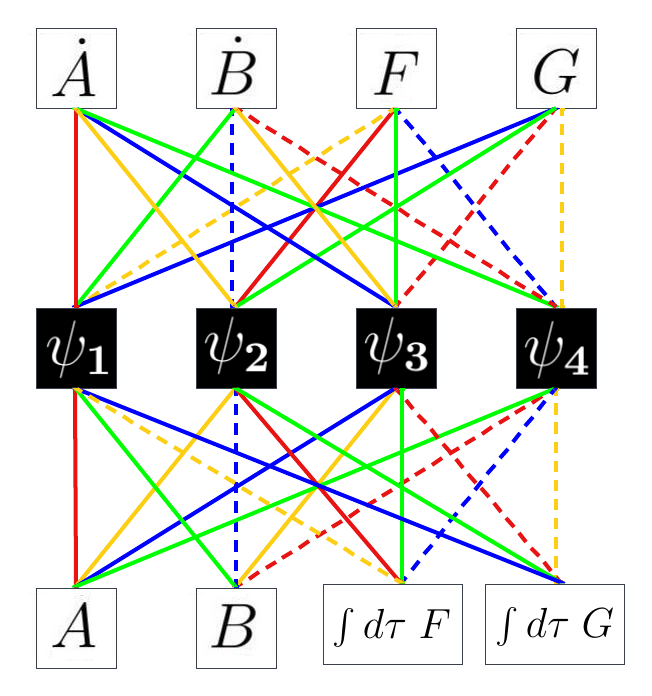}	\caption{A 4D, $\mathcal{N}=1$ chiral multiplet unfolded adinkra with solely ascending edges.}
	\label{fig012}
\end{figure}

However the edges in Fig. \ref{fig002} are bidirectional, which means the action of the supercovariant derivatives corresponds to either upward or downward motion for bosons and fermions. By adding derivatives and integrals of the fields as new nodes, we can define the action of the supercovariant derivative operators work to solely an upward motion in the Adinkra as Fig. \ref{fig012}

In sections \ref{ch2.1}-\ref{ch2.3}, we show the unfolded adinkra for 4D, $\mathcal{N}=1$ chiral, vector, and tensor supermultiplets, along with some additional details regarding their structure.

\subsection{Garden Algebra and Unfolded Adinkra: Chiral Supermultiplet}
\label{ch2.1}
The 4D, $\cal N$ = 1 chiral multiplet contains a scalar $A$, a pseudoscalar $B$, a Majorana fermion $\psi_a$, a scalar auxiliary field $F$, and a pseudoscalar auxiliary field $G$.  

Additionally, the Lagrangian for the chiral supermultiplet is \cite{8}

\begin{equation}
\mathcal{L}_{CS}=
-\frac{1}{2}(\partial_\mu A)(\partial^\mu A)
-\frac{1}{2}(\partial_\mu B)(\partial^\mu B)
+i\frac{1}{2}(\gamma^\mu)^{ab}\psi_a(\partial_\mu\psi_b)
+\frac{1}{2}F^2
+\frac{1}{2}G^2 ~~~.
\end{equation}
with transformations under the supercovariant derivative for each field given as \cite{3}

\begin{equation}
\eqalign{
{\rm D}_a A ~&=~ \psi_a  {~~~~~~~~~~~~~~~~~}~~~, {~~~~~~~~~~~~~~}
{\rm D}_a B ~=~ i \, (\gamma^5){}_a{}^b \, \psi_b  ~~~~~~~~~~, \cr
{\rm D}_a \psi_b ~&=~ i\, (\gamma^\mu){}_{a \,b}\,  \partial_\mu A 
\,-\,  (\gamma^5\gamma^\mu){}_{a \,b} \, \partial_\mu B \,-\, i \, C_{a\, b} 
\,F  \,+\,  (\gamma^5){}_{ a \, b} G  ~~~~~~~~, \cr
{\rm D}_a F ~&=~  (\gamma^\mu){}_a{}^b \, \partial_\mu \, \psi_b  {~~~~~~}~~~, {~~~~~~~~~~~~~~}
{\rm D}_a G ~=~ i \,(\gamma^5\gamma^\mu){}_a{}^b \, \partial_\mu \,  
\psi_b  ~~~.
} \label{chi1}
\end{equation}

so that the supersymmetric covariant derivative ${\rm D}_a$ action on the Lagrangian yields:
\[\begin{array}{l}
{\rm D}{}_a \mathcal{L}_{CS}= \partial^\mu \mathcal{J}_{\mu a}{}^{(CS)} 
\numberthis\end{array} ~~~,\]
where the supercurrent, \(\mathcal{J}_{\mu a}{}^{(CS)}\), is given as,
\[\begin{array}{l}
\mathcal{J}_{\mu a}{}^{(CS)} =-\frac{1}{2} \, [\, 
(\gamma_\mu\gamma^\nu)_a{}^{c}( \partial_\nu A )
+ i (\gamma^5\gamma_\mu\gamma^\nu)_a{}^{c}( \partial_\nu B )
- (\gamma_\mu)_a{}^{c} F
- i(\gamma^5\gamma_\mu)_a{}^{c}G \, ] \, \psi_c  ~~~.
\numberthis\end{array} ~~~\]

\begin{figure}[H]
\centering	\includegraphics[scale=0.35]{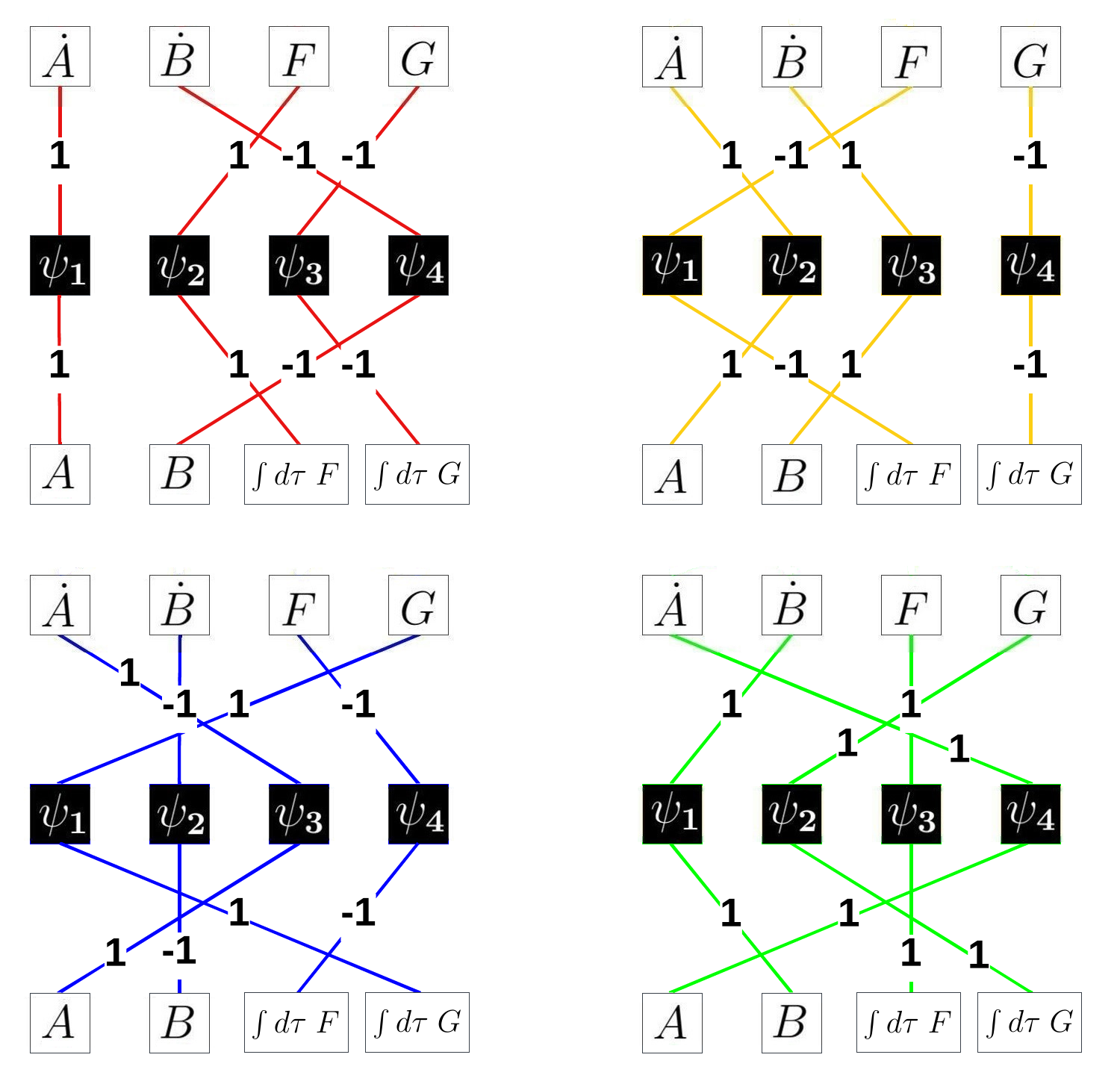}
	\caption{An unfolded adinkra for the 4D $\mathcal{N}=1$ chiral multiplet with ascending edges.}
	\label{fig1a}
\end{figure}

Based on Eq. (\ref{chi1}) we can apply a 0-brane reduction process \cite{4} for each field and draw the relations between each field as a graph, as seen in Fig. \ref{fig1a}.
Here, the edges between the nodes represent the coefficients that appear after applying the supercovariant derivative on lower height-positioned fields.

Furthermore, we can define

\[ \Phi_{i} = \left[\begin{array}{c}
    A  \\ 
    B   \\
    F  \\
    G  
\end{array}\right] ~~~,~~~
i\Psi{}_{\hat k} = \left[\begin{array}{c}
    \psi_1  \\ 
    \psi_2  \\
    \psi_3  \\
    \psi_4  
\end{array}\right]  ~~~,~~~
{\rm D}_{\rm I} = \left[\begin{array}{c}
    {\rm D}_1  \\ 
    {\rm D}_2  \\     
    {\rm D}_3  \\ 
    {\rm D}_4 
\numberthis\end{array}\right] ~~~.
\]

and in terms of these new quantities, Eq. (\ref{chi1}) can be written as
\[ 
{{\rm D}_{\rm I} \Phi{}_i = i\, [ {\bm {\rL}}{}_{\rI}^{(1)}]_{i}{}^{\hat{k}} \, \Psi{}_{\hat k} ~+~ i\, [ {\bm {\rL}}{}_{\rI}^{(2)}]_{i}{}^{\hat{k}}
\, \frac{d}{d \tau} \Psi{}_{\hat k} ~~~,
\numberthis }
\label{CSX1}
\]

\[ 
{{\rm D}_{\rm I} \Psi{}_{\hat k} = [ {\bm {\rR}}{}_{\rI}^{(0)}]_{\hat
k}{}^i \,
\Phi{}_{i }
~+~ [ {\bm {\rR}}{}_{\rI}^{(1)}]_{\hat k}{}^i \, \frac{d ~}{d \tau}
\Phi{}_{i} ~~~.
\numberthis}
\label{CSX2}
\]
where we have introduced four constant matrices (actually tensors) denoted by $[ {\bm {\rL}}{}_{\rI}^{(1)}]_{i}{}^{\hat{k}}$, $[ {\bm {\rL}}{}_{\rI}^{(2)}]_{i}{}^{\hat{k}}$,
$[ {\bm {\rR}}{}_{\rI}^{(0)}]_{\hat k}{}^i$ and $[ {\bm {\rR}}{}_{\rI}^{(1)}]_{\hat k}{}^i$.

As we briefly mentioned in section 2.2, by including the derivatives of the $A$ and $B$ fields and the integrals of the
$ F$ and $ G$ fields, we can create unidirectional, infinite-dimensional ascending-link only adinkras which are also ``unfolded adinkras.''

We can ``disaggregate'' the adinkra shown in Fig. \ref{fig012} into its four distinct color link components, 

These $\bm \rL$ and $\bm \rR$ matrices are given explicitly as

\[[{{\bm \rL}_1^{(1)}}]_{i}{}^{\hat k} = \left[\begin{array}{cccc}
    1 & 0 & 0 & 0 \\ 
    0 & 0 & 0 & -1 \\ 
    0 & 0 & 0 & 0 \\ 
    0 & 0 & 0 & 0
\end{array}\right]
,\quad
[{{\bm \rL}_1^{(2)}}]_{i}{}^{\hat k} = \left[\begin{array}{cccc}
    0 & 0 & 0 & 0 \\ 
    0 & 0 & 0 & 0 \\ 
    0 & 1 & 0 & 0 \\ 
    0 & 0 & -1 & 0
\numberthis
\label{eq211}
\end{array}\right],
\]

\[[{{\bm \rL}_2^{(1)}}]_{i}{}^{\hat k} = \left[\begin{array}{cccc}
    0 & 1 & 0 & 0 \\ 
    0 & 0 & 1 & 0 \\ 
    0 & 0 & 0 & 0 \\ 
    0 & 0 & 0 & 0
\end{array}\right] 
,\quad
[{{\bm \rL}_2^{(2)}}]_{i}{}^{\hat k} = \left[\begin{array}{cccc}
    0 & 0 & 0 & 0 \\ 
    0 & 0 & 0 & 0 \\ 
    -1 & 0 & 0 & 0 \\ 
    0 & 0 & 0 & -1
\numberthis
\label{eq4-10}
\end{array}\right],
\]

\[[{{\bm \rL}_3^{(1)}}]_{i}{}^{\hat k} = \left[\begin{array}{cccc}
    0 & 0 & 1 & 0 \\ 
    0 & -1 & 0 & 0 \\ 
    0 & 0 & 0 & 0 \\ 
    0 & 0 & 0 & 0
\end{array}\right] 
,\quad
[{{\bm \rL}_3^{(2)}}]_{i}{}^{\hat k} = \left[\begin{array}{cccc}
    0 & 0 & 0 & 0 \\ 
    0 & 0 & 0 & 0 \\ 
    0 & 0 & 0 & -1 \\ 
    1 & 0 & 0 & 0
\numberthis
\label{eq4-10}
\end{array}\right],
\]

\[[{{\bm \rL}_4^{(1)}}]_{i}{}^{\hat k} = \left[\begin{array}{cccc}
    0 & 0 & 0 & 1 \\ 
    1 & 0 & 0 & 0 \\ 
    0 & 0 & 0 & 0 \\ 
    0 & 0 & 0 & 0
\end{array}\right]
,\quad
[{{\bm \rL}_4^{(2)}}]_{i}{}^{\hat k} = \left[\begin{array}{cccc}
    0 & 0 & 0 & 0 \\ 
    0 & 0 & 0 & 0 \\ 
    0 & 0 & 1 & 0 \\ 
    0 & 1 & 0 & 0
\numberthis
\label{eq4-10}
\end{array}\right],
\]

\[[{{\bm \rR}_1^{(0)}}]_{\hat i}{}^{k} = \left[\begin{array}{cccc}
    0 & 0 & 0 & 0 \\ 
    0 & 0 & 1 & 0 \\ 
    0 & 0 & 0 & -1 \\ 
    0 & 0 & 0 & 0
\end{array}\right]
,\quad
[{{\bm \rR}_1^{(1)}}]_{\hat i}{}^{k}  = \left[\begin{array}{cccc}
    1 & 0 & 0 & 0 \\ 
    0 & 0 & 0 & 0 \\ 
    0 & 0 & 0 & 0 \\ 
    0 & -1 & 0 & 0
\numberthis
\label{eq4-10}
\end{array}\right],
\]

\[
[{{\bm \rR}_2^{(0)}}]_{\hat i}{}^{k}  = \left[\begin{array}{cccc}
    0 & 0 & -1 & 0 \\ 
    0 & 0 & 0 & 0 \\ 
    0 & 0 & 0 & 0 \\ 
    0 & 0 & 0 & -1
\end{array}\right] 
,\quad
[{{\bm \rR}_2^{(1)}}]_{\hat i}{}^{k}  = \left[\begin{array}{cccc}
    0 & 0 & 0 & 0 \\ 
    1 & 0 & 0 & 0 \\ 
    0 & 1 & 0 & 0 \\ 
    0 & 0 & 0 & 0
\numberthis
\label{eq4-10}
\end{array}\right],
\]

\[
[{{\bm \rR}_3^{(0)}}]_{\hat i}{}^{k}  = \left[\begin{array}{cccc}
    0 & 0 & 0 & 1 \\ 
    0 & 0 & 0 & 0 \\ 
    0 & 0 & 0 & 0 \\ 
    0 & 0 & -1 & 0
\end{array}\right] 
,\quad
[{{\bm \rR}_3^{(1)}}]_{\hat i}{}^{k}  = \left[\begin{array}{cccc}
    0 & 0 & 0 & 0 \\ 
    0 & -1 & 0 & 0 \\ 
    1 & 0 & 0 & 0 \\ 
    0 & 0 & 0 & 0
\numberthis
\label{eq4-10}
\end{array}\right],
\]

\[
[{{\bm \rR}_4^{(0)}}]_{\hat i}{}^{k}  = \left[\begin{array}{cccc}
    0 & 0 & 0 & 0 \\ 
    0 & 0 & 0 & 1 \\ 
    0 & 0 & 1 & 0 \\ 
    0 & 0 & 0 & 0
\end{array}\right]
,\quad
[{{\bm \rR}_4^{(1)}}]_{\hat i}{}^{k}  = \left[\begin{array}{cccc}
    0 & 1 & 0 & 0 \\ 
    0 & 0 & 0 & 0 \\ 
    0 & 0 & 0 & 0 \\ 
    1 & 0 & 0 & 0
\numberthis
\label{eq225}
\end{array}\right].
\]

By using these $\bm \rL$ and $\bm \rR$ matrices we can show that the SUSY closure relation holds for both the bosonic and fermionic fields by imposing the conditions below.  For the bosons one finds,
\begin{eqnarray}
\{{\rm D}_{\rm I},{\rm D}_{\rm J}\}\Phi{}_i&=&
\left[\left(
[{\bm {\rL}}{}_{\rI}^{(1)}]_{{i}}{}^{\hat j}
[{\bm {\rR}}{}_{\rJ}^{(0)}]_{{\hat j}}{}^{k}
+
[{\bm {\rL}}{}_{\rJ}^{(1)}]_{{i}}{}^{\hat j}
[{\bm {\rR}}{}_{\rI}^{(0)}]_{{\hat j}}{}^{k}
\right)\right.
\nonumber\\&{~~~~~~~}+&
\left(
[{\bm {\rL}}{}_{\rI}^{(1)}]_{{i}}{}^{\hat j}
[{\bm {\rR}}{}_{\rJ}^{(1)}]_{{\hat j}}{}^{k}
+
[{\bm {\rL}}{}_{\rJ}^{(1)}]_{{i}}{}^{\hat j}
[{\bm {\rR}}{}_{\rI}^{(1)}]_{{\hat j}}{}^{k}
+
[{\bm {\rL}}{}_{\rI}^{(2)}]_{{i}}{}^{\hat j}
[{\bm {\rR}}{}_{\rJ}^{(0)}]_{{\hat j}}{}^{k}
+
[{\bm {\rL}}{}_{\rJ}^{(2)}]_{{i}}{}^{\hat j}
[{\bm {\rR}}{}_{\rI}^{(0)}]_{{\hat j}}{}^{k}
\right)\frac{d}{d\tau}
\nonumber\\&{~~~~~~~}+&
\left(
[{\bm {\rL}}{}_{\rI}^{(2)}]_{{i}}{}^{\hat j}
[{\bm {\rR}}{}_{\rJ}^{(1)}]_{{\hat j}}{}^{k}
+
[{\bm {\rL}}{}_{\rJ}^{(2)}]_{{i}}{}^{\hat j}
[{\bm {\rR}}{}_{\rI}^{(1)}]_{{\hat j}}{}^{k}
\right)\frac{d^2}{d\tau^2}
\left.\frac{}{}\right] i\Phi{}_{k}
\nonumber\\&=&2i\delta_{{\rm IJ}}
\frac{d}{d\tau}\Phi{}_{i} ~~~,
\label{CSx1}
\end{eqnarray}

We see several results follow that distinguish them
from the discussion in the case of the valise adinkra.
From Eq. (\ref{CSx1}) we obtain three separate conditions,
\begin{eqnarray}
0 &=& ~
[{\bm {\rL}}{}_{( \rI}^{(1)}]_{{i}}{}^{\hat j}
[{\bm {\rR}}{}_{\rJ )}^{(0)}]_{{\hat j}}{}^{k}
 ~~~,
\label{CSx1a}
\end{eqnarray}
\begin{eqnarray}
2 \,\d{}_{\rI \rJ} \, \d_{{i}}{}^{k} &=&
[{\bm {\rL}}{}_{( \rI}^{(1)}]_{{i}}{}^{\hat j}
[{\bm {\rR}}{}_{\rJ )}^{(1)}]_{{\hat j}}{}^{k}
+
[{\bm {\rL}}{}_{( \rI}^{(2)}]_{{i}}{}^{\hat j}
[{\bm {\rR}}{}_{\rJ )}^{(0)}]_{{\hat j}}{}^{k}
\label{CSx1b}
\end{eqnarray}
and
\begin{eqnarray}
0 &=&
[{\bm {\rL}}{}_{( \rI}^{(2)}]_{{i}}{}^{\hat j}
[{\bm {\rR}}{}_{\rJ )}^{(1)}]_{{\hat j}}{}^{k} 
~~~,
\label{CSx1c}
\end{eqnarray}
where the notation $( {\rm I}$ and $ {\rm J ) }$ denote symmetrization of the pair of indices.

A similar calculation for the fermionic fields yields
\begin{eqnarray}
\{{\rm D}_{\rm I},{\rm D}_{\rm J}\}\Psi{}_{\hat i} &=&
\left[\left(
[{\bm {\rR}}{}_{\rI}^{(0)}]_{{\hat i}}{}^{j}
[{\bm {\rL}}{}_{\rJ}^{(1)}]_{{j}}{}^{\hat k}
+
[{\bm {\rR}}{}_{\rJ}^{(0)}]_{{\hat i}}{}^{j}
[{\bm {\rL}}{}_{\rI}^{(1)}]_{{j}}{}^{\hat k}
\right)\right.
\nonumber\\&{~~~~~~~}+&
\left(
[{\bm {\rR}}{}_{\rI}^{(1)}]_{{\hat i}}{}^{j}
[{\bm {\rL}}{}_{\rJ}^{(1)}]_{{j}}{}^{\hat k}
+
[{\bm {\rR}}{}_{\rJ}^{(1)}]_{{\hat i}}{}^{j}
[{\bm {\rL}}{}_{\rI}^{(1)}]_{{j}}{}^{\hat k}
+
[{\bm {\rR}}{}_{\rI}^{(0)}]_{{\hat i}}{}^{j}
[{\bm {\rL}}{}_{\rJ}^{(2)}]_{{j}}{}^{\hat k}
+
[{\bm {\rR}}{}_{\rJ}^{(0)}]_{{\hat i}}{}^{j}
[{\bm {\rL}}{}_{\rI}^{(2)}]_{{j}}{}^{\hat k}
\right)\frac{d}{d\tau}
\nonumber\\&{~~~~~~~}+&
\left(
[{\bm {\rR}}{}_{\rI}^{(1)}]_{{\hat i}}{}^{j}
[{\bm {\rL}}{}_{\rJ}^{(2)}]_{{j}}{}^{\hat k}
+
[{\bm {\rR}}{}_{\rJ}^{(1)}]_{{\hat i}}{}^{j}
[{\bm {\rL}}{}_{\rI}^{(2)}]_{{j}}{}^{\hat k}
\right)\frac{d^2}{d\tau^2}
\left.\frac{}{}\right] i\Psi{}_{\hat k}
\nonumber\\&=&2i\delta_{{\rm IJ}}
\frac{d}{d\tau}\Psi{}_{\hat i} ~~~.
\label{CSx2}
\end{eqnarray}
and from Eq.\ (\ref{CSx2}) we find the three
conditions
\begin{eqnarray}
0 &=&
[{\bm {\rR}}{}_{( \rI}^{(0)}]_{{\hat i}}{}^{j}
[{\bm {\rL}}{}_{\rJ )}^{(1)}]_{{j}}{}^{\hat k}
~~~,
\label{CSx2a}
\end{eqnarray}
\begin{eqnarray}
2 \, \d{}_{\rI \rJ} \, \d_{{\hat i}}{}^{\hat k} &=&
[{\bm {\rR}}{}_{( \rI}^{(1)}]_{{\hat i}}{}^{j}
[{\bm {\rL}}{}_{\rJ ) }^{(1)}]_{{j}}{}^{\hat k}
+
[{\bm {\rR}}{}_{( \rI}^{(0)}]_{{\hat i}}{}^{j}
[{\bm {\rL}}{}_{\rJ )}^{(2)}]_{{j}}{}^{\hat k}
~~~,
\label{CSx2b}
\end{eqnarray}
\begin{eqnarray}
0 &=& ~
[{\bm {\rR}}{}_{( \rI}^{(1)}]_{{\hat i}}{}^{j}
[{\bm {\rL}}{}_{\rJ )}^{(2)}]_{{j}}{}^{\hat k}
~~~.
\label{CSx2c}
\end{eqnarray}

\subsection{Garden Algebra and Unfolded Adinkra: Vector Supermultiplet}
The 4D, $\cal N$ = 1 vector multiplet is described by a vector $A{}_{\mu}$, a Majorana fermion $\l_a$, and a pseudoscalar auxiliary field $d$.

The Lagrangian for the vector supermultiplet \cite{8} is given as 

\begin{equation}
\mathcal{L}_{VS}=
-\frac{1}{4}F_{\mu\nu}F^{\mu\nu}
+i\frac{1}{2}(\gamma^\mu)^{ab}\lambda_a(\partial_\mu\lambda_b)
+\frac{1}{2}d^2 ~~~,
\end{equation}
where
\begin{equation}
F_{\mu\nu}=\partial_\mu A_\nu-\partial_\nu A_\mu ~~~.
\end{equation}

The familiar transformation rules under a supercovariant derivation for each field are given as \cite{3}
\be \eqalign{
{\rm D}_a \, A{}_{\mu} ~&=~  (\gamma_\mu){}_a {}^b \,  \l_b  ~~~, \cr
{\rm D}_a \l_b ~&=~   - \,i \, \fracm 14 ( [\, \gamma^{\mu}\, , \,  \gamma^{\nu} 
\,]){}_a{}_b \, (\,  \partial_\mu  \, A{}_{\nu}    ~-~  \partial_\nu \, A{}_{\mu}  \, )
~+~  (\gamma^5){}_{a \,b} \,    {\rm d} ~~,  \cr
{\rm D}_a \, {\rm d} ~&=~  i \, (\gamma^5\gamma^\mu){}_a {}^b \, 
\,  \partial_\mu \l_b  ~~~. \cr
} \label{V1}
\ee
so the action is invariant when acted upon by the supersymmetric covariant derivative, since it satisfies
\[\begin{array}{l}
{\rm D}{}_a \mathcal{L}_{VS}= \partial^\mu \mathcal{J}_{\mu a}{}^{(VS)} 
\numberthis\end{array} ~~~,\]
where the supercurrent \(\mathcal{J}_{\mu a} {}^{(VS)} \) is given as
\[\begin{array}{l}
\mathcal{J}_{\mu a}{}^{(VS)} = i \, \left[ \, 
\frac{1}{4}(\gamma_\mu\sigma^{\alpha\beta})_a{}^{c} F_{\alpha\beta}
+\frac{1}{2}(\gamma^5\gamma_\mu)_a{}^{c} \, d \,  \right] \,\lambda_c
\numberthis\end{array} ~~~.\]

\begin{figure}[H]
\centering	\includegraphics[scale=0.35]{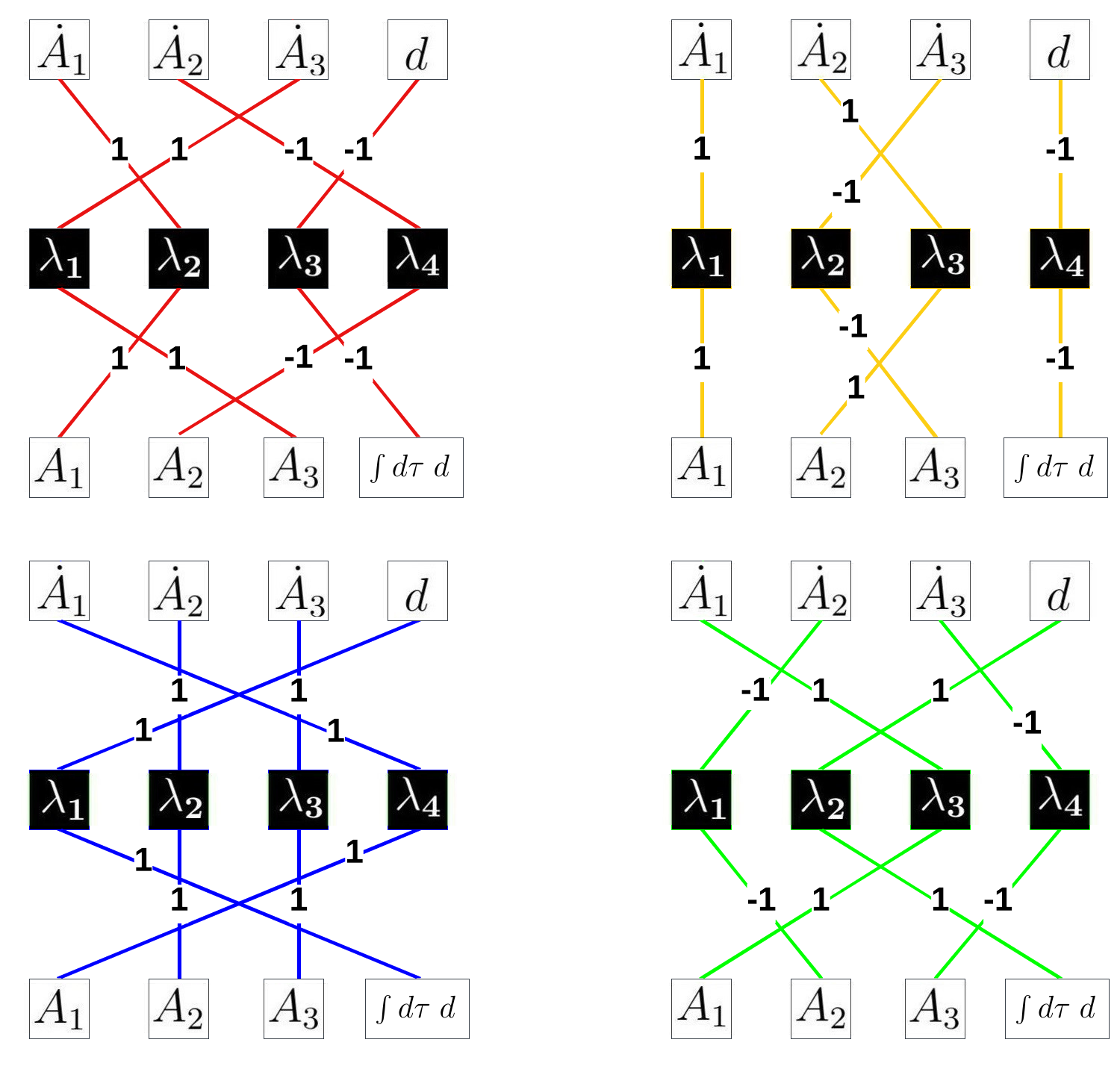}	\caption{An unfolded adinkra for the 4D $\mathcal{N}=1$ vector multiplet with ascending edges.}
	\label{fig2}
\end{figure}

Then based on Eq. (\ref{V1}) we can apply the 0-brane reduction process \cite{3} for each field and draw the relations between each field as a graph, as seen in Fig. \ref{fig2}.
The color-disaggregated component unfolded adinkras are obtained as in the example of the chiral supermultiplet.

In obtaining these we define
\[ \Phi_{i} = \left[\begin{array}{c}
    A_1  \\ 
    A_2  \\
    A_3  \\
    d  
\end{array}\right] ~~~,~~~
i\Psi{}_{\hat k} = \left[\begin{array}{c}
    \lambda_1  \\ 
    \lambda_2  \\
    \lambda_3  \\
    \lambda_4 
\end{array}\right]  ~~~,~~~
{\rm D}_{\rm I} = \left[\begin{array}{c}
    {\rm D}_1  \\ 
    {\rm D}_2  \\     
    {\rm D}_3  \\ 
    {\rm D}_4 
\numberthis\end{array}\right] ~~~.
\]

and in terms of these new quantities, Eq. (\ref{V1}) can be written as
\[ {\rm D}_{\rm I} \Phi{}_i = i \, [ {\bm {\rL}}{}_{\rI}^{(1)}]_{i}{}^{\hat 
k} \, \Psi{}_{\hat k} ~+~ i \, [ {\bm {\rL}}{}_{\rI}^{(2)}]_{i}{}^{\hat k} 
\, \frac{d}{d \tau} \Psi{}_{\hat k} ~~~,
\numberthis
\]

\[ {\rm D}_{\rm I} \Psi{}_{\hat k} = [ {\bm {\rR}}{}_{\rI}^{(0)}]_{\hat
k}{}^i \,
\Phi{}_{i }
~+~ [ {\bm {\rR}}{}_{\rI}^{(1)}]_{\hat k}{}^i \, \frac{d ~}{d \tau}
\Phi{}_{i} ~~~.
\numberthis
\]

These $\bm \rL$ and $\bm \rR$ matrices are given explicitly as
\[
[{{\bm \rL}_1^{(1)}}]_{i}{}^{\hat k}  = \left[\begin{array}{cccc}
    0 & 1 & 0 & 0 \\ 
    0 & 0 & 0 & -1 \\ 
    1 & 0 & 0 & 0 \\ 
    0 & 0 & 0 & 0
\end{array}\right]
,\quad
[{{\bm \rL}_1^{(2)}}]{i}{}^{\hat k} = \left[\begin{array}{cccc}
    0 & 0 & 0 & 0 \\ 
    0 & 0 & 0 & 0 \\ 
    0 & 0 & 0 & 0 \\ 
    0 & 0 & -1 & 0
\numberthis
\label{eq245}
\end{array}\right],
\]

\[
[{{\bm \rL}_2^{(1)}}]_{i}{}^{\hat k} = \left[\begin{array}{cccc}
    1 & 0 & 0 & 0 \\ 
    0 & 0 & 1 & 0 \\ 
    0 & -1 & 0 & 0 \\ 
    0 & 0 & 0 & 0
\end{array}\right] 
,\quad
[{{\bm \rL}_2^{(2)}}]_{i}{}^{\hat k} = \left[\begin{array}{cccc}
    0 & 0 & 0 & 0 \\ 
    0 & 0 & 0 & 0 \\ 
    0 & 0 & 0 & 0 \\ 
    0 & 0 & 0 & -1
\numberthis
\label{eq4-10}
\end{array}\right],
\]

\[
[{{\bm \rL}_3^{(1)}}]_{i}{}^{\hat k} = \left[\begin{array}{cccc}
    0 & 0 & 0 & 1 \\ 
    0 & 1 & 0 & 0 \\ 
    0 & 0 & 1 & 0 \\ 
    0 & 0 & 0 & 0
\end{array}\right] 
,\quad
[{{\bm \rL}_3^{(2)}}]_{i}{}^{\hat k} = \left[\begin{array}{cccc}
    0 & 0 & 0 & 0 \\ 
    0 & 0 & 0 & 0 \\ 
    0 & 0 & 0 & 0 \\ 
    1 & 0 & 0 & 0
\numberthis
\label{eq4-10}
\end{array}\right],
\]

\[
[{{\bm \rL}_4^{(1)}}]_{i}{}^{\hat k} = \left[\begin{array}{cccc}
    0 & 0 & 1 & 0 \\ 
    -1 & 0 & 0 & 0 \\ 
    0 & 0 & 0 & -1 \\ 
    0 & 0 & 0 & 0
\end{array}\right] 
,\quad
[{{\bm \rL}_4^{(2)}}]_{i}{}^{\hat k} = \left[\begin{array}{cccc}
    0 & 0 & 0 & 0 \\ 
    0 & 0 & 0 & 0 \\ 
    0 & 0 & 0 & 0 \\ 
    0 & 1 & 0 & 0
\numberthis
\label{eq4-10}
\end{array}\right],
\]

\[
[{{\bm \rR}_1^{(0)}}]_{\hat i}{}^{k} = \left[\begin{array}{cccc}
    0 & 0 & 0 & 0 \\ 
    0 & 0 & 0 & 0 \\ 
    0 & 0 & 0 & -1 \\ 
    0 & 0 & 0 & 0
\end{array}\right]
,\quad
[{{\bm \rR}_1^{(1)}}]_{\hat i}{}^{k} = \left[\begin{array}{cccc}
    0 & 0 & 1 & 0 \\ 
    1 & 0 & 0 & 0 \\ 
    0 & 0 & 0 & 0 \\ 
    0 & -1 & 0 & 0
\numberthis
\label{eq4-10}
\end{array}\right],
\]

\[
[{{\bm \rR}_2^{(0)}}]_{\hat i}{}^{k} = \left[\begin{array}{cccc}
    0 & 0 & 0 & 0 \\ 
    0 & 0 & 0 & 0 \\ 
    0 & 0 & 0 & 0 \\ 
    0 & 0 & 0 & -1
\end{array}\right] 
,\quad
[{{\bm \rR}_2^{(1)}}]_{\hat i}{}^{k} = \left[\begin{array}{cccc}
    1 & 0 & 0 & 0 \\ 
    0 & 0 & -1 & 0 \\ 
    0 & 1 & 0 & 0 \\ 
    0 & 0 & 0 & 0
\numberthis
\label{eq4-10}
\end{array}\right],
\]

\[
[{{\bm \rR}_3^{(0)}}]_{\hat i}{}^{k} = \left[\begin{array}{cccc}
    0 & 0 & 0 & 1 \\ 
    0 & 0 & 0 & 0 \\ 
    0 & 0 & 0 & 0 \\ 
    0 & 0 & 0 & 0
\end{array}\right] 
,\quad
[{{\bm \rR}_3^{(1)}}]_{\hat i}{}^{k} = \left[\begin{array}{cccc}
    0 & 0 & 0 & 0 \\ 
    0 & 1 & 0 & 0 \\ 
    0 & 0 & 1 & 0 \\ 
    1 & 0 & 0 & 0
\numberthis
\label{eq4-10}
\end{array}\right],
\]

\[
[{{\bm \rR}_4^{(0)}}]_{\hat i}{}^{k} = \left[\begin{array}{cccc}
    0 & 0 & 0 & 0 \\ 
    0 & 0 & 0 & 1 \\ 
    0 & 0 & 0 & 0 \\ 
    0 & 0 & 0 & 0
\end{array}\right] 
,\quad
[{{\bm \rR}_4^{(1)}}]_{\hat i}{}^{k} = \left[\begin{array}{cccc}
    0 & -1 & 0 & 0 \\ 
    0 & 0 & 0 & 0 \\ 
    1 & 0 & 0 & 0 \\ 
    0 & 0 & -1 & 0
\numberthis
\label{eq259}
\end{array}\right].
\]

Using this set of $[{\bm {\rR}}{}_{\rJ }^{(0)}]$, $[{\bm {\rL}}{}_{ \rI}^{(1)}]$, $[{\bm {\rR}}{}_{\rI}^{(1)}]$, $[{\bm {\rL}}{}_{\rJ}^{(2)}]$, $[{\bm {\rR}}{}_{\rJ }^{(1)}]$ matrices leads exactly
back to the same equations seen in Eqs. (\ref{CSx1a})-(\ref{CSx1c}), 
(\ref{CSx2a})-(\ref{CSx2c}).
This is the case even though the $[{\bm {\rR}}{}_{\rJ }^{(0)}]$, $[{\bm {\rL}}{}_{ \rI}^{(1)}]$, $[{\bm {\rR}}{}_{\rI}^{(1)}]$, $[{\bm {\rL}}{}_{\rJ}^{(2)}]$, $[{\bm {\rR}}{}_{\rJ }^{(1)}]$ matrices for the chiral 
supermultiplet are completely different from those 
that appear for the vector supermultiplet

\subsection{Garden Algebra and Unfolded Adinkra: Tensor Supermultiplet}
\label{ch2.3}

The 4D, $\cal N$ = 1 tensor multiplet consists of a scalar $\varphi$, a second-rank skew-symmetric tensor $B{}_{\mu \, \nu }$, and a Majorana fermion $\chi_a$.

The Lagrangian for tensor supermultiplet \cite{10} is the following:

\begin{equation}
\mathcal{L}_{TS}=
-\frac{1}{2}(\partial_\mu \varphi)(\partial^\mu \varphi)
-\frac{1}{3}H_{\mu\nu \alpha}H^{\mu\nu \alpha}
+i\frac{1}{2}(\gamma^\mu)^{bc}\chi_b(\partial_\mu\chi_c) ~~~,
\end{equation}
where
\begin{equation}
H_{\mu\nu \alpha}=
\partial_\mu B_{\nu \alpha}
+\partial_\nu B_{\alpha\mu}
+\partial_\alpha B_{\mu\nu} ~~~.
\end{equation}
with the transformation rules under the supercovariant derivative operators for each field \cite{3} given as:
\be \eqalign{
{\rm D}_a \varphi ~&=~ \chi_a  ~~~, \cr
{\rm D}_a B{}_{\mu \, \nu } ~&=~ -\, \fracm 14 ( [\, \gamma_{\mu}
\, , \,  \gamma_{\nu} \,]){}_a{}^b \, \chi_b  ~~~, \cr
{\rm D}_a \chi_b ~&=~ i\, (\gamma^\mu){}_{a \,b}\,  \partial_\mu \varphi 
~-~  (\gamma^5\gamma^\mu){}_{a \,b} \, \e{}_{\mu}{}^{\r \, \s \, \t}
\partial_\r B {}_{\s \, \t}~~.
} \label{ten1}
\ee
The action is invariant when acted on by the supersymmetric covariant derivative:
\[\begin{array}{l}
{\rm D}{}_a \mathcal{L}_{TS}= \partial^\mu \mathcal{J}_{\mu a}{}^{(TS)} 
\numberthis\end{array} ~~~,\]
where the supercurrent \(\mathcal{J}_{\mu a}{}^{(TS)} \) is given as,
\[\begin{array}{l}
\mathcal{J}_{\mu a}{}^{(TS)} = 
-\frac{1}{2} \left[ \,(\gamma_\mu\gamma^\nu)_a{}^{c} \,(\partial_\nu\varphi ) + i\,
(\sigma^{\sigma\tau})_a{}^{c}
H_{\mu\sigma\tau} - i \, \frac{1}{3}\epsilon_\mu {}^{\rho \sigma \tau} (\gamma^5)_a {}^c  H_{\rho \sigma \tau} \, \right] \, \chi_c
\numberthis\end{array} ~~~.\]

\begin{figure}[H]
	\centering
	\includegraphics[scale=0.6]{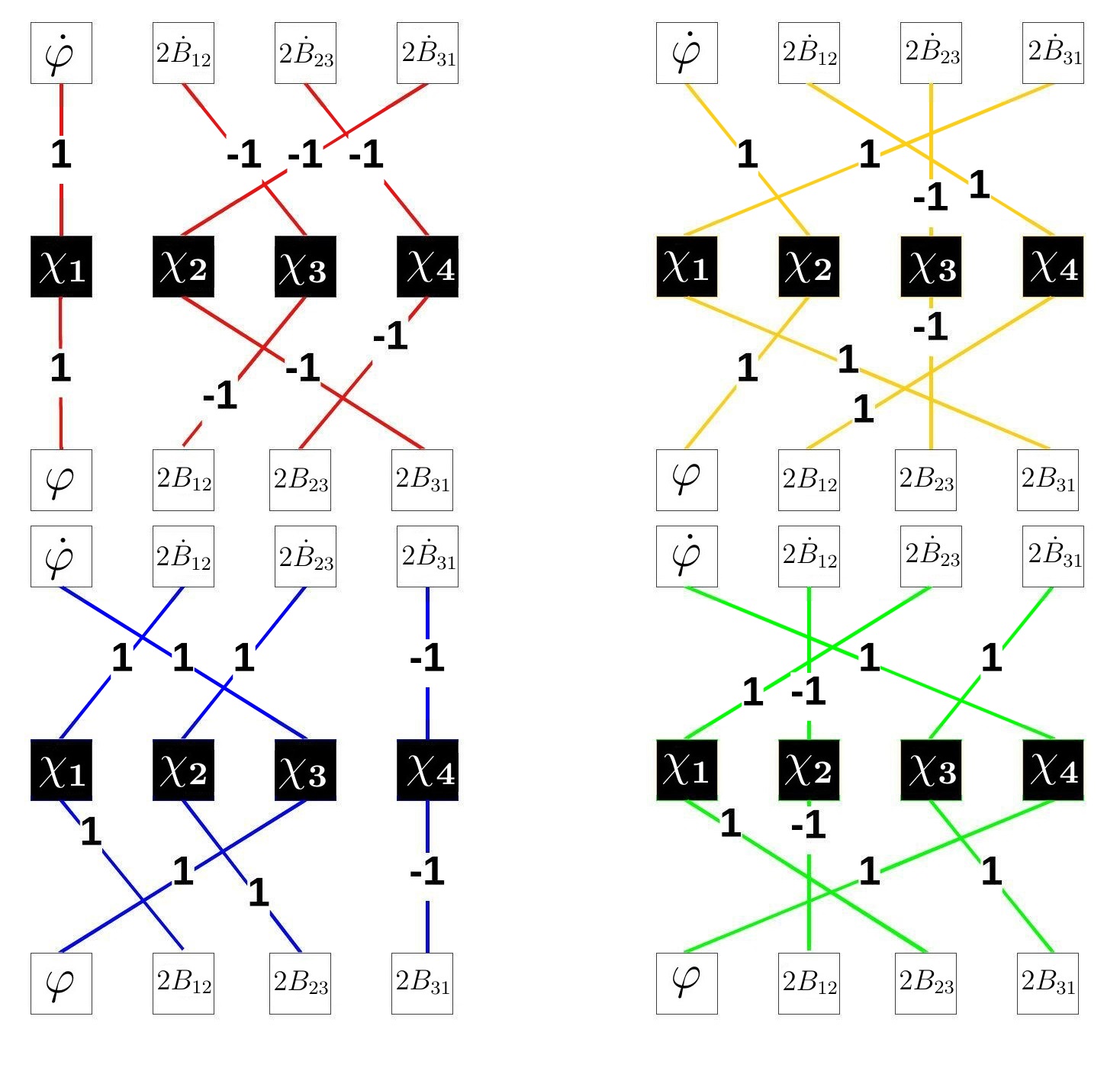}	\caption{An unfolded adinkra for the 4D $\mathcal{N}=1$ tensor multiplet with ascending edges.}	\label{fig3}
\end{figure}

Then based on Eq. (\ref{ten1}) we can apply the 0-brane reduction process \cite{3} for each field and draw the relations between each field as a graph, as seen in Fig. \ref{fig3}.

Furthermore, we can define

\[ \Phi_{i} = \left[\begin{array}{c}
    \varphi  \\ 
    2B_{12}   \\
    2B_{23}   \\
    2B_{31}  
\end{array}\right] ~~~,~~~
i\Psi{}_{\hat k} = \left[\begin{array}{c}
    \chi_1  \\ 
    \chi_2  \\
    \chi_3  \\
    \chi_4 
\end{array}\right]  ~~~,~~~
{\rm D}_{\rm I} = \left[\begin{array}{c}
    {\rm D}_1  \\ 
    {\rm D}_2  \\     
    {\rm D}_3  \\ 
    {\rm D}_4 
\numberthis\end{array}\right] ~~~.
\]

and in terms of these new quantities, Eq. (\ref{ten1}) can be written as
\[ {\rm D}_{\rm I} \Phi{}_i = i \, [ {\bm {\rL}}{}_{\rI}^{(1)}]_i \, ^{\hat 
k} \, \Psi{}_{\hat k}
\numberthis
\]

\[ {\rm D}_{\rm I} \Psi{}_{\hat k} = [ {\bm {\rR}}{}_{\rI}^{(1)}]_{\hat
k}\,{}^i
\,\frac{d}{d\tau}\Phi{}_{i }
\numberthis
\]

These $\bm \rL$ and $\bm \rR$ matrices are given explicitly as
\[
[{{\bm \rL}_1^{(1)}}]_{i}{}^{\hat k} = \left[\begin{array}{cccc}
    1 & 0 & 0 & 0 \\ 
    0 & 0 & -1 & 0 \\ 
    0 & 0 & 0 & -1 \\ 
    0 & -1 & 0 & 0
   \end{array}\right]
,\quad
[{{\bm \rL}_2^{(1)}}]_{i}{}^{\hat k} = \left[\begin{array}{cccc}
    0 & 1 & 0 & 0 \\ 
    0 & 0 & 0 & 1 \\ 
    0 & 0 & -1 & 0 \\ 
    1 & 0 & 0 & 0
\numberthis
\label{eq4-10}
\end{array}\right],
\]

\[
[{{\bm \rL}_3^{(1)}}]_{i}{}^{\hat k} = \left[\begin{array}{cccc}
    0 & 0 & 1 & 0 \\ 
    1 & 0 & 0 & 0 \\ 
    0 & 1 & 0 & 0 \\ 
    0 & 0 & 0 & -1
\end{array}\right]
,\quad
[{{\bm \rL}_4^{(1)}}]_{i}{}^{\hat k} = \left[\begin{array}{cccc}
    0 & 0 & 0 & 1 \\ 
    0 & -1 & 0 & 0 \\ 
    1 & 0 & 0 & 0 \\ 
    0 & 0 & 1 & 0
\numberthis
\label{eq4-10}
\end{array}\right],
\]

\[
[{{\bm \rR}_1^{(1)}}]_{\hat i}{}^{k} = \left[\begin{array}{cccc}
    1 & 0 & 0 & 0 \\ 
    0 & 0 & 0 & -1 \\ 
    0 & -1 & 0 & 0 \\ 
    0 & 0 & -1 & 0
   \end{array}\right]
,\quad
[{{\bm \rR}_2^{(1)}}]_{\hat i}{}^{k} = \left[\begin{array}{cccc}
    0 & 0 & 0 & 1 \\ 
    1 & 0 & 0 & 0 \\ 
    0 & 0 & -1 & 0 \\ 
    0 & 1 & 0 & 0
\numberthis
\label{eq4-10}
\end{array}\right],
\]

\[
[{{\bm \rR}_3^{(1)}}]_{\hat i}{}^{k} = \left[\begin{array}{cccc}
    0 & 1 & 0 & 0 \\ 
    0 & 0 & 1 & 0 \\ 
    1 & 0 & 0 & 0 \\ 
    0 & 0 & 0 & -1
   \end{array}\right]
,\quad
[{{\bm \rR}_4^{(1)}}]_{\hat i}{}^{k} = \left[\begin{array}{cccc}
    0 & 0 & 1 & 0 \\ 
    0 & -1 & 0 & 0 \\ 
    0 & 0 & 0 & 1 \\ 
    1 & 0 & 0 & 0
\numberthis
\label{eq4-10}
\end{array}\right].
\]

By using these $\bm \rL$ and $\bm \rR$ matrices we can show that the SUSY closure relation holds for both the bosonic and fermionic fields:
\begin{eqnarray}
\{{\rm D}_{\rm I},{\rm D}_{\rm J}\}\Phi{}_i=
\left[\left(
[{\bm {\rL}}{}_{\rI}^{(1)}]_{{i}}{}^{\hat j}
[{\bm {\rR}}{}_{\rJ}^{(1)}]_{{\hat j}}{}^{k}
+
[{\bm {\rL}}{}_{\rJ}^{(1)}]_{{i}}{}^{\hat j}
[{\bm {\rR}}{}_{\rI}^{(1)}]_{{\hat j}}{}^{k}
\right)\right] i\frac{d}{d\tau}\Phi{}_{k}= 2i\delta_{{\rm IJ}}
\frac{d}{d\tau}\Phi{}_{i}~~~,
\end{eqnarray}

\begin{eqnarray}
\{{\rm D}_{\rm I},{\rm D}_{\rm J}\}\Psi{}_{\hat i}=
\left[\left(
[{\bm {\rR}}{}_{\rI}^{(1)}]_{{\hat i}}{}^{j}
[{\bm {\rL}}{}_{\rJ}^{(1)}]_{{j}}{}^{\hat k}
+
[{\bm {\rR}}{}_{\rJ}^{(1)}]_{{\hat i}}{}^{j}
[{\bm {\rL}}{}_{\rI}^{(1)}]_{{j}}{}^{\hat k}
\right)\right] i\frac{d}{d\tau}\Psi{}_{\hat k}= 2i\delta_{{\rm IJ}}
\frac{d}{d\tau}\Psi{}_{\hat i}~~~.
\end{eqnarray}

\section{CLS: Field Definitions, Properties}

The complex linear supermultiplet (CLS) contains scalar $K$, pseudoscalar $L$, Majorana spinor $\zeta_a$ and auxiliary scalar $M$, auxiliary pseudoscalar $N$, auxiliary vector $V_\mu$, auxiliary axial-vector $U_\mu$, and auxiliary Majorana spinors $\rho_a$ and $\beta_a$. 

The Lagrangian for the CLS field is given as, \cite{6}

\[\begin{array}{l}
\mathcal{L}_{CLS}=-\frac{1}{2} (\partial_\mu K)( \partial^\mu K)-\frac{1}{2}(\partial_\mu L)(\partial^\mu L)
-\frac{1}{2}(M^2)-\frac{1}{2}(N^2) 
\\[10pt]
\qquad\qquad+\frac{1}{4}(V_\mu V^\mu)+ \frac{1}{4} (U_\mu U^\mu)
+\frac{i}{2}(\gamma^\mu)^{ab} (\zeta_a \partial_\mu \zeta_b)+ iC^{ab}(\rho_a\beta_b) 
\label{ACT-CLS}
\numberthis
\end{array}.\]

The supersymmetry transformation laws are given as, \cite{6,5}
\newpage
\begin{eqnarray}
{\rm D}{}_a K &=& \rho_a - \zeta_a~~~, \nonumber\\
{\rm D}{}_a L &=& -i{(\gamma^5)_a}^b(\rho_b + \zeta_b)~~~, \nonumber\\
{\rm D}{}_a M &=&\beta_{a}-\frac{1}{2}\left(\gamma^{\mu}\right)_a{}^{b}\left(\partial_{\mu} \rho_{b}\right)~~~, \nonumber\\
{\rm D}{}_a N &=& -i\left(\gamma^{5}\right)_a{}^{b}\left(\beta_{b}\right)+\frac{i}{2}\left(\gamma^{5} \gamma^{\mu}\right)_{a}{}^{b}\left(\partial_{\mu} \rho_{b}\right)~~~, \nonumber\\
{\rm D}{}_a V_{\mu}&=&-\left(\gamma_{\mu}\right)_a{}^{b}\left( \beta_{b}\right)-\left(\gamma_{\mu} \gamma^{\nu}\right)_a{}^{b}\left(\partial_{\nu} \zeta_{b}\right)+\left(\partial_{\mu}  \rho_{a}\right)+\frac{1}{2}\left(\gamma^{\nu} \gamma_{\mu}\right)_{a}{}^{b}\left(\partial_{\nu} \rho_{b}\right)~~~, \nonumber\\
{\rm D}{}_a U_{\mu}&=&i\left(\gamma^{5} \gamma_{\mu}\right)_{a}{}^{b}\left( \beta_{b}\right)-i\left(\gamma^{5} \gamma_{\mu} \gamma^{\nu}\right)_{a}{}^{b}\left(\partial_{\nu}  \zeta_{b}\right)-i\left(\gamma^{5}\right)_{a}{ }^{b}\left(\partial_{\mu} \rho_{b}\right)-\frac{i}{2}\left(\gamma^{5} \gamma^{\nu} \gamma_{\mu}\right)_{a}{ }^{b}\left(\partial_{\nu}  \rho_{b}\right) ~~~,\nonumber\\
{\rm D}{}_a \zeta_{b}&=&-i\left(\gamma^{\mu}\right)_{ab}\left(\partial_{\mu}  K\right)+\left(\gamma^{5} \gamma^{\mu}\right)_{ab}\left(\partial_{\mu} L\right)+\frac{i}{2}\left(\gamma^{\mu}\right)_{ab}\left(V_{\mu}\right)-\frac{1}{2}\left(\gamma^{5} \gamma^{\mu}\right)_{ab}\left( U_{\mu}\right)~~~, \nonumber\\
{\rm D}{}_a \rho_{b}&=&i C_{a b}\left( M\right)+\left(\gamma^{5}\right)_{a b}\left( N\right)+\frac{i}{2}\left(\gamma^{\mu}\right)_{a b}\left( V_{\mu}\right)+\frac{1}{2}\left(\gamma^{5} \gamma^{\mu}\right)_{a b}\left( U_{\mu}\right)~~~, \nonumber\\
{\rm D}{}_a \beta_{b}&=&-\eta^{\mu \nu} \partial_{\mu} \partial_{\nu}\left\{i C_{a b}\left( K\right)+\left(\gamma^{5}\right)_{ab}\left( L\right)\right\}+\frac{i}{2}\left(\gamma^{\mu}\right)_{ab}\left(\partial_{\mu} M\right)\nonumber\\
&&+\frac{1}{2}\left(\gamma^{5} \gamma^{\mu}\right)_{a b} \left(\partial_\mu  N\right)+\frac{i}{2}\left(\gamma^{\mu} \gamma^{\nu}\right)_{ab}\left(\partial_{\mu}  V_{\nu}\right)+\frac{i}{4}\left(\gamma^{\nu} \gamma^{\mu}\right)_{a b}\left(\partial_{\mu}  V_{\nu}\right)\nonumber\\
&&+\frac{1}{2}\left(\gamma^{5} \gamma^{\mu} \gamma^{\nu}\right)_{a b}\left(\partial_{\mu}  U_{\nu}\right)+\frac{1}{4}\left(\gamma^{5} \gamma^{\nu} \gamma^{\mu}\right)_{a b}\left(\partial_{\mu} U_{\nu}\right)~~~.
\end{eqnarray}

We confirmed that for any field $X$ in CLS, it satisfies the field closure relationship when acted on by the anticommutator of two supersymmetry covariant derivatives,
\[\begin{array}{l}
\{{\rm D}{}_a, {\rm D}{}_b\} X = 2{i}\left(\gamma^{\mu}\right)_{a b}\left(\partial_{\mu} X\right)
\numberthis{} ~~~.
\end{array}\]

Additionally, we confirmed that the Lagrangian is gauge invariant when acted on by the supersymmetric covariant derivative:

\[\begin{array}{l}
{\rm D}{}_a \mathcal{L}_{CLS} = \partial^\mu \mathcal{J}_{\mu a}{}^{(CLS)} 
\numberthis\end{array} ~~~,\]

where the current \(\mathcal{J}_{\mu a}{}^{(CLS)}\) is given as,

\[\begin{array}{l}
\mathcal{J}_{\mu a}{}^{(CLS)} = [ \,  - \, (\partial_\mu K) + i{(\gamma^5)_a}^b (\partial_\mu L) + \frac{1}{2}(\gamma_{\mu})_{a}{}^{b}
M -\frac{i}{2}{(\gamma^5\gamma_\mu)_a}^b  N  \, ] \rho_a\\[10pt]
\qquad\quad  {~~~~~~} + [ \,
\frac{1}{2}(\gamma_{\mu} \gamma_{\nu})_a{}^{b} V^{\nu}+\frac{1}{4}(\gamma_{\nu} \gamma_{\mu})_{a}{}^{b}  V^{\nu}
-\frac{i}{2}(\gamma^{5} \gamma_{\mu} \gamma_{\nu})_{a}{}^{b} U^\nu 
-\frac{i}{4}(\gamma^{5} \gamma_{\nu} \gamma_{\mu})_a{}^{b} U^{\nu} \, ] \, \rho_{b}\\[10pt]
\qquad\quad {~~~~~~} + [ \,
\frac{1}{2}(\gamma_\mu\gamma_\nu)_a{}^b (\partial^\nu K)
+ \frac{i}{2}(\gamma^5\gamma_\mu\gamma_\nu)_a{}^b (\partial^\nu L)
- \frac{1}{4}(\gamma_\nu \gamma_\mu)_a{}^b  V^\nu 
- \frac{i}{4} (\gamma^5 \gamma_\nu \gamma_\mu)_a{}^b U^\nu \, ] \, \zeta_b
\numberthis\end{array} ~~~.\]

\section{Valise Adinkras for CLS Model}

By using the 0-brane reduction, \cite{3} and in order for the $\bm \rL$ and $\bm \rR$ matrices associated with a valise adinkra
formulation to have entries among
\{-1,0,1\} , we must utilize the following new fields which are defined as linear combinations of the original fields and their derivatives \cite{5,2}.

\begin{alignat}{4}
    &\dot{\Phi}_1 = M \qquad &&\dot{\Phi}_2 = V_0 - \dot{K} \qquad &&\dot{\Phi}_3 = U_0 - \dot{L} \qquad &&\dot{\Phi}_4 = N \nonumber\\[10pt]
    &\dot{\Phi}_5 = U_2 \qquad &&\dot{\Phi}_6 = V_0 - 2\dot{K} \qquad &&\dot{\Phi}_7 = -U_1 \qquad &&\dot{\Phi}_8 = U_3 \nonumber\\[10pt]
    &\dot{\Phi}_9 = -V_3 \qquad &&\dot{\Phi}_{10} = -V_1 \qquad &&\dot{\Phi}_{11} = U_0 - 2\dot{L} \qquad &&\dot{\Phi}_{12} = V_2 
    \label{Bsa1}
\end{alignat}

\begin{alignat}{4}
    &i\dot{\Psi}_1 = \beta_1 - \frac{1}{2} \dot{\rho}_2 \qquad &&i\dot{\Psi}_2 = \beta_2 + \frac{1}{2} \dot{\rho}_1 \nonumber \\[10pt] &i\dot{\Psi}_3 = \beta_3 + \frac{1}{2} \dot{\rho}_4 \qquad &&i\dot{\Psi}_4 = \beta_4 - \frac{1}{2} \dot{\rho}_3 \nonumber \\[10pt]
    &i\dot{\Psi}_5 = \beta_1 - \dot{\zeta}_2 + \frac{1}{2} \dot{\rho}_2 \qquad && i\dot{\Psi}_6 = \beta_2 + \dot{\zeta}_1 - \frac{1}{2} \dot{\rho}_1 \nonumber \\[10pt] &i\dot{\Psi}_7 = \beta_3 + \dot{\zeta}_4 - \frac{1}{2} \dot{\rho}_4 \qquad &&i\dot{\Psi}_8 = \beta_4 - \dot{\zeta}_3 + \frac{1}{2} \dot{\rho}_3 \nonumber \\[10pt]
    &i\dot{\Psi}_9 = \beta_1 + \dot{\zeta}_2 + \frac{1}{2} \dot{\rho}_2 \qquad && i\dot{\Psi}_{10} = \beta_2 - \dot{\zeta}_1 - \frac{1}{2} \dot{\rho}_1 \nonumber \\[10pt] &i\dot{\Psi}_{11} = \beta_3 - \dot{\zeta}_4 - \frac{1}{2} \dot{\rho}_4 \qquad &&i\dot{\Psi}_{12} = \beta_4 + \dot{\zeta}_3 + \frac{1}{2} \dot{\rho}_3 
    \label{Bsa2}
\end{alignat}

\subsection{Garden Algebra}
Given the above new bosonic field transformations, we can construct the $\bm \rL$ matrices as follows after 0-brane reduction \cite{3}, where the matrices satisfy:

\[ {\rm D}_{\rm I} \Phi{}_i = i[ {\bm {\rL}}{}_{\rI}]_i \, ^{\hat 
k} \, \Psi{}_{\hat k} ~~~.
\numberthis \label{cls_boson}
\]

Explicitly, these matrices are given by:

\be  {
[{{\bm \rL}_1}]_{i}{}^{\hat k}  = \left[\begin{array}{cccc|cccc|cccc}
    1 & 0 & 0 & 0 & 0 & 0 & 0 & 0 & 0 & 0 & 0 & 0 \\ 
    0 & 1 & 0 & 0 & 0 & 0 & 0 & 0 & 0 & 0 & 0 & 0 \\ 
    0 & 0 & 1 & 0 & 0 & 0 & 0 & 0 & 0 & 0 & 0 & 0 \\ 
    0 & 0 & 0 & 1 & 0 & 0 & 0 & 0 & 0 & 0 & 0 & 0 \\ 
    \hline
    0 & 0 & 0 & 0 & 1 & 0 & 0 & 0 & 0 & 0 & 0 & 0 \\ 
    0 & 0 & 0 & 0 & 0 & 1 & 0 & 0 & 0 & 0 & 0 & 0 \\ 
    0 & 0 & 0 & 0 & 0 & 0 & 1 & 0 & 0 & 0 & 0 & 0 \\ 
    0 & 0 & 0 & 0 & 0 & 0 & 0 & 1 & 0 & 0 & 0 & 0 \\ 
    \hline
    0 & 0 & 0 & 0 & 0 & 0 & 0 & 0 & 1 & 0 & 0 & 0 \\ 
    0 & 0 & 0 & 0 & 0 & 0 & 0 & 0 & 0 & 1 & 0 & 0 \\ 
    0 & 0 & 0 & 0 & 0 & 0 & 0 & 0 & 0 & 0 & 1 & 0 \\ 
    0 & 0 & 0 & 0 & 0 & 0 & 0 & 0 & 0 & 0 & 0 & 1
\numberthis\end{array}\right] ~~~,
\label{Ell1}
} \ee

\be  {
[{{\bm \rL}_2}]_{i}{}^{\hat k}  = \left[\begin{array}{cccc|cccc|cccc}
    0 & 1 & 0 & 0 & 0 & 0 & 0 & 0 & 0 & 0 & 0 & 0 \\  
    -1 & 0 & 0 & 0 & 0 & 0 & 0 & 0 & 0 & 0 & 0 & 0 \\ 
    0 & 0 & 0 & 1 & 0 & 0 & 0 & 0 & 0 & 0 & 0 & 0 \\  
    0 & 0 & -1 & 0 & 0 & 0 & 0 & 0 & 0 & 0 & 0 & 0 \\
    \hline
    0 & 0 & 0 & 0 & 0 & 1 & 0 & 0 & 0 & 0 & 0 & 0 \\
    0 & 0 & 0 & 0 & -1 & 0 & 0 & 0 & 0 & 0 & 0 & 0 \\
    0 & 0 & 0 & 0 & 0 & 0 & 0 & -1 & 0 & 0 & 0 & 0 \\
    0 & 0 & 0 & 0 & 0 & 0 & 1 & 0 & 0 & 0 & 0 & 0 \\
    \hline
    0 & 0 & 0 & 0 & 0 & 0 & 0 & 0 & 0 & -1 & 0 & 0 \\
    0 & 0 & 0 & 0 & 0 & 0 & 0 & 0 & 1 & 0 & 0 & 0 \\
    0 & 0 & 0 & 0 & 0 & 0 & 0 & 0 & 0 & 0 & 0 & 1 \\
    0 & 0 & 0 & 0 & 0 & 0 & 0 & 0 & 0 & 0 & -1 & 0
\numberthis\end{array}\right] ~~~, 
\label{Ell2}
} \ee

\be  {
[{{\bm \rL}_3}]_{i}{}^{\hat k}  = \left[\begin{array}{cccc|cccc|cccc}
    0 & 0 & 1 & 0 & 0 & 0 & 0 & 0 & 0 & 0 & 0 & 0 \\  
    0 & 0 & 0 & -1 & 0 & 0 & 0 & 0 & 0 & 0 & 0 & 0 \\ 
    -1 & 0 & 0 & 0 & 0 & 0 & 0 & 0 & 0 & 0 & 0 & 0 \\ 
    0 & 1 & 0 & 0 & 0 & 0 & 0 & 0 & 0 & 0 & 0 & 0 \\  
    \hline
    0 & 0 & 0 & 0 & 0 & 0 & -1 & 0 & 0 & 0 & 0 & 0 \\ 
    0 & 0 & 0 & 0 & 0 & 0 & 0 & -1 & 0 & 0 & 0 & 0 \\ 
    0 & 0 & 0 & 0 & 1 & 0 & 0 & 0 & 0 & 0 & 0 & 0 \\  
    0 & 0 & 0 & 0 & 0 & 1 & 0 & 0 & 0 & 0 & 0 & 0 \\  
    \hline
    0 & 0 & 0 & 0 & 0 & 0 & 0 & 0 & 0 & 0 & 1 & 0 \\  
    0 & 0 & 0 & 0 & 0 & 0 & 0 & 0 & 0 & 0 & 0 & 1 \\  
    0 & 0 & 0 & 0 & 0 & 0 & 0 & 0 & -1 & 0 & 0 & 0 \\ 
    0 & 0 & 0 & 0 & 0 & 0 & 0 & 0 & 0 & -1 & 0 & 0
\numberthis\end{array}\right] ~~~,
\label{Ell3}
} \ee

\be {
[{{\bm \rL}_4}]_{i}{}^{\hat k}  = \left[\begin{array}{cccc|cccc|cccc}
    0 & 0 & 0 & 1 & 0 & 0 & 0 & 0 & 0 & 0 & 0 & 0 \\
    0 & 0 & 1 & 0 & 0 & 0 & 0 & 0 & 0 & 0 & 0 & 0 \\
    0 & -1 & 0 & 0 & 0 & 0 & 0 & 0 & 0 & 0 & 0 & 0 \\
    -1 & 0 & 0 & 0 & 0 & 0 & 0 & 0 & 0 & 0 & 0 & 0 \\
    \hline
    0 & 0 & 0 & 0 & 0 & 0 & 0 & -1 & 0 & 0 & 0 & 0 \\
    0 & 0 & 0 & 0 & 0 & 0 & 1 & 0 & 0 & 0 & 0 & 0 \\
    0 & 0 & 0 & 0 & 0 & -1 & 0 & 0 & 0 & 0 & 0 & 0 \\
    0 & 0 & 0 & 0 & 1 & 0 & 0 & 0 & 0 & 0 & 0 & 0 \\
    \hline
    0 & 0 & 0 & 0 & 0 & 0 & 0 & 0 & 0 & 0 & 0 & -1 \\
    0 & 0 & 0 & 0 & 0 & 0 & 0 & 0 & 0 & 0 & 1 & 0 \\
    0 & 0 & 0 & 0 & 0 & 0 & 0 & 0 & 0 & -1 & 0 & 0 \\
    0 & 0 & 0 & 0 & 0 & 0 & 0 & 0 & 1 & 0 & 0 & 0
\numberthis\end{array}\right] ~~~.
\label{Ell4}
} \ee

By the same logic, given the above new fermionic field transformations, we can construct the $\bm \rR$ matrices as follows, where the matrices satisfy:

\[{\rm D}_{\rm I} \Psi{}_{\hat k} = \, [ {\bm {\rR}}{}_{\rI}]_{\hat
k}\,{}^i
\,\frac{d}{d\tau}\Phi{}_{i } ~~~. \numberthis \label{cls_fermion}\]

Explicitly, these matrices are given by:
\be  {  
[{{\bm \rR}_1}]_{\hat i}{}^{k} = \left[\begin{array}{cccc|cccc|cccc}
    1 & 0 & 0 & 0 & 0 & 0 & 0 & 0 & 0 & 0 & 0 & 0 \\ 
    0 & 1 & 0 & 0 & 0 & 0 & 0 & 0 & 0 & 0 & 0 & 0 \\ 
    0 & 0 & 1 & 0 & 0 & 0 & 0 & 0 & 0 & 0 & 0 & 0 \\ 
    0 & 0 & 0 & 1 & 0 & 0 & 0 & 0 & 0 & 0 & 0 & 0 \\ 
    \hline
    0 & 0 & 0 & 0 & 1 & 0 & 0 & 0 & 0 & 0 & 0 & 0 \\ 
    0 & 0 & 0 & 0 & 0 & 1 & 0 & 0 & 0 & 0 & 0 & 0 \\ 
    0 & 0 & 0 & 0 & 0 & 0 & 1 & 0 & 0 & 0 & 0 & 0 \\ 
    0 & 0 & 0 & 0 & 0 & 0 & 0 & 1 & 0 & 0 & 0 & 0 \\ 
    \hline
    0 & 0 & 0 & 0 & 0 & 0 & 0 & 0 & 1 & 0 & 0 & 0 \\ 
    0 & 0 & 0 & 0 & 0 & 0 & 0 & 0 & 0 & 1 & 0 & 0 \\ 
    0 & 0 & 0 & 0 & 0 & 0 & 0 & 0 & 0 & 0 & 1 & 0 \\ 
    0 & 0 & 0 & 0 & 0 & 0 & 0 & 0 & 0 & 0 & 0 & 1
\numberthis\end{array}\right]~~~,    
}     \label{Rrrs1} \ee

\be  {  
[{{\bm \rR}_2}]_{\hat i}{}^{k} = \left[\begin{array}{cccc|cccc|cccc}
    0 & -1 & 0 & 0 & 0 & 0 & 0 & 0 & 0 & 0 & 0 & 0 \\ 
    1 & 0 & 0 & 0 & 0 & 0 & 0 & 0 & 0 & 0 & 0 & 0 \\  
    0 & 0 & 0 & -1 & 0 & 0 & 0 & 0 & 0 & 0 & 0 & 0 \\ 
    0 & 0 & 1 & 0 & 0 & 0 & 0 & 0 & 0 & 0 & 0 & 0 \\  
    \hline
    0 & 0 & 0 & 0 & 0 & -1 & 0 & 0 & 0 & 0 & 0 & 0 \\ 
    0 & 0 & 0 & 0 & 1 & 0 & 0 & 0 & 0 & 0 & 0 & 0 \\  
    0 & 0 & 0 & 0 & 0 & 0 & 0 & 1 & 0 & 0 & 0 & 0 \\  
    0 & 0 & 0 & 0 & 0 & 0 & -1 & 0 & 0 & 0 & 0 & 0 \\ 
    \hline
    0 & 0 & 0 & 0 & 0 & 0 & 0 & 0 & 0 & 1 & 0 & 0 \\  
    0 & 0 & 0 & 0 & 0 & 0 & 0 & 0 & -1 & 0 & 0 & 0 \\ 
    0 & 0 & 0 & 0 & 0 & 0 & 0 & 0 & 0 & 0 & 0 & -1 \\ 
    0 & 0 & 0 & 0 & 0 & 0 & 0 & 0 & 0 & 0 & 1 & 0
\numberthis\end{array}\right]~~~,    
}     \label{Rrrs2} \ee

\be  {   
[{{\bm \rR}_3}]_{\hat i}{}^{k} = \left[\begin{array}{cccc|cccc|cccc}
    0 & 0 & -1 & 0 & 0 & 0 & 0 & 0 & 0 & 0 & 0 & 0 \\ 
    0 & 0 & 0 & 1 & 0 & 0 & 0 & 0 & 0 & 0 & 0 & 0 \\  
    1 & 0 & 0 & 0 & 0 & 0 & 0 & 0 & 0 & 0 & 0 & 0 \\  
    0 & -1 & 0 & 0 & 0 & 0 & 0 & 0 & 0 & 0 & 0 & 0 \\ 
    \hline
    0 & 0 & 0 & 0 & 0 & 0 & 1 & 0 & 0 & 0 & 0 & 0 \\  
    0 & 0 & 0 & 0 & 0 & 0 & 0 & 1 & 0 & 0 & 0 & 0 \\  
    0 & 0 & 0 & 0 & -1 & 0 & 0 & 0 & 0 & 0 & 0 & 0 \\ 
    0 & 0 & 0 & 0 & 0 & -1 & 0 & 0 & 0 & 0 & 0 & 0 \\ 
    \hline
    0 & 0 & 0 & 0 & 0 & 0 & 0 & 0 & 0 & 0 & -1 & 0 \\ 
    0 & 0 & 0 & 0 & 0 & 0 & 0 & 0 & 0 & 0 & 0 & -1 \\ 
    0 & 0 & 0 & 0 & 0 & 0 & 0 & 0 & 1 & 0 & 0 & 0 \\  
    0 & 0 & 0 & 0 & 0 & 0 & 0 & 0 & 0 & 1 & 0 & 0
\numberthis\end{array}\right]~~~,    
}     \label{Rrrs3} \ee

\be  {  
[{{\bm \rR}_4}]_{\hat i}{}^{k} = \left[\begin{array}{cccc|cccc|cccc}
    0 & 0 & 0 & -1 & 0 & 0 & 0 & 0 & 0 & 0 & 0 & 0 \\ 
    0 & 0 & -1 & 0 & 0 & 0 & 0 & 0 & 0 & 0 & 0 & 0 \\ 
    0 & 1 & 0 & 0 & 0 & 0 & 0 & 0 & 0 & 0 & 0 & 0 \\  
    1 & 0 & 0 & 0 & 0 & 0 & 0 & 0 & 0 & 0 & 0 & 0 \\  
    \hline
    0 & 0 & 0 & 0 & 0 & 0 & 0 & 1 & 0 & 0 & 0 & 0 \\  
    0 & 0 & 0 & 0 & 0 & 0 & -1 & 0 & 0 & 0 & 0 & 0 \\ 
    0 & 0 & 0 & 0 & 0 & 1 & 0 & 0 & 0 & 0 & 0 & 0 \\  
    0 & 0 & 0 & 0 & -1 & 0 & 0 & 0 & 0 & 0 & 0 & 0 \\ 
    \hline
    0 & 0 & 0 & 0 & 0 & 0 & 0 & 0 & 0 & 0 & 0 & 1 \\  
    0 & 0 & 0 & 0 & 0 & 0 & 0 & 0 & 0 & 0 & -1 & 0 \\ 
    0 & 0 & 0 & 0 & 0 & 0 & 0 & 0 & 0 & 1 & 0 & 0 \\  
    0 & 0 & 0 & 0 & 0 & 0 & 0 & 0 & -1 & 0 & 0 & 0
\numberthis\end{array}\right]~~~,    
}   \label{Rrrs4} \ee

We can see that for each ${\bm \rL}_\rI$ matrix, the associated ${\bm \rR}_\rI$ matrix is its transpose. This property of the ${\bm \rL}_\rI$ and ${\bm \rR}_\rI$ matrices is apparently related to the
fact that the basis of fields defined in Eq.\ (\ref{Bsa1}) and Eq.\ (\ref{Bsa2})
causes the 12 $\times $ 12 matrices to be of forms that are block diagonal,
Additionally, these matrices satisfy the garden algebra, Eq. (\ref{eq3-4}):

\begin{equation}
    [{{\bm \rL}_\rI}]_{i}{}^{\hat j}[{{\bm \rR}_\rJ}]_{\hat j}{}^{k}+[{{\bm \rL}_\rJ}]_{i}{}^{\hat j}[{{\bm \rR}_\rI}]_{\hat j}{}^{k}=2\delta_{\rI\rJ}\delta_{i}{}^{k}~~~,\quad
    [{{\bm \rR}_\rI}]_{\hat i}{}^{j}[{{\bm \rL}_\rJ}]_{j}{}^{\hat k}+[{{\bm \rR}_\rJ}]_{\hat i}{}^{j}[{{\bm \rL}_\rI}]_{j}{}^{\hat k}=2\delta_{\rI\rJ}\delta_{\hat i}{}^{\hat k}
    \label{eq3-4} ~~~.
\end{equation}

As a direct result, the SUSY closure relation holds for each bosonic and fermionic field: 

\begin{equation}
    \{{{\rm D_\rI},{\rm D_J}\}\Phi_i=2i\delta_{\rI\rJ}\frac{d}{d\tau}\Phi_{i}~~~,\quad  \{{\rm D_\rI},{\rm D_\rJ}\}\Psi_{\hat i}=2i\delta_{\rI\rJ}\frac{d}{d\tau}\Psi_{\hat i}}~~~.
    \label{eq3-5}
\end{equation}

\subsection{Valise Adinkras}
We can draw the three separate adinkras for the CLS based on calculated $\bm \rL$ and $\bm \rR$ matrices:\footnote{Here, red, yellow, blue, green indicate $\rD_1, \rD_2, \rD_3, \rD_4$ respectively. A dotted line represents a negative relationship and a solid line represents a positive relationship between the two fields.}

\begin{figure}[H]
	\centering
	\includegraphics[scale=0.8]{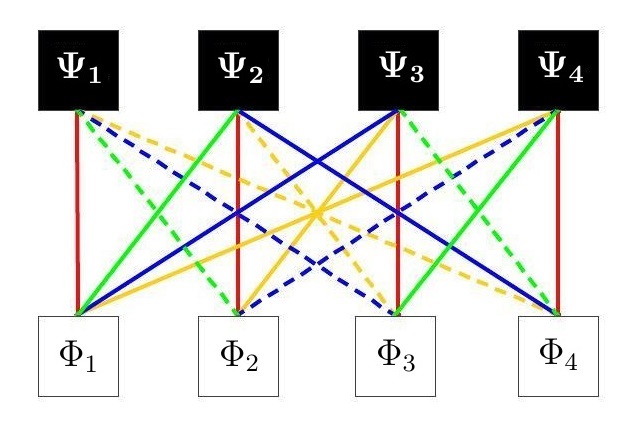}
	\caption{A valise adinkra for CLS, containing the first four bosonic and fermionic fields.}
	\label{fig4}
\end{figure}

\begin{figure}[H]
	\centering
	\includegraphics[scale=0.8]{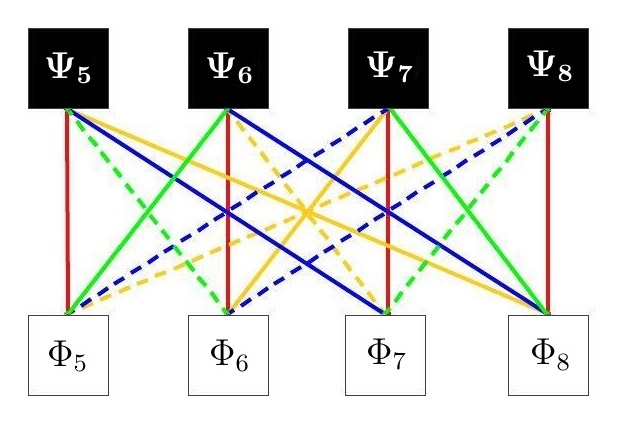}
	\caption{A valise adinkra for CLS, containing the middle four bosonic and fermionic fields.}
	\label{fig5}
\end{figure}

\begin{figure}[H]
	\centering
	\includegraphics[scale=0.8]{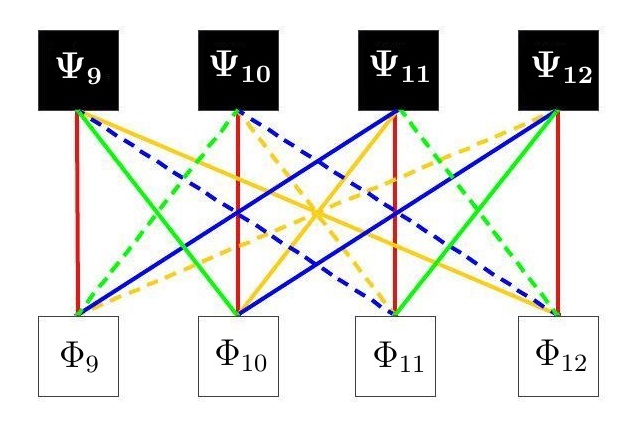}
	\caption{A valise adinkra for CLS, containing the last four bosonic and fermionic fields.}
	\label{fig6}
\end{figure}

\section{Unfolded Adinkras for CLS Model}
We can define the total collection of bosonic fields, fermionic fields, and the
supercovariant derivatives in a basis where
\[ \Phi_{i} = \left[\begin{array}{c}
    K  \\ 
    L   \\
    M  \\
    N  \\ 
    \hline
    V_0  \\ 
    V_1  \\ 
    V_2  \\ 
    V_3  \\ 
    \hline
     U_0  \\ 
     U_1  \\ 
     U_2  \\ 
     U_3  
\end{array}\right] ~~~,~~~
i\Psi{}_{\hat k} = \left[\begin{array}{c}
    \z_1  \\ 
    \z_2  \\
    \z_3  \\
    \z_4  \\
    \hline
    \rho_1  \\ 
    \rho_2  \\ 
    \rho_3  \\ 
    \rho_4  \\ 
    \hline
    \b_1  \\ 
    \b_2  \\ 
    \b_3  \\ 
    \b_4 
\end{array}\right]  ~~~,~~~
{\rm D}_{\rm I} = \left[\begin{array}{c}
    {\rm D}_1  \\ 
    {\rm D}_2  \\     
    {\rm D}_3  \\ 
    {\rm D}_4 
\numberthis\end{array}\right] ~~~.
\]
In terms of these new quantities, Eq. (\ref{cls_boson}) and Eq. (\ref{cls_fermion}) take the forms:
\[ {\rm D}_{\rm I} \Phi{}_i = i \, [ {\bm {\rL}}{}_{\rI}^{(1)}]_i \, ^{\hat 
k} \, \Psi{}_{\hat k} ~+~ i \, [ {\bm {\rL}}{}_{\rI}^{(2)}]_i{}^{\hat k}
\, \frac{d}{d \tau} \Psi{}_{\hat k}~~~,
\label{eq54}
\numberthis\]
\[ {\rm D}_{\rm I} \Psi{}_{\hat k} = [ {\bm {\rR}}{}_{\rI}^{(0)}]_{\hat
k} \,{}^ i \,
\Phi{}_{i }
~+~ [ {\bm {\rR}}{}_{\rI}^{(1)}]_{\hat k} \, {}^i \, \frac{d ~}{d \tau}
\Phi{}_{i}
~+~ [ {\bm {\rR}}{}_{\rI}^{(2)}]_{\hat k} \, {}^i \, \frac{d^2 ~}{d \tau^2}
\Phi{}_{i}~~~.
\label{eq55}
\numberthis\]

We can now derive all $\bm \rL$ matrices as follows:

\[
[{{\bm \rL}_1^{(1)}}]_{i}{}^{\hat k}  = \left[\begin{array}{cccc|cccc|cccc}
    -1 & 0 & 0 & 0 & 1 & 0 & 0 & 0 & 0 & 0 & 0 & 0 \\ 
    0 & 0 & 0 & 1 & 0 & 0 & 0 & 1 & 0 & 0 & 0 & 0 \\ 
    0 & 0 & 0 & 0 & 0 & 0 & 0 & 0 & 1 & 0 & 0 & 0 \\ 
    0 & 0 & 0 & 0 & 0 & 0 & 0 & 0 & 0 & 0 & 0 & 1 \\ 
    \hline
    0 & 0 & 0 & 0 & 0 & 0 & 0 & 0 & 0 & 1 & 0 & 0 \\ 
    0 & 0 & 0 & 0 & 0 & 0 & 0 & 0 & 0 & -1 & 0 & 0 \\ 
    0 & 0 & 0 & 0 & 0 & 0 & 0 & 0 & 0 & 0 & 0 & 1 \\ 
    0 & 0 & 0 & 0 & 0 & 0 & 0 & 0 & -1 & 0 & 0 & 0 \\ 
    \hline
    0 & 0 & 0 & 0 & 0 & 0 & 0 & 0 & 0 & 0 & 1 & 0 \\ 
    0 & 0 & 0 & 0 & 0 & 0 & 0 & 0 & 0 & 0 & -1 & 0 \\ 
    0 & 0 & 0 & 0 & 0 & 0 & 0 & 0 & 1 & 0 & 0 & 0 \\ 
    0 & 0 & 0 & 0 & 0 & 0 & 0 & 0 & 0 & 0 & 0 & 1
\numberthis
\label{eq4-10}
\end{array}\right], \]

\[
[{{\bm \rL}_1^{(2)}}]_{i}{}^{\hat k}  = \left[\begin{array}{cccc|cccc|cccc}
    0 & 0 & 0 & 0 & 0 & 0 & 0 & 0 & 0 & 0 & 0 & 0 \\ 
    0 & 0 & 0 & 0 & 0 & 0 & 0 & 0 & 0 & 0 & 0 & 0 \\ 
    0 & 0 & 0 & 0 & 0 & -\frac{1}{2} & 0 & 0 & 0 & 0 & 0 & 0 \\ 
    0 & 0 & 0 & 0 & 0 & 0 & -\tfrac{1}{2} & 0 & 0 & 0 & 0 & 0 \\ 
    \hline
    -1 & 0 & 0 & 0 & \frac{3}{2} & 0 & 0 & 0 & 0 & 0 & 0 & 0 \\ 
    1 & 0 & 0 & 0 & \frac{1}{2} & 0 & 0 & 0 & 0 & 0 & 0 & 0 \\ 
    0 & 0 & 1 & 0 & 0 & 0 & \frac{1}{2} & 0 & 0 & 0 & 0 & 0 \\ 
    0 & -1 & 0 & 0 & 0 & -\frac{1}{2} & 0 & 0 & 0 & 0 & 0 & 0 \\ 
    \hline
    0 & 0 & 0 & 1 & 0 & 0 & 0 & \frac{3}{2} & 0 & 0 & 0 & 0 \\ 
    0 & 0 & 0 & -1 & 0 & 0 & 0 & \frac{1}{2} & 0 & 0 & 0 & 0 \\ 
    0 & -1 & 0 & 0 & 0 & \frac{1}{2} & 0 & 0 & 0 & 0 & 0 & 0 \\ 
    0 & 0 & -1 & 0 & 0 & 0 & \frac{1}{2} & 0 & 0 & 0 & 0 & 0
\numberthis\end{array}\right], \]

\[
[{{\bm \rL}_2^{(1)}}]_{i}{}^{\hat k}  = \left[\begin{array}{cccc|cccc|cccc}
    0 & -1 & 0 & 0 & 0 & 1 & 0 & 0 & 0 & 0 & 0 & 0 \\ 
    0 & 0 & -1 & 0 & 0 & 0 & -1 & 0 & 0 & 0 & 0 & 0 \\ 
    0 & 0 & 0 & 0 & 0 & 0 & 0 & 0 & 0 & 1 & 0 & 0 \\ 
    0 & 0 & 0 & 0 & 0 & 0 & 0 & 0 & 0 & 0 & -1 & 0 \\ 
    \hline
    0 & 0 & 0 & 0 & 0 & 0 & 0 & 0 & -1 & 0 & 0 & 0 \\ 
    0 & 0 & 0 & 0 & 0 & 0 & 0 & 0 & -1 & 0 & 0 & 0 \\ 
    0 & 0 & 0 & 0 & 0 & 0 & 0 & 0 & 0 & 0 & -1 & 0 \\ 
    0 & 0 & 0 & 0 & 0 & 0 & 0 & 0 & 0 & 1 & 0 & 0 \\ 
    \hline
    0 & 0 & 0 & 0 & 0 & 0 & 0 & 0 & 0 & 0 & 0 & 1 \\ 
    0 & 0 & 0 & 0 & 0 & 0 & 0 & 0 & 0 & 0 & 0 & 1 \\ 
    0 & 0 & 0 & 0 & 0 & 0 & 0 & 0 & 0 & 1 & 0 & 0 \\ 
    0 & 0 & 0 & 0 & 0 & 0 & 0 & 0 & 0 & 0 & 1 & 0
\numberthis\end{array}\right], \]

\[
[{{\bm \rL}_2^{(2)}}]_{i}{}^{\hat k}  = \left[\begin{array}{cccc|cccc|cccc}
    0 & 0 & 0 & 0 & 0 & 0 & 0 & 0 & 0 & 0 & 0 & 0 \\ 
    0 & 0 & 0 & 0 & 0 & 0 & 0 & 0 & 0 & 0 & 0 & 0 \\ 
    0 & 0 & 0 & 0 & \frac{1}{2} & 0 & 0 & 0 & 0 & 0 & 0 & 0 \\ 
    0 & 0 & 0 & 0 & 0 & 0 & 0 & -\frac{1}{2} & 0 & 0 & 0 & 0 \\ 
    \hline
    0 & -1 & 0 & 0 & 0 & \frac{3}{2} & 0 & 0 & 0 & 0 & 0 & 0 \\ 
    0 & -1 & 0 & 0 & 0 & -\frac{1}{2} & 0 & 0 & 0 & 0 & 0 & 0 \\ 
    0 & 0 & 0 & 1 & 0 & 0 & 0 & \frac{1}{2} & 0 & 0 & 0 & 0 \\ 
    -1 & 0 & 0 & 0 & -\frac{1}{2} & 0 & 0 & 0 & 0 & 0 & 0 & 0 \\ 
    \hline
    0 & 0 & -1 & 0 & 0 & 0 & -\frac{3}{2} & 0 & 0 & 0 & 0 & 0 \\ 
    0 & 0 & -1 & 0 & 0 & 0 & \frac{1}{2} & 0 & 0 & 0 & 0 & 0 \\ 
    1 & 0 & 0 & 0 & -\frac{1}{2} & 0 & 0 & 0 & 0 & 0 & 0 & 0 \\ 
    0 & 0 & 0 & 1 & 0 & 0 & 0 & -\frac{1}{2} & 0 & 0 & 0 & 0
\numberthis\end{array}\right], \]

\[
[{{\bm \rL}_3^{(1)}}]_{i}{}^{\hat k}  = \left[\begin{array}{cccc|cccc|cccc}
    0 & 0 & -1 & 0& 0& 0 & 1 & 0 & 0 & 0 & 0 & 0 \\ 
    0 & 1 & 0 & 0 & 0 & 1 & 0 & 0 & 0 & 0 & 0 & 0 \\ 
    0 & 0 & 0 & 0 & 0 & 0 & 0 & 0 & 0 & 0 & 1 & 0 \\ 
    0 & 0 & 0 & 0 & 0 & 0 & 0 & 0 & 0 & 1 & 0 & 0 \\ 
    \hline
    0 & 0 & 0 & 0 & 0 & 0 & 0 & 0 & 0 & 0 & 0 & -1 \\ 
    0 & 0 & 0 & 0 & 0 & 0 & 0 & 0 & 0 & 0 & 0 & -1 \\ 
    0 & 0 & 0 & 0 & 0 & 0 & 0 & 0 & 0 & -1 & 0 & 0 \\ 
    0 & 0 & 0 & 0 & 0 & 0 & 0 & 0 & 0 & 0 & -1 & 0 \\ 
    \hline
    0 & 0 & 0 & 0 & 0 & 0 & 0 & 0 & -1 & 0 & 0 & 0 \\ 
    0 & 0 & 0 & 0 & 0 & 0 & 0 & 0 & -1 & 0 & 0 & 0 \\ 
    0 & 0 & 0 & 0 & 0 & 0 & 0 & 0 & 0 & 0 & -1 & 0 \\ 
    0 & 0 & 0 & 0 & 0 & 0 & 0 & 0 & 0 & 1 & 0 & 0
\numberthis\end{array}\right], \]

\[
[{{\bm \rL}_3^{(2)}}]_{i}{}^{\hat k}  = \left[\begin{array}{cccc|cccc|cccc}
    0 & 0 & 0 & 0 & 0 & 0 & 0 & 0 & 0 & 0 & 0 & 0 \\ 
    0 & 0 & 0 & 0 & 0 & 0 & 0 & 0 & 0 & 0 & 0 & 0 \\ 
    0 & 0 & 0 & 0 & 0 & 0 & 0 & \frac{1}{2} & 0 & 0 & 0 & 0 \\ 
    0 & 0 & 0 & 0 & \frac{1}{2} & 0 & 0 & 0 & 0 & 0 & 0 & 0 \\ 
    \hline
    0 & 0 & -1 & 0 & 0 & 0 & \frac{3}{2} & 0 & 0 & 0 & 0 & 0 \\ 
    0 & 0 & -1 & 0 & 0 & 0 & -\frac{1}{2} & 0 & 0 & 0 & 0 & 0 \\ 
    1 & 0 & 0 & 0 & \frac{1}{2} & 0 & 0 & 0 & 0 & 0 & 0 & 0 \\ 
    0 & 0 & 0 & 1 & 0 & 0 & 0 & \frac{1}{2} & 0 & 0 & 0 & 0 \\ 
    \hline
    0 & 1 & 0 & 0 & 0 & \frac{3}{2} & 0 & 0 & 0 & 0 & 0 & 0 \\ 
    0 & 1 & 0 & 0 & 0 & -\frac{1}{2} & 0 & 0 & 0 & 0 & 0 & 0 \\ 
    0 & 0 & 0 & -1 & 0 & 0 & 0 & \frac{1}{2} & 0 & 0 & 0 & 0 \\ 
    1 & 0 & 0 & 0 & -\frac{1}{2} & 0 & 0 & 0 & 0 & 0 & 0 & 0
\numberthis\end{array}\right], \]

\[
[{{\bm \rL}_4^{(1)}}]_{i}{}^{\hat k}  = \left[\begin{array}{cccc|cccc|cccc}
    0 & 0 & 0 & -1 & 0 & 0 & 0 & 1 & 0 & 0 & 0 & 0 \\ 
    -1 & 0 & 0 & 0 & -1 & 0 & 0 & 0 & 0 & 0 & 0 & 0 \\ 
    0 & 0 & 0 & 0 & 0 & 0 & 0 & 0 & 0 & 0 & 0 & 1 \\ 
    0 & 0 & 0 & 0 & 0 & 0 & 0 & 0 & -1 & 0 & 0 & 0 \\ 
    \hline
    0 & 0 & 0 & 0 & 0 & 0 & 0 & 0 & 0 & 0 & 1 & 0 \\ 
    0 & 0 & 0 & 0 & 0 & 0 & 0 & 0 & 0 & 0 & -1 & 0 \\ 
    0 & 0 & 0 & 0 & 0 & 0 & 0 & 0 & 1 & 0 & 0 & 0 \\ 
    0 & 0 & 0 & 0 & 0 & 0 & 0 & 0 & 0 & 0 & 0 & 1 \\ 
    \hline
    0 & 0 & 0 & 0 & 0 & 0 & 0 & 0 & 0 & -1 & 0 & 0 \\ 
    0 & 0 & 0 & 0 & 0 & 0 & 0 & 0 & 0 & 1 & 0 & 0 \\ 
    0 & 0 & 0 & 0 & 0 & 0 & 0 & 0 & 0 & 0 & 0 & -1 \\ 
    0 & 0 & 0 & 0 & 0 & 0 & 0 & 0 & 1 & 0 & 0 & 0
\numberthis\end{array}\right], \]

\[
[{{\bm \rL}_4^{(2)}}]_{i}{}^{\hat k}  = \left[\begin{array}{cccc|cccc|cccc}
    0 & 0 & 0 & 0 & 0 & 0 & 0 & 0 & 0 & 0 & 0 & 0 \\ 
    0 & 0 & 0 & 0 & 0 & 0 & 0 & 0 & 0 & 0 & 0 & 0 \\ 
    0 & 0 & 0 & 0 & 0 & 0 & -\frac{1}{2} & 0 & 0 & 0 & 0 & 0 \\ 
    0 & 0 & 0 & 0 & 0 & \frac{1}{2} & 0 & 0 & 0 & 0 & 0 & 0 \\ 
    \hline
    0 & 0 & 0 & -1 & 0 & 0 & 0 & \frac{3}{2} & 0 & 0 & 0 & 0 \\ 
    0 & 0 & 0 & 1 & 0 & 0 & 0 & \frac{1}{2} & 0 & 0 & 0 & 0 \\ 
    0 & 1 & 0 & 0 & 0 & \frac{1}{2} & 0 & 0 & 0 & 0 & 0 & 0 \\ 
    0 & 0 & 1 & 0 & 0 & 0 & \frac{1}{2} & 0 & 0 & 0 & 0 & 0 \\ 
    \hline
    -1 & 0 & 0 & 0 & -\frac{3}{2} & 0 & 0 & 0 & 0 & 0 & 0 & 0 \\ 
    1 & 0 & 0 & 0 & -\frac{1}{2} & 0 & 0 & 0 & 0 & 0 & 0 & 0 \\ 
    0 & 0 & 1 & 0 & 0 & 0 & -\frac{1}{2} & 0 & 0 & 0 & 0 & 0 \\ 
    0 & -1 & 0 & 0 & 0 & \frac{1}{2} & 0 & 0 & 0 & 0 & 0 & 0
\numberthis\end{array}\right]. \]

Similarly, for the $\bm \rR$ matrices:

\[
[{{\bm \rR}_1^{(0)}}]_{\hat i}{}^{k} = \left[\begin{array}{cccc|cccc|cccc}
    0 & 0 & 0 & 0 & \frac{1}{2} & \frac{1}{2} & 0 & 0 & 0 & 0 & 0 & 0 \\ 
    0 & 0 & 0 & 0 & 0 & 0 & 0 & -\frac{1}{2} & 0 & 0 & -\frac{1}{2} & 0 \\ 
    0 & 0 & 0 & 0 & 0 & 0 & \frac{1}{2} & 0 & 0 & 0 & 0 & -\frac{1}{2} \\ 
    0 & 0 & 0 & 0 & 0 & 0 & 0 & 0 & -\frac{1}{2} & -\frac{1}{2} & 0 & 0 \\ 
    \hline
    0 & 0 & 0 & 0 & \frac{1}{2} & \frac{1}{2} & 0 & 0 & 0 & 0 & 0 & 0 \\ 
    0 & 0 & -1 & 0 & 0 & 0 & 0 & -\frac{1}{2} & 0 & 0 & \frac{1}{2} & 0 \\ 
    0 & 0 & 0 & -1 & 0 & 0 & \frac{1}{2} & 0 & 0 & 0 & 0 & \frac{1}{2} \\ 
    0 & 0 & 0 & 0 & 0 & 0 & 0 & 0 & \frac{1}{2} & \frac{1}{2} & 0 & 0 \\ 
    \hline
    0 & 0 & 0 & 0 & 0 & 0 & 0 & 0 & 0 & 0 & 0 & 0 \\ 
    0 & 0 & 0 & 0 & 0 & 0 & 0 & 0 & 0 & 0 & 0 & 0 \\ 
    0 & 0 & 0 & 0 & 0 & 0 & 0 & 0 & 0 & 0 & 0 & 0 \\ 
    0 & 0 & 0 & 0 & 0 & 0 & 0 & 0 & 0 & 0 & 0 & 0
\numberthis\end{array}\right], \]

\[
[{{\bm \rR}_1^{(1)}}]_{\hat i}{}^{k} = \left[\begin{array}{cccc|cccc|cccc}
    -1 & 0 & 0 & 0 & 0 & 0 & 0 & 0 & 0 & 0 & 0 & 0 \\ 
    0 & 0 & 0 & 0 & 0 & 0 & 0 & 0 & 0 & 0 & 0 & 0 \\ 
    0 & 0 & 0 & 0 & 0 & 0 & 0 & 0 & 0 & 0 & 0 & 0 \\ 
    0 & 1 & 0 & 0 & 0 & 0 & 0 & 0 & 0 & 0 & 0 & 0 \\ 
    \hline
    0 & 0 & 0 & 0 & 0 & 0 & 0 & 0 & 0 & 0 & 0 & 0 \\ 
    0 & 0 & 0 & 0 & 0 & 0 & 0 & 0 & 0 & 0 & 0 & 0 \\ 
    0 & 0 & 0 & 0 & 0 & 0 & 0 & 0 & 0 & 0 & 0 & 0 \\ 
    0 & 0 & 0 & 0 & 0 & 0 & 0 & 0 & 0 & 0 & 0 & 0 \\ 
    \hline
    0 & 0 & \frac{1}{2} & 0 & 0 & 0 & 0 & -\frac{1}{4} & 0 & 0 & \frac{1}{4} & 0 \\ 
    0 & 0 & 0 & 0 & \frac{3}{4} & -\frac{1}{4} & 0 & 0 & 0 & 0 & 0 & 0 \\ 
    0 & 0 & 0 & 0 & 0 & 0 & 0 & 0 & \frac{3}{4} & -\frac{1}{4} & 0 & 0 \\ 
    0 & 0 & 0 & \frac{1}{2} & 0 & 0 & \frac{1}{4} & 0 & 0 & 0 & 0 & \frac{1}{4}
\numberthis\end{array}\right], \]

\[
[{{\bm \rR}_1^{(2)}}]_{\hat i}{}^{k} = \left[\begin{array}{cccc|cccc|cccc}
    0 & 0 & 0 & 0 & 0 & 0 & 0 & 0 & 0 & 0 & 0 & 0 \\ 
    0 & 0 & 0 & 0 & 0 & 0 & 0 & 0 & 0 & 0 & 0 & 0 \\ 
    0 & 0 & 0 & 0 & 0 & 0 & 0 & 0 & 0 & 0 & 0 & 0 \\ 
    0 & 0 & 0 & 0 & 0 & 0 & 0 & 0 & 0 & 0 & 0 & 0 \\ 
    \hline
    0 & 0 & 0 & 0 & 0 & 0 & 0 & 0 & 0 & 0 & 0 & 0 \\ 
    0 & 0 & 0 & 0 & 0 & 0 & 0 & 0 & 0 & 0 & 0 & 0 \\ 
    0 & 0 & 0 & 0 & 0 & 0 & 0 & 0 & 0 & 0 & 0 & 0 \\ 
    0 & 0 & 0 & 0 & 0 & 0 & 0 & 0 & 0 & 0 & 0 & 0 \\ 
    \hline
    0 & 0 & 0 & 0 & 0 & 0 & 0 & 0 & 0 & 0 & 0 & 0 \\ 
    -1 & 0 & 0 & 0 & 0 & 0 & 0 & 0 & 0 & 0 & 0 & 0 \\ 
    0 & -1 & 0 & 0 & 0 & 0 & 0 & 0 & 0 & 0 & 0 & 0 \\ 
    0 & 0 & 0 & 0 & 0 & 0 & 0 & 0 & 0 & 0 & 0 & 0
\numberthis\end{array}\right], \]

\[
[{{\bm \rR}_2^{(0)}}]_{\hat i}{}^{k} = \left[\begin{array}{cccc|cccc|cccc}
    0 & 0 & 0 & 0 & 0 & 0 & 0 & -\frac{1}{2} & 0 & 0 & \frac{1}{2} & 0 \\ 
    0 & 0 & 0 & 0 & \frac{1}{2} & -\frac{1}{2} & 0 & 0 & 0 & 0 & 0 & 0 \\ 
    0 & 0 & 0 & 0 & 0 & 0 & 0 & 0 & \frac{1}{2} & -\frac{1}{2} & 0 & 0 \\ 
    0 & 0 & 0 & 0 & 0 & 0 & \frac{1}{2} & 0 & 0 & 0 & 0 & \frac{1}{2} \\ 
    \hline
    0 & 0 & 1 & 0 & 0 & 0 & 0 & -\frac{1}{2} & 0 & 0 & -\frac{1}{2} & 0 \\ 
    0 & 0 & 0 & 0 & \frac{1}{2} & -\frac{1}{2} & 0 & 0 & 0 & 0 & 0 & 0 \\ 
    0 & 0 & 0 & 0 & 0 & 0 & 0 & 0 & -\frac{1}{2} & \frac{1}{2} & 0 & 0 \\ 
    0 & 0 & 0 & -1 & 0 & 0 & \frac{1}{2} & 0 & 0 & 0 & 0 & -\frac{1}{2} \\ 
    \hline
    0 & 0 & 0 & 0 & 0 & 0 & 0 & 0 & 0 & 0 & 0 & 0 \\ 
    0 & 0 & 0 & 0 & 0 & 0 & 0 & 0 & 0 & 0 & 0 & 0 \\ 
    0 & 0 & 0 & 0 & 0 & 0 & 0 & 0 & 0 & 0 & 0 & 0 \\ 
    0 & 0 & 0 & 0 & 0 & 0 & 0 & 0 & 0 & 0 & 0 & 0
\numberthis\end{array}\right], \]

\[
[{{\bm \rR}_2^{(1)}}]_{\hat i}{}^{k} = \left[\begin{array}{cccc|cccc|cccc}
    0 & 0 & 0 & 0 & 0 & 0 & 0 & 0 & 0 & 0 & 0 & 0 \\ 
    -1 & 0 & 0 & 0 & 0 & 0 & 0 & 0 & 0 & 0 & 0 & 0 \\ 
    0 & -1 & 0 & 0 & 0 & 0 & 0 & 0 & 0 & 0 & 0 & 0 \\ 
    0 & 0 & 0 & 0 & 0 & 0 & 0 & 0 & 0 & 0 & 0 & 0 \\ 
    \hline
    0 & 0 & 0 & 0 & 0 & 0 & 0 & 0 & 0 & 0 & 0 & 0 \\ 
    0 & 0 & 0 & 0 & 0 & 0 & 0 & 0 & 0 & 0 & 0 & 0 \\ 
    0 & 0 & 0 & 0 & 0 & 0 & 0 & 0 & 0 & 0 & 0 & 0 \\ 
    0 & 0 & 0 & 0 & 0 & 0 & 0 & 0 & 0 & 0 & 0 & 0 \\ 
    \hline
    0 & 0 & 0 & 0 & -\frac{3}{4} & -\frac{1}{4} & 0 & 0 & 0 & 0 & 0 & 0 \\ 
    0 & 0 & \frac{1}{2} & 0 & 0 & 0 & 0 & \frac{1}{4} & 0 & 0 & \frac{1}{4} & 0 \\ 
    0 & 0 & 0 & -\frac{1}{2} & 0 & 0 & -\frac{1}{4} & 0 & 0 & 0 & 0 & \frac{1}{4} \\ 
    0 & 0 & 0 & 0 & 0 & 0 & 0 & 0 & \frac{3}{4} & \frac{1}{4} & 0 & 0
\numberthis\end{array}\right], \]

\[
[{{\bm \rR}_2^{(2)}}]_{\hat i}{}^{k} = \left[\begin{array}{cccc|cccc|cccc}
    0 & 0 & 0 & 0 & 0 & 0 & 0 & 0 & 0 & 0 & 0 & 0 \\ 
    0 & 0 & 0 & 0 & 0 & 0 & 0 & 0 & 0 & 0 & 0 & 0 \\ 
    0 & 0 & 0 & 0 & 0 & 0 & 0 & 0 & 0 & 0 & 0 & 0 \\ 
    0 & 0 & 0 & 0 & 0 & 0 & 0 & 0 & 0 & 0 & 0 & 0 \\ 
    \hline
    0 & 0 & 0 & 0 & 0 & 0 & 0 & 0 & 0 & 0 & 0 & 0 \\ 
    0 & 0 & 0 & 0 & 0 & 0 & 0 & 0 & 0 & 0 & 0 & 0 \\ 
    0 & 0 & 0 & 0 & 0 & 0 & 0 & 0 & 0 & 0 & 0 & 0 \\ 
    0 & 0 & 0 & 0 & 0 & 0 & 0 & 0 & 0 & 0 & 0 & 0 \\ 
    \hline
    1 & 0 & 0 & 0 & 0 & 0 & 0 & 0 & 0 & 0 & 0 & 0 \\ 
    0 & 0 & 0 & 0 & 0 & 0 & 0 & 0 & 0 & 0 & 0 & 0 \\ 
    0 & 0 & 0 & 0 & 0 & 0 & 0 & 0 & 0 & 0 & 0 & 0 \\ 
    0 & -1 & 0 & 0 & 0 & 0 & 0 & 0 & 0 & 0 & 0 & 0
\numberthis\end{array}\right], \]

\[
[{{\bm \rR}_3^{(0)}}]_{\hat i}{}^{k} = \left[\begin{array}{cccc|cccc|cccc}
    0 & 0 & 0 & 0 & 0 & 0 & \frac{1}{2} & 0 & 0 & 0 & 0 & \frac{1}{2} \\ 
    0 & 0 & 0 & 0 & 0 & 0 & 0 & 0 & -\frac{1}{2} & \frac{1}{2} & 0 & 0 \\ 
    0 & 0 & 0 & 0 & \frac{1}{2} & -\frac{1}{2} & 0 & 0 & 0 & 0 & 0 & 0 \\ 
    0 & 0 & 0 & 0 & 0 & 0 & 0 & \frac{1}{2} & 0 & 0 & -\frac{1}{2} & 0 \\ 
    \hline
    0 & 0 & 0 & 1 & 0 & 0 & \frac{1}{2} & 0 & 0 & 0 & 0 & -\frac{1}{2} \\ 
    0 & 0 & 0 & 0 & 0 & 0 & 0 & 0 & \frac{1}{2} & -\frac{1}{2} & 0 & 0 \\ 
    0 & 0 & 0 & 0 & \frac{1}{2} & -\frac{1}{2} & 0 & 0 & 0 & 0 & 0 & 0 \\ 
    0 & 0 & 1 & 0 & 0 & 0 & 0 & \frac{1}{2} & 0 & 0 & \frac{1}{2} & 0 \\ 
    \hline
    0 & 0 & 0 & 0 & 0 & 0 & 0 & 0 & 0 & 0 & 0 & 0 \\ 
    0 & 0 & 0 & 0 & 0 & 0 & 0 & 0 & 0 & 0 & 0 & 0 \\ 
    0 & 0 & 0 & 0 & 0 & 0 & 0 & 0 & 0 & 0 & 0 & 0 \\ 
    0 & 0 & 0 & 0 & 0 & 0 & 0 & 0 & 0 & 0 & 0 & 0
\numberthis\end{array}\right], \]

\[
[{{\bm \rR}_3^{(1)}}]_{\hat i}{}^{k} = \left[\begin{array}{cccc|cccc|cccc}
    0 & 0 & 0 & 0 & 0 & 0 & 0 & 0 & 0 & 0 & 0 & 0 \\ 
    0 & 1 & 0 & 0 & 0 & 0 & 0 & 0 & 0 & 0 & 0 & 0 \\ 
    -1 & 0 & 0 & 0 & 0 & 0 & 0 & 0 & 0 & 0 & 0 & 0 \\ 
    0 & 0 & 0 & 0 & 0 & 0 & 0 & 0 & 0 & 0 & 0 & 0 \\ 
    \hline
    0 & 0 & 0 & 0 & 0 & 0 & 0 & 0 & 0 & 0 & 0 & 0 \\ 
    0 & 0 & 0 & 0 & 0 & 0 & 0 & 0 & 0 & 0 & 0 & 0 \\ 
    0 & 0 & 0 & 0 & 0 & 0 & 0 & 0 & 0 & 0 & 0 & 0 \\ 
    0 & 0 & 0 & 0 & 0 & 0 & 0 & 0 & 0 & 0 & 0 & 0 \\ 
    \hline
    0 & 0 & 0 & 0 & 0 & 0 & 0 & 0 & -\frac{3}{4} & -\frac{1}{4} & 0 & 0 \\ 
    0 & 0 & 0 & \frac{1}{2} & 0 & 0 & -\frac{1}{4} & 0 & 0 & 0 & 0 & \frac{1}{4} \\ 
    0 & 0 & \frac{1}{2} & 0 & 0 & 0 & 0 & -\frac{1}{4} & 0 & 0 & -\frac{1}{4} & 0 \\ 
    0 & 0 & 0 & 0 & -\frac{3}{4} & -\frac{1}{4} & 0 & 0 & 0 & 0 & 0 & 0
\numberthis\end{array}\right], \]

\[
[{{\bm \rR}_3^{(2)}}]_{\hat i}{}^{k} = \left[\begin{array}{cccc|cccc|cccc}
    0 & 0 & 0 & 0 & 0 & 0 & 0 & 0 & 0 & 0 & 0 & 0 \\ 
    0 & 0 & 0 & 0 & 0 & 0 & 0 & 0 & 0 & 0 & 0 & 0 \\ 
    0 & 0 & 0 & 0 & 0 & 0 & 0 & 0 & 0 & 0 & 0 & 0 \\ 
    0 & 0 & 0 & 0 & 0 & 0 & 0 & 0 & 0 & 0 & 0 & 0 \\ 
    \hline
    0 & 0 & 0 & 0 & 0 & 0 & 0 & 0 & 0 & 0 & 0 & 0 \\ 
    0 & 0 & 0 & 0 & 0 & 0 & 0 & 0 & 0 & 0 & 0 & 0 \\ 
    0 & 0 & 0 & 0 & 0 & 0 & 0 & 0 & 0 & 0 & 0 & 0 \\ 
    0 & 0 & 0 & 0 & 0 & 0 & 0 & 0 & 0 & 0 & 0 & 0 \\ 
    \hline
    0 & 1 & 0 & 0 & 0 & 0 & 0 & 0 & 0 & 0 & 0 & 0 \\ 
    0 & 0 & 0 & 0 & 0 & 0 & 0 & 0 & 0 & 0 & 0 & 0 \\ 
    0 & 0 & 0 & 0 & 0 & 0 & 0 & 0 & 0 & 0 & 0 & 0 \\ 
    1 & 0 & 0 & 0 & 0 & 0 & 0 & 0 & 0 & 0 & 0 & 0
\numberthis\end{array}\right], \]

\[
[{{\bm \rR}_4^{(0)}}]_{\hat i}{}^{k} = \left[\begin{array}{cccc|cccc|cccc}
    0 & 0 & 0 & 0 & 0 & 0 & 0 & 0 & \frac{1}{2} & \frac{1}{2} & 0 & 0 \\ 
    0 & 0 & 0 & 0 & 0 & 0 & \frac{1}{2} & 0 & 0 & 0 & 0 & -\frac{1}{2} \\ 
    0 & 0 & 0 & 0 & 0 & 0 & 0 & \frac{1}{2} & 0 & 0 & \frac{1}{2} & 0 \\ 
    0 & 0 & 0 & 0 & \frac{1}{2} & \frac{1}{2} & 0 & 0 & 0 & 0 & 0 & 0 \\ 
    \hline
    0 & 0 & 0 & 0 & 0 & 0 & 0 & 0 & -\frac{1}{2} & -\frac{1}{2} & 0 & 0 \\ 
    0 & 0 & 0 & 1 & 0 & 0 & \frac{1}{2} & 0 & 0 & 0 & 0 & \frac{1}{2} \\ 
    0 & 0 & -1 & 0 & 0 & 0 & 0 & \frac{1}{2} & 0 & 0 & -\frac{1}{2} & 0 \\ 
    0 & 0 & 0 & 0 & \frac{1}{2} & \frac{1}{2} & 0 & 0 & 0 & 0 & 0 & 0 \\ 
    \hline
    0 & 0 & 0 & 0 & 0 & 0 & 0 & 0 & 0 & 0 & 0 & 0 \\ 
    0 & 0 & 0 & 0 & 0 & 0 & 0 & 0 & 0 & 0 & 0 & 0 \\ 
    0 & 0 & 0 & 0 & 0 & 0 & 0 & 0 & 0 & 0 & 0 & 0 \\ 
    0 & 0 & 0 & 0 & 0 & 0 & 0 & 0 & 0 & 0 & 0 & 0
\numberthis\end{array}\right], \]

\[
[{{\bm \rR}_4^{(1)}}]_{\hat i}{}^{k} = \left[\begin{array}{cccc|cccc|cccc}
    0 & -1 & 0 & 0 & 0 & 0 & 0 & 0 & 0 & 0 & 0 & 0 \\ 
    0 & 0 & 0 & 0 & 0 & 0 & 0 & 0 & 0 & 0 & 0 & 0 \\ 
    0 & 0 & 0 & 0 & 0 & 0 & 0 & 0 & 0 & 0 & 0 & 0 \\ 
    -1 & 0 & 0 & 0 & 0 & 0 & 0 & 0 & 0 & 0 & 0 & 0 \\ 
    \hline
    0 & 0 & 0 & 0 & 0 & 0 & 0 & 0 & 0 & 0 & 0 & 0 \\ 
    0 & 0 & 0 & 0 & 0 & 0 & 0 & 0 & 0 & 0 & 0 & 0 \\ 
    0 & 0 & 0 & 0 & 0 & 0 & 0 & 0 & 0 & 0 & 0 & 0 \\ 
    0 & 0 & 0 & 0 & 0 & 0 & 0 & 0 & 0 & 0 & 0 & 0 \\ 
    \hline
    0 & 0 & 0 & -\frac{1}{2} & 0 & 0 & \frac{1}{4} & 0 & 0 & 0 & 0 & \frac{1}{4} \\ 
    0 & 0 & 0 & 0 & 0 & 0 & 0 & 0 & -\frac{3}{4} & \frac{1}{4} & 0 & 0 \\ 
    0 & 0 & 0 & 0 & \frac{3}{4} & -\frac{1}{4} & 0 & 0 & 0 & 0 & 0 & 0 \\ 
    0 & 0 & \frac{1}{2} & 0 & 0 & 0 & 0 & \frac{1}{4} & 0 & 0 & -\frac{1}{4} & 0
\numberthis\end{array}\right], \]

\[
[{{\bm \rR}_4^{(2)}}]_{\hat i}{}^{k} = \left[\begin{array}{cccc|cccc|cccc}
    0 & 0 & 0 & 0 & 0 & 0 & 0 & 0 & 0 & 0 & 0 & 0 \\ 
    0 & 0 & 0 & 0 & 0 & 0 & 0 & 0 & 0 & 0 & 0 & 0 \\ 
    0 & 0 & 0 & 0 & 0 & 0 & 0 & 0 & 0 & 0 & 0 & 0 \\ 
    0 & 0 & 0 & 0 & 0 & 0 & 0 & 0 & 0 & 0 & 0 & 0 \\ 
    \hline
    0 & 0 & 0 & 0 & 0 & 0 & 0 & 0 & 0 & 0 & 0 & 0 \\ 
    0 & 0 & 0 & 0 & 0 & 0 & 0 & 0 & 0 & 0 & 0 & 0 \\ 
    0 & 0 & 0 & 0 & 0 & 0 & 0 & 0 & 0 & 0 & 0 & 0 \\ 
    0 & 0 & 0 & 0 & 0 & 0 & 0 & 0 & 0 & 0 & 0 & 0 \\ 
    \hline
    0 & 0 & 0 & 0 & 0 & 0 & 0 & 0 & 0 & 0 & 0 & 0 \\ 
    0 & 1 & 0 & 0 & 0 & 0 & 0 & 0 & 0 & 0 & 0 & 0 \\ 
    -1 & 0 & 0 & 0 & 0 & 0 & 0 & 0 & 0 & 0 & 0 & 0 \\ 
    0 & 0 & 0 & 0 & 0 & 0 & 0 & 0 & 0 & 0 & 0 & 0
\numberthis
\label{eq4-29}
\end{array}\right]. \]

\newpage
\subsection{Garden Algebra}

From the supersymmetric transformation rules,
\[ {\rm D}_{\rm I} \Phi{}_i = [ {\bm {\rL}}{}_{\rI}^{(1)}]_i \, {}^{\hat 
k} \, \Psi{}_{\hat k} ~+~ [ {\bm {\rL}}{}_{\rI}^{(2)}]_i \, ^{\hat k}
\, \frac{d}{d \tau} \Psi{}_{\hat k}~~~,
\numberthis\]
\[ {\rm D}_{\rm I} \Psi{}_{\hat k} = i \, [ {\bm {\rR}}{}_{\rI}^{(0)}]_{\hat
k} \, ^i 
\Phi{}_{i }
~+~ i\, [ {\bm {\rR}}{}_{\rI}^{(1)}]_{\hat k} \, ^i \, \frac{d ~}{d \tau}
\Phi{}_{i}
~+~ i\, [ {\bm {\rR}}{}_{\rI}^{(2)}]_{\hat k} \, ^i \, \frac{d^2 ~}{d \tau^2}
\Phi{}_{i}~~~.
\numberthis\]
The form of the final equation points toward the emergence of new mathematical structures 
associated with the CLS system that were not present for the CS, VS, nor TS networks.  This is
clear from the appearance of the matrics denoted by $ [ {\bm {\rR}}{}_{\rI}^{(2)}]$ that have 
not appeared previously.

Additionally, these unfolded $\bm \rL$ and $\bm \rR$ matrices are such that the bosonic fields satisfy the SUSY closure relation:
\begin{eqnarray}
\{{\rm D}_{\rm I},{\rm D}_{\rm J}\}\Phi{}_i&=&
\left[\left(
[{\bm {\rL}}{}_{\rI}^{(1)}]_{{i}}{}^{\hat j}
[{\bm {\rR}}{}_{\rJ}^{(0)}]_{{\hat j}}{}^{k}
+
[{\bm {\rL}}{}_{\rJ}^{(1)}]_{{i}}{}^{\hat j}
[{\bm {\rR}}{}_{\rI}^{(0)}]_{{\hat j}}{}^{k}
\right)\right.
\nonumber\\&+&
\left(
[{\bm {\rL}}{}_{\rI}^{(1)}]_{{i}}{}^{\hat j}
[{\bm {\rR}}{}_{\rJ}^{(1)}]_{{\hat j}}{}^{k}
+
[{\bm {\rL}}{}_{\rJ}^{(1)}]_{{i}}{}^{\hat j}
[{\bm {\rR}}{}_{\rI}^{(1)}]_{{\hat j}}{}^{k}
+
[{\bm {\rL}}{}_{\rI}^{(2)}]_{{i}}{}^{\hat j}
[{\bm {\rR}}{}_{\rJ}^{(0)}]_{{\hat j}}{}^{k}
+
[{\bm {\rL}}{}_{\rJ}^{(2)}]_{{i}}{}^{\hat j}
[{\bm {\rR}}{}_{\rI}^{(0)}]_{{\hat j}}{}^{k}
\right)\frac{d}{d\tau}
\nonumber\\&+&
\left(
[{\bm {\rL}}{}_{\rI}^{(1)}]_{{i}}{}^{\hat j}
[{\bm {\rR}}{}_{\rJ}^{(2)}]_{{\hat j}}{}^{k}
+
[{\bm {\rL}}{}_{\rJ}^{(1)}]_{{i}}{}^{\hat j}
[{\bm {\rR}}{}_{\rI}^{(2)}]_{{\hat j}}{}^{k}
+
[{\bm {\rL}}{}_{\rI}^{(2)}]_{{i}}{}^{\hat j}
[{\bm {\rR}}{}_{\rJ}^{(1)}]_{{\hat j}}{}^{k}
+
[{\bm {\rL}}{}_{\rJ}^{(2)}]_{{i}}{}^{\hat j}
[{\bm {\rR}}{}_{\rI}^{(1)}]_{{\hat j}}{}^{k}
\right)\frac{d^2}{d\tau^2}
\nonumber\\&+&
\left.\left(
[{\bm {\rL}}{}_{\rI}^{(2)}]_{{i}}{}^{\hat j}
[{\bm {\rR}}{}_{\rJ}^{(2)}]_{{\hat j}}{}^{k}
+
[{\bm {\rL}}{}_{\rJ}^{(2)}]_{{i}}{}^{\hat j}
[{\bm {\rR}}{}_{\rI}^{(2)}]_{{\hat j}}{}^{k}
\right)\frac{d^3}{d\tau^3}\right] i\Phi{}_{k}
\nonumber\\&=&2i\delta_{{\rm IJ}}
\frac{d}{d\tau}\Phi{}_{i}~~~.
\label{eq5.28}
\end{eqnarray}

It is clear from these equations that the conditions for the realization of supersymmetry in the unfolded adinkra are very different than is seen in the case of folded adinkra.  In particular, the following conditions are separately found to be satisfied:
\begin{eqnarray}
0 &=&
[{\bm {\rL}}{}_{( \rI}^{(1)}]_{{i}}{}^{\hat j}
[{\bm {\rR}}{}_{\rJ )}^{(0)}]_{{\hat j}}{}^{k}
~~~,
\label{eq5.28A}
\end{eqnarray}
\begin{eqnarray}
2 \, \d{}_{\rI}{}_{\rJ} \, \d{}_{i}{}^{k} &=&
[{\bm {\rL}}{}_{( \rI}^{(1)}]_{{i}}{}^{\hat j}
[{\bm {\rR}}{}_{\rJ )}^{(1)}]_{{\hat j}}{}^{k}
+
[{\bm {\rL}}{}_{( \rI}^{(2)}]_{{i}}{}^{\hat j}
[{\bm {\rR}}{}_{\rJ )}^{(0)}]_{{\hat j}}{}^{k}
~~~,
\label{eq5.28B}
\end{eqnarray}
\begin{eqnarray}
0 &=&
[{\bm {\rL}}{}_{( \rI}^{(1)}]_{{i}}{}^{\hat j}
[{\bm {\rR}}{}_{\rJ )}^{(2)}]_{{\hat j}}{}^{k}
+
[{\bm {\rL}}{}_{( \rI}^{(2)}]_{{i}}{}^{\hat j}
[{\bm {\rR}}{}_{\rJ )}^{(1)}]_{{\hat j}}{}^{k}
~~~,
\label{eq5.28C}
\end{eqnarray}
\begin{eqnarray}
0 &=&
[{\bm {\rL}}{}_{( \rI}^{(2)}]_{{i}}{}^{\hat j}
[{\bm {\rR}}{}_{\rJ )}^{(2)}]_{{\hat j}}{}^{k}
~~~.
\label{eq5.28D}
\end{eqnarray}

The result shown in Eq. (\ref{eq5.28B}) controls the single time derivative term. The structure of this relation is similar to that of the familiar garden algebra. However, whereas the garden algebra is the sum of two matrix products, this relation is the sum of four matrix products. In particular, each product is such that the superscripts on the $\bm \rL$ and $\bm \rR$ matrices sum to $2$.
Thus, as in the usual manner, these matrices can be used to form a Clifford
algebra representation that is  twice as long as the length of the garden algebra
$[{\bm {\rL}}{}_{\rI}^{(1)}]_{{i}}{}^{\hat j}$-$[{\bm {\rR}}{}_{\rJ}^{(1)}]_{{\hat i}}{}^{j}$ that appear above.

Meanwhile, for the portions of Eq. (\ref{eq5.28}) that involve zero, two, or three time derivatives, the conditions are quite remarkably different, where the sums matrix products as seen in Eq. (\ref{eq5.28A}), (\ref{eq5.28C}), and (\ref{eq5.28D}) each evaluate to zero.

Similarly for the fermionic fields, we can write,
\begin{eqnarray}
\{{\rm D}_{\rm I},{\rm D}_{\rm J}\}\Psi{}_{\hat i}&=&
i \, \left[\left(
[{\bm {\rR}}{}_{\rI}^{(0)}]_{{\hat i}}{}^{j}
[{\bm {\rL}}{}_{\rJ}^{(1)}]_{{j}}{}^{\hat k}
+
[{\bm {\rR}}{}_{\rJ}^{(0)}]_{{\hat i}}{}^{j}
[{\bm {\rL}}{}_{\rI}^{(1)}]_{{j}}{}^{\hat k}
\right)\right.
\nonumber\\&{~}&{~~~}+
\left(
[{\bm {\rR}}{}_{\rI}^{(1)}]_{{\hat i}}{}^{j}
[{\bm {\rL}}{}_{\rJ}^{(1)}]_{{j}}{}^{\hat k}
+
[{\bm {\rR}}{}_{\rJ}^{(1)}]_{{\hat i}}{}^{j}
[{\bm {\rL}}{}_{\rI}^{(1)}]_{{j}}{}^{\hat k}
+
[{\bm {\rR}}{}_{\rI}^{(0)}]_{{\hat i}}{}^{j}
[{\bm {\rL}}{}_{\rJ}^{(2)}]_{{j}}{}^{\hat k}
+
[{\bm {\rR}}{}_{\rJ}^{(0)}]_{{\hat i}}{}^{j}
[{\bm {\rL}}{}_{\rI}^{(2)}]_{{j}}{}^{\hat k}
\right)\frac{d}{d\tau}
\nonumber\\&{~}&{~~~}+
\left(
[{\bm {\rR}}{}_{\rI}^{(2)}]_{{\hat i}}{}^{j}
[{\bm {\rL}}{}_{\rJ}^{(1)}]_{{j}}{}^{\hat k}
+
[{\bm {\rR}}{}_{\rJ}^{(2)}]_{{\hat i}}{}^{j}
[{\bm {\rL}}{}_{\rI}^{(1)}]_{{j}}{}^{\hat k}
+
[{\bm {\rR}}{}_{\rI}^{(1)}]_{{\hat i}}{}^{j}
[{\bm {\rL}}{}_{\rJ}^{(2)}]_{{j}}{}^{\hat k}
+
[{\bm {\rR}}{}_{\rJ}^{(1)}]_{{\hat i}}{}^{j}
[{\bm {\rL}}{}_{\rI}^{(2)}]_{{j}}{}^{\hat k}
\right)\frac{d^2}{d\tau^2}
\nonumber\\&{~}&{~~~}+
\left.\left(
[{\bm {\rR}}{}_{\rI}^{(2)}]_{{\hat i}}{}^{j}
[{\bm {\rL}}{}_{\rJ}^{(2)}]_{{j}}{}^{\hat k}
+
[{\bm {\rR}}{}_{\rJ}^{(2)}]_{{\hat i}}{}^{j}
[{\bm {\rL}}{}_{\rI}^{(2)}]_{{j}}{}^{\hat k}
\right)\frac{d^3}{d\tau^3}\right] i\Psi{}_{\hat k}
\nonumber\\&=&2i\delta_{{\rm IJ}}\frac{d}{d\tau}\Psi{}_{\hat i} ~~~,
\label{eq5.29}
\end{eqnarray}

And thus, the fermionic fields also satisfy the SUSY closure relation. As with the closure of the bosonic fields, we can also separate Eq. (\ref{eq5.29}) into four relations as follows:

\begin{eqnarray}
0 &=&
[{\bm {\rR}}{}_{( \rI}^{(0)}]_{{\hat i}}{}^{j}
[{\bm {\rL}}{}_{\rJ  )}^{(1)}]_{{j}}{}^{\hat k}
~~~,
\label{eq5.29A}
\end{eqnarray}
\begin{eqnarray}
2 \, \d{}_{\rI}{}_{\rJ} \, \d{}_{\hat i}{}^{\hat k}
&=&
[{\bm {\rR}}{}_{( \rI}^{(1)}]_{{\hat i}}{}^{j}
[{\bm {\rL}}{}_{\rJ )}^{(1)}]_{{j}}{}^{\hat k}
+
[{\bm {\rR}}{}_{( \rI}^{(0)}]_{{\hat i}}{}^{j}
[{\bm {\rL}}{}_{\rJ )}^{(2)}]_{{j}}{}^{\hat k}
~~~,
\label{eq5.29B}
\end{eqnarray}
\begin{eqnarray}
0 &=&
[{\bm {\rR}}{}_{( \rI}^{(2)}]_{{\hat i}}{}^{j}
[{\bm {\rL}}{}_{\rJ  )}^{(1)}]_{{j}}{}^{\hat k}
+
[{\bm {\rR}}{}_{( \rI}^{(1)}]_{{\hat i}}{}^{j}
[{\bm {\rL}}{}_{\rJ )}^{(2)}]_{{j}}{}^{\hat k}
~~~,
\label{eq5.29C}
\end{eqnarray}
\begin{eqnarray}
0 &=&
[{\bm {\rR}}{}_{( \rI}^{(2)}]_{{\hat i}}{}^{j}
[{\bm {\rL}}{}_{\rJ )}^{(2)}]_{{j}}{}^{\hat k}
~~~.
\label{eq5.29D}
\end{eqnarray}

These relations are similar to those derived from Eq. (\ref{eq5.28}) for the bosonic fields. As you can see, the 
matrix products corresponding to zero, two, or three-time derivatives sum to zero, while those corresponding to 
a single time derivative satisfy a relation similar to the familiar garden algebra, but which again contains four matrix products rather than the usual two.

Furthermore, these relations can be related to the notion of BRST charges \cite{7,brst1,brst2,brst3,brst4}.  To 
see this, a new conceptual framework is needed.  The basic idea is that the abstract operators ${\rm D}_{\rm I}$
themselves can be expanded in terms of abstract operators ${\rm D}_{\rm I}^{(\ell)}$ where the index $(\ell)$ takes on values 0, 1, $...$, $\infty$, so that one is permitted to write and expansion of
the form
\be   {
{\rm D}_{\rm I}  ~=~ {\rm D}_{\rm I}^{(0)} ~+~ {\rm D}_{\rm I}^{(1)} ~+~ 
{\rm D}_{\rm I}^{(2)} ~+~ {\rm D}_{\rm I}^{(3)} ~+~ \dots ~=~ \sum_{\ell \,=\, 0}^{\infty} \, {\rm D}_{\rm I}^{(\ell)} ~~~,
}
\label{Dexp}
\ee
where each of the summand operators is associated with the action of lifting nodes at a single specified
level related to the $\ell$-parameter.  The explicit realization of each of the summand operators is
fixed by the requirement that when their anticommutator is calculated on any bosonic node the results must be
with Eqs.\ (\ref{eq5.28A}) - (\ref{eq5.28D}).  Similarly, when their anticommutator is calculated on any fermionic node the results must agree with Eqs.\ (\ref{eq5.29A}) - (\ref{eq5.29D}).
These conventions fix how each ${\rm D}_{\rm I}^{(\ell)}$ operator is related to the $[{\bm {\rL}}{}_{\rI}^{(\ell)}]$ and $[{\bm {\rR}}{}_{\rI}^{(\ell)}]$ matrix realizations.  

To see how this works, it is useful to illustrate these concepts in
the fields of the CS valise network.  We can return to the Eq. (\ref{CSX1}) and
Eq.\ (\ref{CSX2}).  On the LHS of these two equations, we substitute the
expansion shown in Eq.\ (\ref{Dexp}).  The definitions of the 
${\rm D}_{\rm I}^{(\ell)}$ operators are chosen so that
\[   { {~~~~~\,~~~~~~~}
{\rm D}_{\rm I}^{(0)} \Phi{}_i = 0  ~~~~~~~~~~~~~~,~~~
 {\rm D}_{\rm I}^{(1)} \Phi{}_i =  i\, [ {\bm {\rL}}{}_{\rI}^{(1)}]_{i}{}^{\hat{k}} \, \Psi{}_{\hat k}
~~~~~\,~,~~
 {\rm D}_{\rm I}^{(2)} \Phi{}_i = i\, [ {\bm {\rL}}{}_{\rI}^{(2)}]_{i}{}^{\hat{k}}
\, \frac{d}{d \tau} \Psi{}_{\hat k} ~~~~,~~
}
\label{Dexp1cs}
\]
\[   {
 {\rm D}_{\rm I}^{(0)} \Psi{}_{\hat k} = [ {\bm {\rR}}{}_{\rI}^{(0)}]_{\hat k}{}^i \, \Phi{}_{i } \,~~,~~
 {\rm D}_{\rm I}^{(1)} \Psi{}_{\hat k} =
 [ {\bm {\rR}}{}_{\rI}^{(1)}]_{\hat k}{}^i \, \frac{d ~}{d \tau} \Phi{}_{i}
~~~~~,~~~ \,
 {\rm D}_{\rm I}^{(2)} \Psi{}_{\hat k} =  0  ~~~. {~~\,~~~~}
\numberthis}
\label{Dexp2cs}
\]

\subsection{Unfolded Adinkras}
\begin{figure}[H]
	\centering	\includegraphics[scale=0.35]{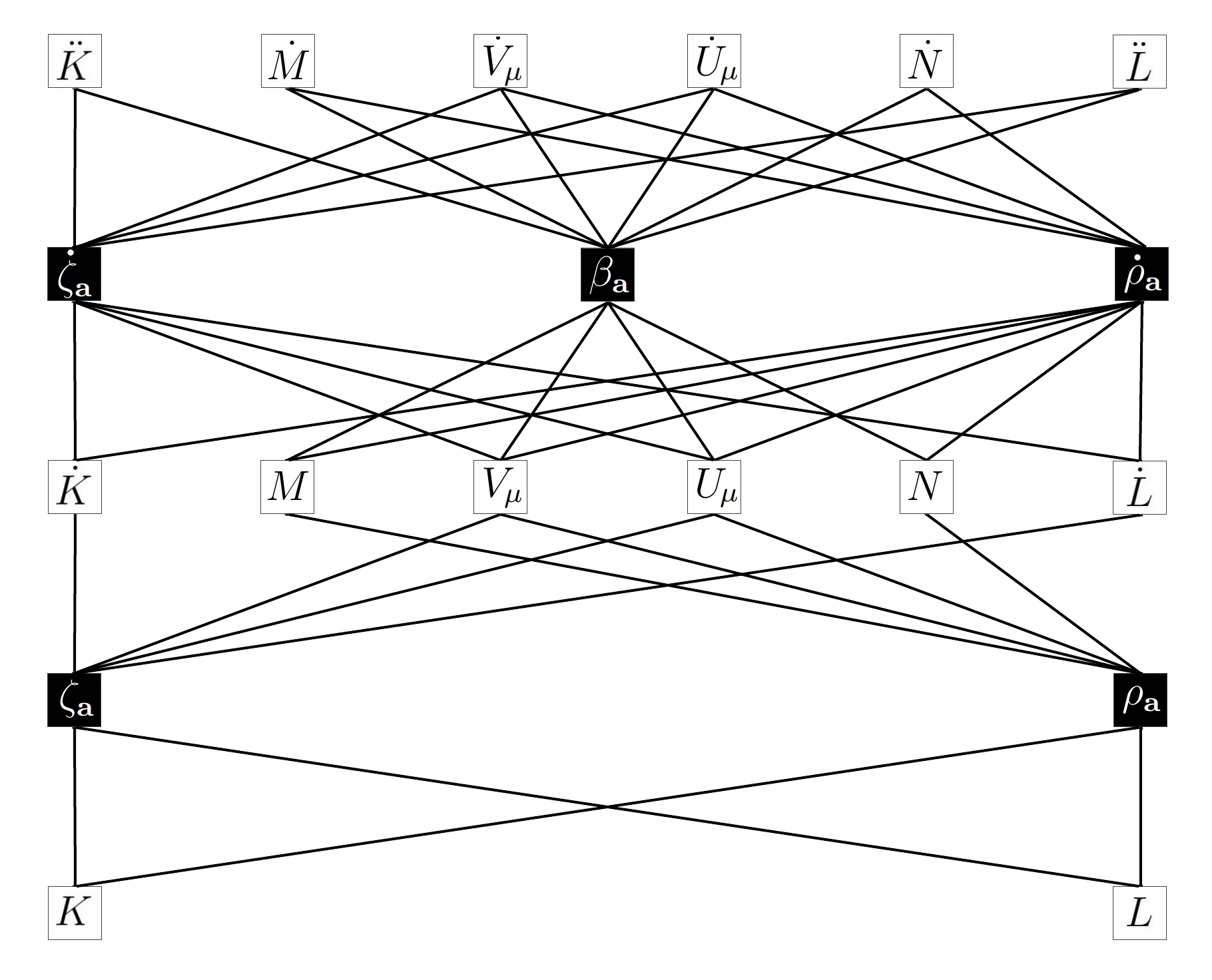}
	\caption{An unfolded adinkra for 4D $\mathcal{N}=1$ CLS.}
	\label{fig12}
\end{figure}

Fig. \ref{fig12} shows the unfolded adinkra for CLS. This graph includes each of the $\rD_1$ to $\rD_4$ transformations.

\begin{figure}[H]
	\centering	\includegraphics[scale=0.35]{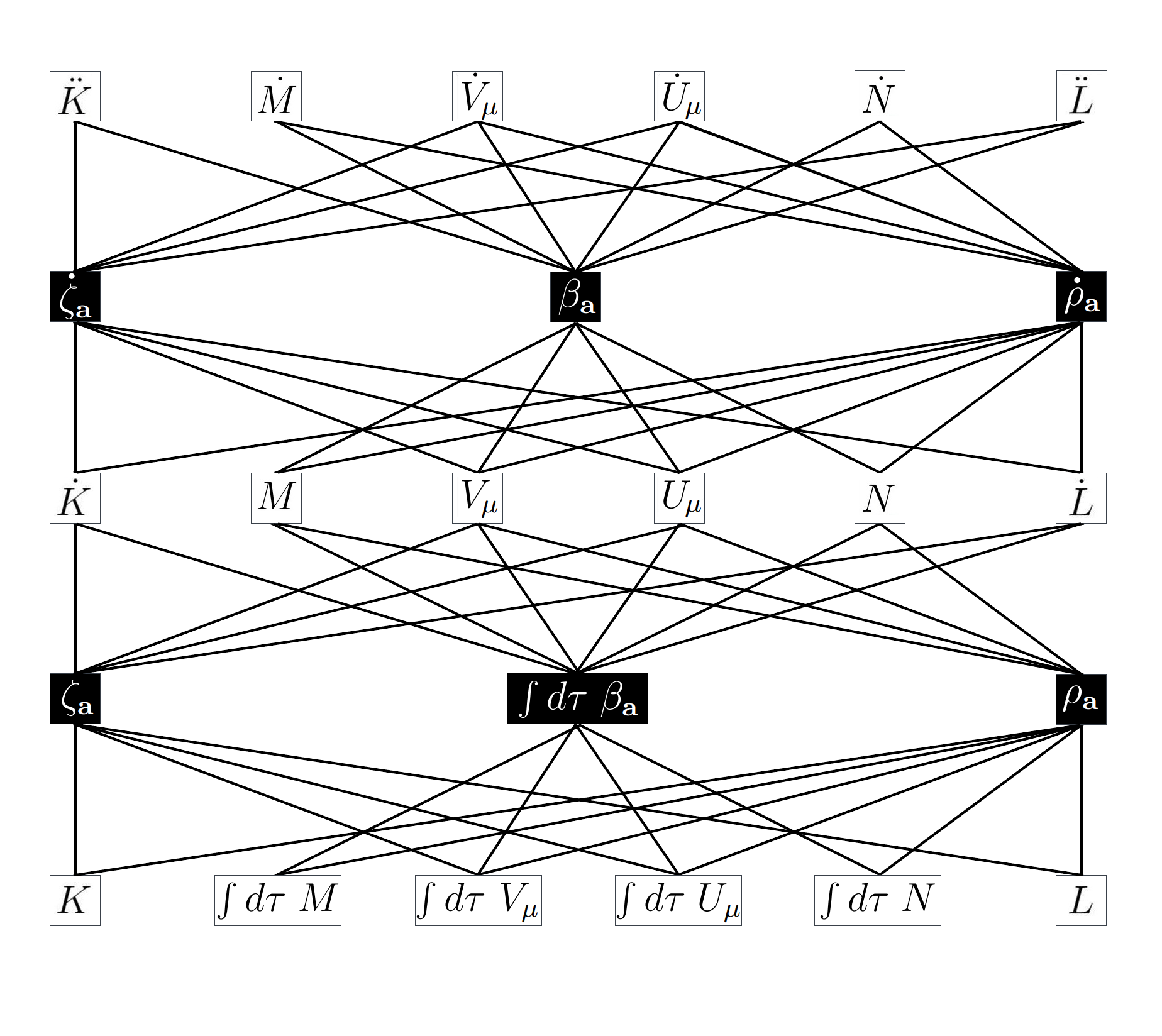}
	\caption{An unfolded adinkra for 4D $\mathcal{N}=1$ CLS with ascending edges.}
	\label{fig13}
\end{figure}

And if we add integrals of bosonic fields to match the engineering dimension in each layer so that to make ascending edges we obtain Fig. \ref{fig13}.

\begin{figure}[H]
	\centering	\includegraphics[scale=0.28]{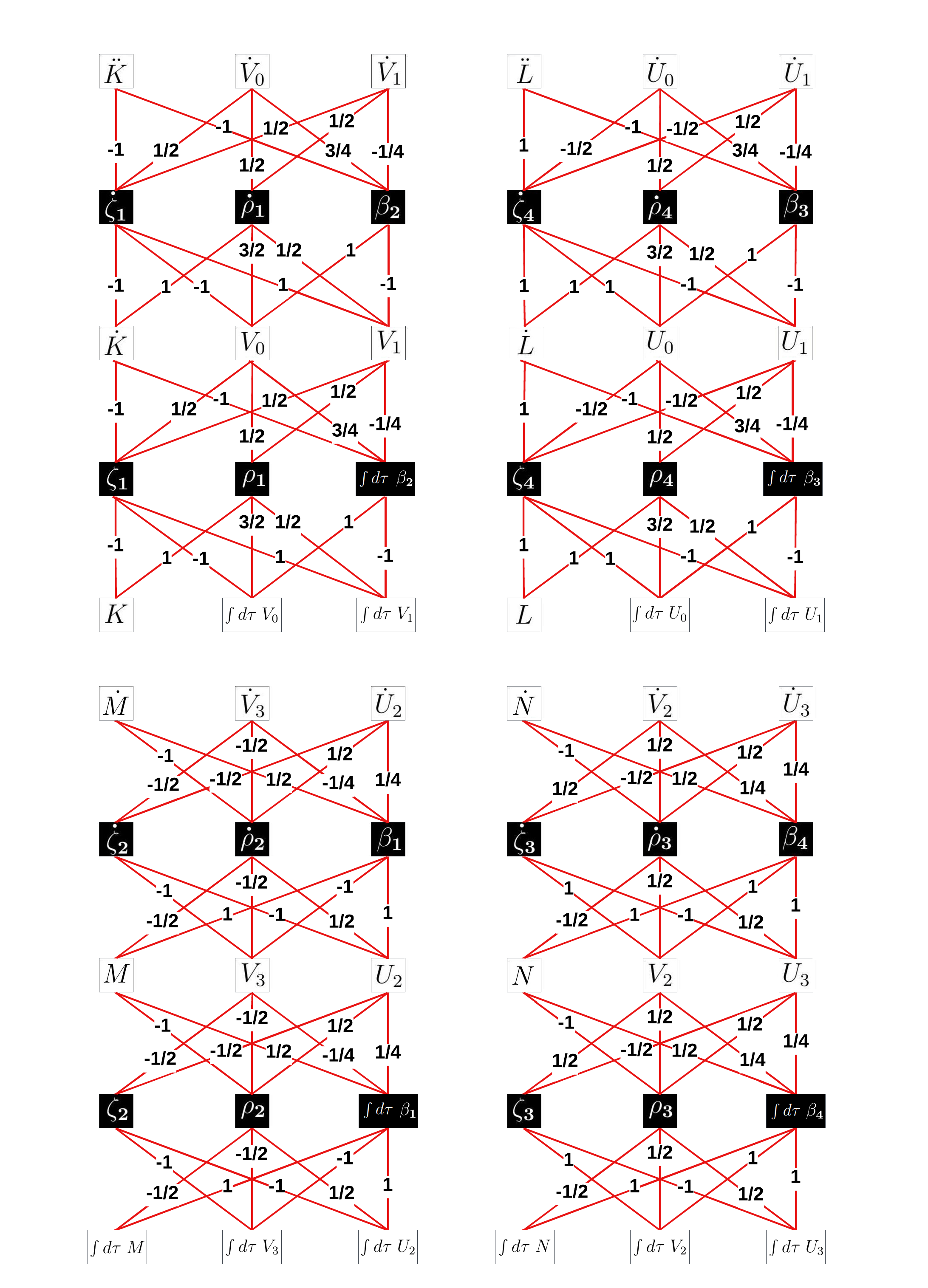}
	\caption{An unfolded adinkra for CLS under $\rD_1$ with ascending edges.}
	\label{fig8}
\end{figure}

\begin{figure}[H]
	\centering	\includegraphics[scale=0.28]{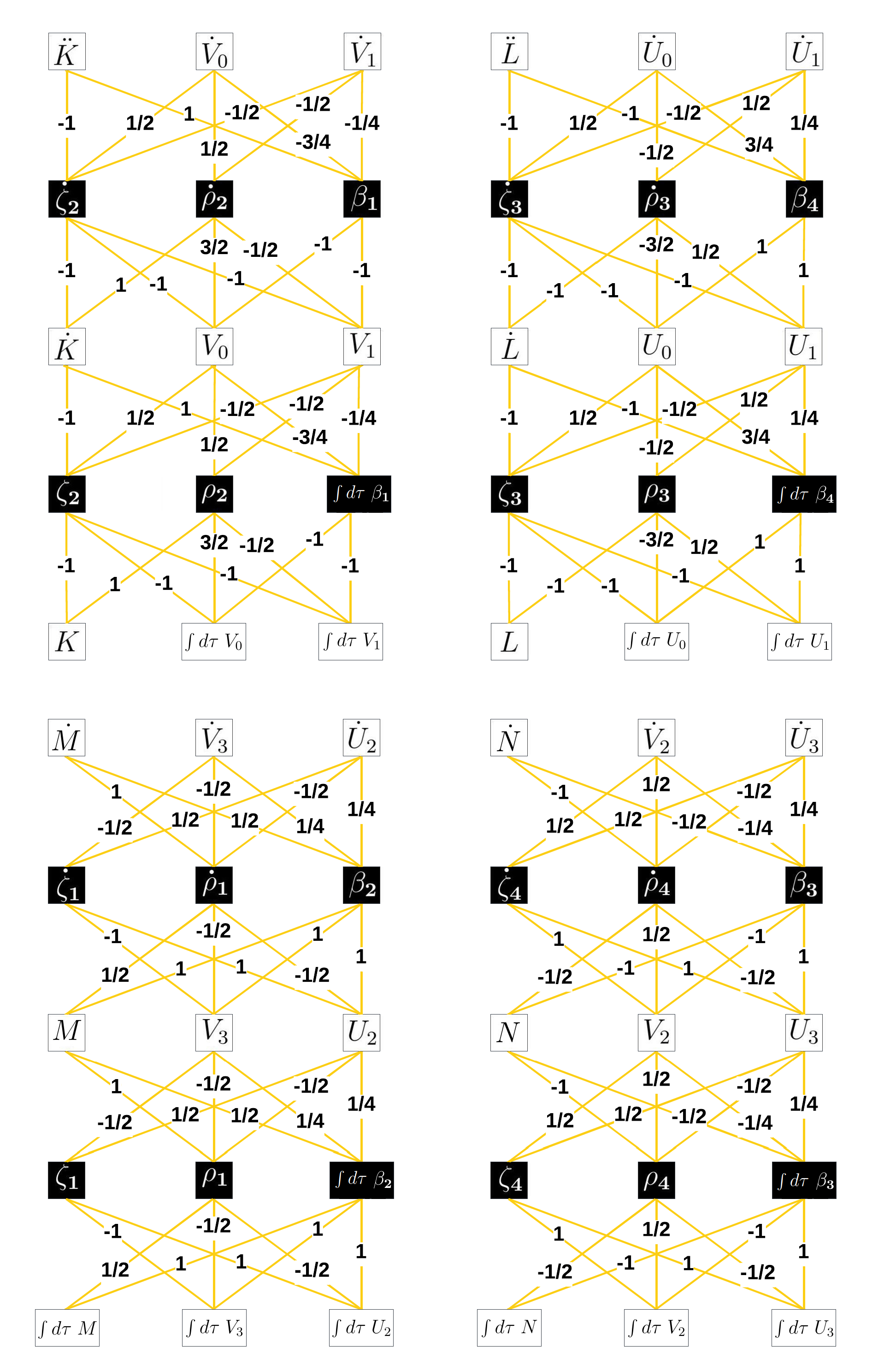}
	\caption{An unfolded adinkra for CLS under $\rD_2$ with ascending edges.}
	\label{fig9}
\end{figure}

\begin{figure}[H]
	\centering	\includegraphics[scale=0.28]{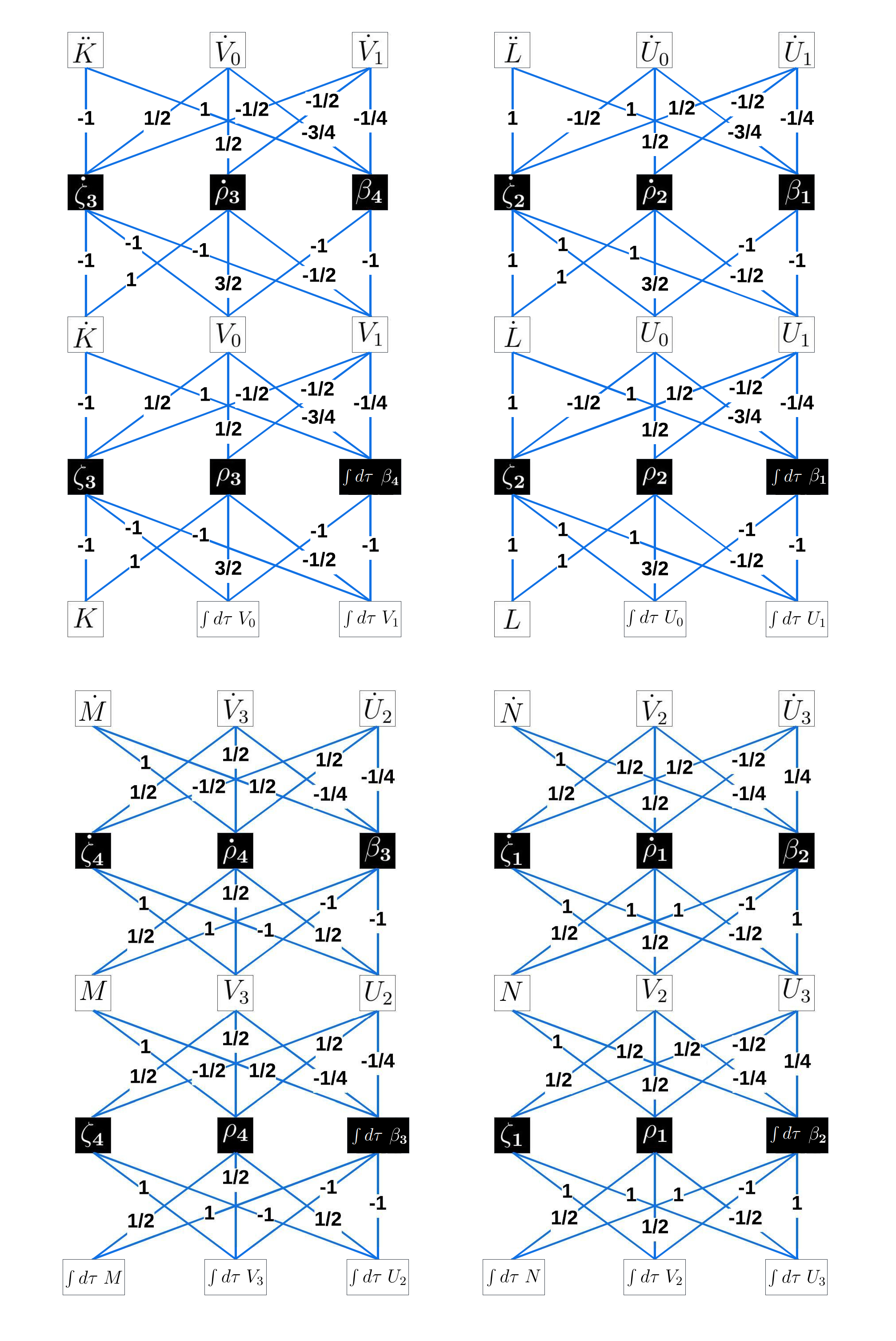}
	\caption{An unfolded adinkra for CLS under $\rD_3$ with ascending edges.}
	\label{fig10}
\end{figure}

\begin{figure}[H]
	\centering	\includegraphics[scale=0.28]{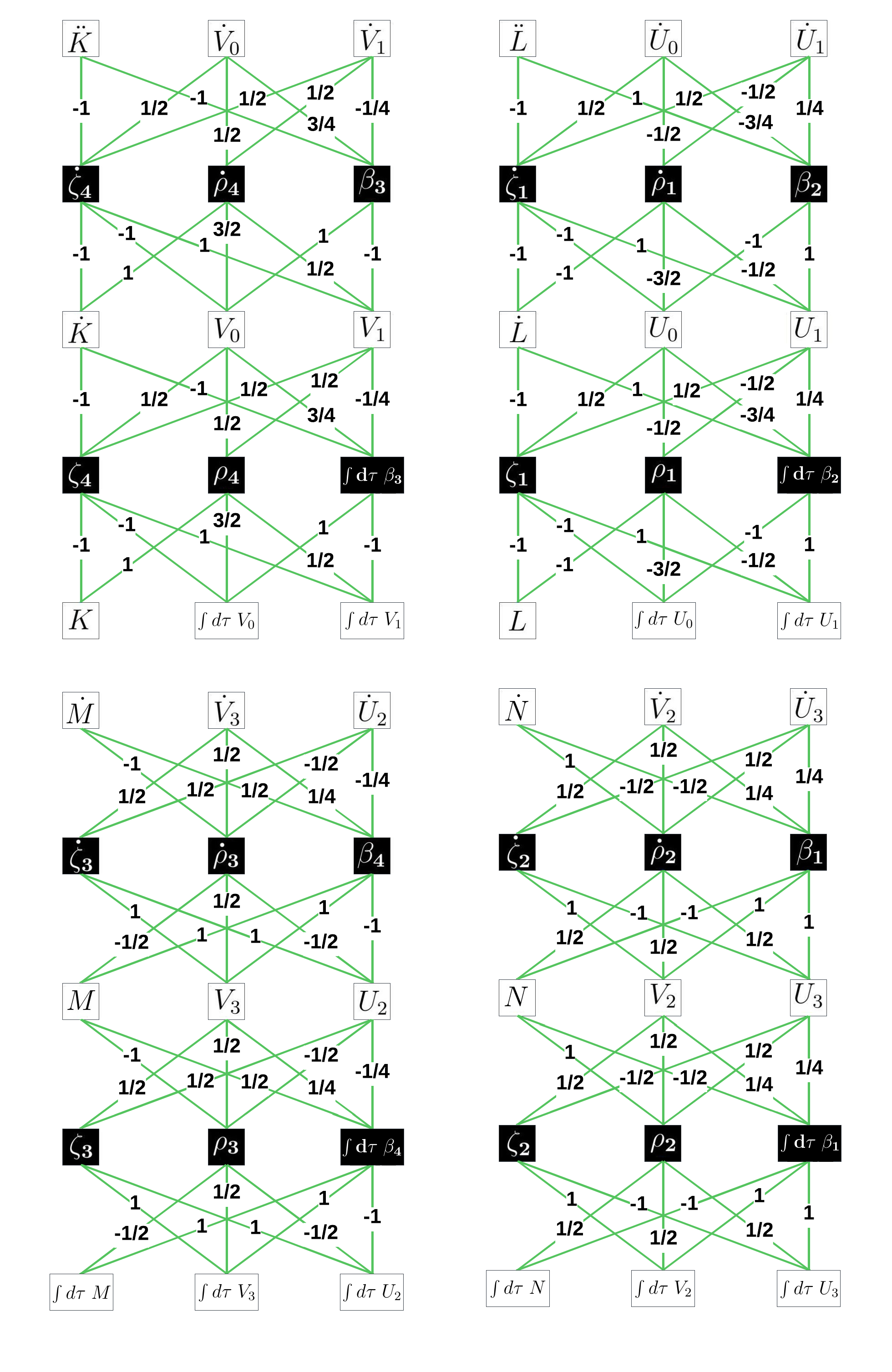}
	\caption{An unfolded adinkra for CLS under $\rD_4$ with ascending edges.}
	\label{fig11}
\end{figure}

Figs. \ref{fig8}-\ref{fig11} show the unfolded adinkras for each $\rD_\rI$ with each given a different color.
For each $\rD_\rI$, we can observe four different unfolded adinkras.
The edge coefficients are different for each $\rD_\rI$, however, the general graph structure is conserved. 
We can see that the bosonic fields are grouped, where $K$, $V_0$, and $V_1$ always appear in the same adinkra, $L$, $U_0$, and $U_1$ appear together, $M$, $V_3$, and $U_2$ appear together, and $N$, $V_2$, and $U_3$ appear together. The fermionic fields are similarly grouped, where $\zeta_1$, $\rho_1$, and $\beta_2$ always appear in the same adinkra, $\zeta_2$, $\rho_2$, and $\beta_1$ appear together, $\zeta_3$, $\rho_3$, and $\beta_4$ appear together, and $\zeta_4$, $\rho_4$, and $\beta_3$ appear together.

The expansion shown in Eq.\ (\ref{Dexp}) can also be applied to the CLS network.  The definitions of the 
${\rm D}_{\rm I}^{(\ell)}$ operators for this system are chosen so that
\[   { {~~~~~~~~~\,~~~~~~~}
{\rm D}_{\rm I}^{(0)} \Phi{}_i = 0  ~~~~~~~~~~~~~~,~~~
 {\rm D}_{\rm I}^{(1)} \Phi{}_i =  i\, [ {\bm {\rL}}{}_{\rI}^{(1)}]_{i}{}^{\hat{k}} \, \Psi{}_{\hat k}
~~~\,~,~~
 {\rm D}_{\rm I}^{(2)} \Phi{}_i = i\, [ {\bm {\rL}}{}_{\rI}^{(2)}]_{i}{}^{\hat{k}}
\, \frac{d}{d \tau} \Psi{}_{\hat k} ~~~,  {~~~~~~~~~~~~~~~~~~~~~~~~~~}
\numberthis}
\label{Dexp3cs}
\]
\[  {~~~~~~~~} {
 {\rm D}_{\rm I}^{(0)} \Psi{}_{\hat k} =   [ {\bm {\rR}}{}_{\rI}^{(0)}]_{\hat k}{}^i \, \Phi{}_{i } \,~~~~,~~
 {\rm D}_{\rm I}^{(1)} \Psi{}_{\hat k} =   
 [ {\bm {\rR}}{}_{\rI}^{(1)}]_{\hat k}{}^i \, \frac{d ~}{d \tau} \Phi{}_{i}
~~~,~ \,
 {\rm D}_{\rm I}^{(2)} \Psi{}_{\hat k} =   
 [ {\bm {\rR}}{}_{\rI}^{(2)}]_{\hat k}{}^i \, \frac{d ~}{d \tau} \Phi{}_{i}  ~~~. {~~\,~~~~}
}
\label{Dexp4cs}
\]
In conventions consistent with these unfolded diagrams with ascending only edges, the basis vectors for the CLS system take the forms
\[  { 
\Phi_{i} = \left[\begin{array}{c}
    K  \\ 
    L   \\
 \int d\t   M  \\
 \int d\t       N  \\ 
    \hline
 \int d\t       V_0  \\ 
 \int d\t       V_1  \\ 
 \int d\t       V_2  \\ 
  \int d\t      V_3  \\ 
    \hline
 \int d\t        U_0  \\ 
 \int d\t        U_1  \\ 
 \int d\t        U_2  \\ 
 \int d\t      U_3  
\end{array}\right] ~~~,~~~
i\Psi{}_{\hat k} = \left[\begin{array}{c}
    \z_1  \\ 
    \z_2  \\
    \z_3  \\
    \z_4  \\
    \hline
    \rho_1  \\ 
    \rho_2  \\ 
    \rho_3  \\ 
    \rho_4  \\ 
    \hline
\int d\t       \b_1  \\ 
\int d\t       \b_2  \\ 
 \int d\t       \b_3  \\ 
 \int d\t       \b_4 
\end{array}\right]  ~~~,~~~
{\rm D}_{\rm I} = \left[\begin{array}{c}
    {\rm D}_1  \\ 
    {\rm D}_2  \\     
    {\rm D}_3  \\ 
    {\rm D}_4 
\numberthis\end{array}\right] ~~~.
}
\label{CLSFNLbas}
\]

At this point, it is useful to remind the reader about a quantity defined for valise
adinkras and given the name ``cycle parity'' by T.\ H\" ubsch \cite{TH}.  Though the
first calculations of the ``cycle parity'' were introduced in the work of \cite{3}, and represented
by the symbol $\chi_{\rm {o}}$. For a given off-shell 4D, $\cal N$ = 1 supermultiplet, it is defined in terms of the $\bm \rL$ and $\bm \rR$ matrices via the definition \cite{2}:
\be
{\rm Tr}\left({\bm \rL}_\rI {\bm \rR}_\rJ {\bm \rL}_\rK {\bm \rR}_\rL \right) = 4\left[(n_c + n_t) \left( \delta_{\rI \rJ} \delta_{\rK \rL} - \delta_{\rI \rK} \delta_{\rJ \rL} + \delta_{\rI \rL} \delta_{\rJ \rK}\right) + \chi_0 ~ \epsilon_{\rI \rJ \rK \rL}\right]
\label{eq544}\ee

Notice that in order to isolate \(\chi_{\rm o}\) one must simply choose \(\rI, \rJ, \rK,\) and \(\rL\) such that \(\epsilon_{\rI \rJ \rK \rL} = 1\). In this case, the product of \(\bm \rL\) and \(\bm \rR\) matrices corresponds to the successive application of four supercovariant derivatives to a bosonic field, where the spinorial indices of the supercovariant derivatives form an even permutation. As an indirect result of the SUSY closure relation, this application will necessarily yield the original bosonic field with a second-time derivative and possibly a sign. Thus, the product reduces to an identity matrix with entries that can either be positive or negative, and \(\chi_{\rm o}\) is then defined as the trace of this matrix divided by 4.

This repeated application of supercovariant derivatives can be done graphically by examining the valise adinkra of a particular supermultiplet. In valise adinkras, there can be identified a number of quadrilaterals composed of edges with each of the four colors. These arise from the behavior of the matrix product just described. To calculate \(\chi_{\rm o}\), one can start at a bosonic vertex and traverse any such a four-color quadrilateral in which the colors are traversed in an order corresponding to an even permutation of the supercovariant derivatives. In the process, one will traverse some dashed lines (each representing a factor of -1) and some non-dashed lines (each representing a factor of +1). By taking the product of these factors, summing these products for each choice of starting bosonic vertex, and dividing by 4, one will arrive at the value for \(\chi_{\rm o}\).  This is the approach advanced by H\" ubsch.

For example, consider the CS adinkra shown in Fig. \ref{fig002}. Starting at the ``A'' node, then successively following the red, yellow, blue, and green edges (corresponding to the even identity permutation of \(\rD_1, \rD_2, \rD_3,\) and \(\rD_4\)), one is returned to the ``A'' node. Along this path, one encounters two dashed lines and two non-dashed lines, yielding a value of +1. In fact, in the case of CS, this value of +1 is attained regardless of the choice of the starting bosonic vertex. Thus we calculate
\be
\chi_{\rm {o}}(CS) ~=~ + 1 ~~~.
\ee
Applying this algorithm to VS and TS, we see that a value of -1 is attained regardless of the choice of starting vertex, yielding
\be
\chi_{\rm {o}}(VS) ~=~ \chi_{\rm {o}}(TS) ~=~  - 1 ~~~.
\ee

A striking point of the unfolded adinkras construction is the appearance of the ``evenness''
versus the ``oddness'' as one moves up the graphs. This has long been understood since the
CLS network system can be interpreted as the result of one CS network and two VS/TS networks. To expand upon this behavior, it is useful to present two different vantage points

\subsection{Block Diagonal and u(3) Generators}

The $\bm \rL$-matrices (Eqs. (\ref{Ell1}) - (\ref{Ell4})) and $\bm \rR$-matrices
(Eqs. (\ref{Rrrs1}) - (\ref{Rrrs4})) are block diagonal which implies that they can be 
written in forms that utilize the diagonal generators of the u(3) Lie algebra in an outer product with 4 $\times $ 4 $\bm \rL$-matrices and $\bm \rR$-matrices.  One can
choose as a basis a linear combination of the diagonal generators of the u(3) Lie algebra the following choice.
\be
{\bm E}_{1} ~=~  
\left[\begin{array}{ccc}
1 & 0 & 0  \\ 
0 & 0 & 0  \\ 
0 & 0 & 0  \\ 
\end{array}\right]      
~~~,~~~
{\bm E}_{2} ~=~  
\left[\begin{array}{ccc}
0 & 0 & 0  \\ 
0 & 1 & 0  \\ 
0 & 0 & 0  \\ 
\end{array}\right]  ~~~,~~~
{\bm E}_{3} ~=~  
\left[\begin{array}{ccc}
0 & 0 & 0  \\ 
0 & 0 & 0  \\ 
0 & 0 & 1  \\ 
\end{array}\right] 
\ee
Once this is done, then the multiplications of the 12 $\times $ 12 $\bm \rL$-matrices and $\bm \rR$-matrices collapses into a set of multiplications of the 4 $\times $ 4 $\bm \rL$-matrices.  Finally, taking the trace of the 12 $\times $ 12 $\bm \rL$-matrices
reduces to a sum of three traces of the 4 $\times $ 4 $\bm \rL$-matrices.

\subsection{Non-Block Diagonal and Unfolded Matrices}

The $\bm \rL$-matrices  and $\bm \rR$-matrices (Eqs. (\ref{eq4-10}) - (\ref{eq4-29})) are {\em {not}} block diagonal and so the calculations of $\chi_{\rm o}$ become more intricate.  For this
purpose a code in \textit{Mathematica} was developed and used to make a brute-force determination.  The
code is described in Appendix C and yields the same result, even though the intermediate
steps are much more intricate and completely different.  It is a very satisfying result
that the $\chi_{\rm o}$ value equal to -1 is the same.

\newpage
\section{Discussion: Properties of Adinkra}

In this chapter, we introduce the new measures that can represent the classification of the various adinkras and their properties.
Specifically, $\chi$-values and vorticity.
 
\subsection{$\chi$-value Definition and Properties}
In our presentation of the unfolded adinkras, we have included numerical data associated 
with the links in the graphs.  These are determined by the entries in the non-vanishing 
link matrices between graph nodes.  Although the resulting images bear a resemblance to 
the graphs that occur in discussions of braid groups, this is misleading as the notion of one link passing over or under a second link has no meaning in the present context.

Let us restrict our attention to the links in each disaggregated unfolded adinkra that connect the very lowest level of each to the one above. We define a numerical quantity denoted by ${\Tilde \chi}_{(1)}$ as the product of the numbers associated with each link of a given color in the first level. For example, in the case of the CS networks shown 
in Fig. \ref{fig1a}, one obtains
\be
{\Tilde \chi}_{(1)}(CS:red) 
 ~=~ {\Tilde \chi}_{(1)}(CS:yellow) ~=~ {\Tilde \chi}_{(1)}(CS:blue) ~=~ {\Tilde \chi}_{(1)}(CS:green) ~=~ +1
 ~~~.
 \label{C1}
\ee
Applying this same analysis to the VS and TS systems yield.
\be
{\Tilde \chi}_{(1)}(VS:red) 
 ~=~ {\Tilde \chi}_{(1)}(VS:yellow) ~=~ {\Tilde \chi}_{(1)}(VS:blue) ~=~ {\Tilde \chi}_{(1)}(VS:green) ~=~ + 1
 ~~~.
 \label{C2}
\ee
\be
{\Tilde \chi}_{(1)}(TS:red) 
 ~=~ {\Tilde \chi}_{(1)}(TS:yellow) ~=~ {\Tilde \chi}_{(1)}(TS:blue) ~=~ {\Tilde \chi}_{(1)}(TS:green) ~=~ -1
 ~~~.
 \label{C3}
\ee

The set of calculations can be carried out at the second level of the disaggregated adinkras to define ${\Tilde \chi}_{(2)}$.  It is easily seen that the same results
are obtained. So in these systems, we have
\be 
{\Tilde \chi}_{(1)} ~=~ {\Tilde \chi}_{(2)} ~~~.
\label{eq6.4}
\ee
We can neatly summarize the differences between the CS, VS, and TS networks, along with the VS${}_1$, VS${}_2$, and VS${}_3$ networks in the following table.
One can find the ${\bm\rL}$ and ${\bm\rR}$ matrices for the last three networks in \cite{permutadnk}.

\vskip3pt
\begin{table}[h]
\begin{center}
\footnotesize
\begin{tabular}{|c||c|c|c|}
\hline 
 {} &  ~~~~\, $ \chi_{\rm {o}}$ &  ${\Tilde \chi}_{(1)}$ ~ &  ${\Tilde \chi}_{(2)}$ ~  \\ \hline   
 \hline
{\rm {CS}} &  ~\,~  1 &  ~~~~\,~ 1 ~   &  ~~~~\,~ 1 ~ \\  \hline
{\rm {VS}}  &  ~\,~ - 1 &  ~~~~\,~ 1 ~  &  ~~~~\,~ 1 ~ \\  \hline  
{\rm {TS}}  &  ~\,~ - 1 &  ~~~~\,~ - 1 ~ &  ~~~~\,~ -1 ~  \\  \hline
{\rm {VS}}${}_1$ &  ~\,~ - 1 &  ~~~~\,~  1 ~   &  ~~~~\,~  1~ \\  \hline
{\rm {VS}}${}_2$  &  ~\,~ 1 &  ~~~~\,~  1~  &  ~~~~\,~  1 ~ \\  \hline  
{\rm {VS}}${}_3$  &  ~\,~ 1 &  ~~~~\,~  1~ &  ~~~~\,~  1 ~  \\  \hline
\end{tabular}
\end{center}
\caption{$\chi$-values \& ${\Tilde\chi}$-values for  
CS, VS, TS, VS${}_1$, VS${}_2$, and VS${}_3$  Networks}
\end{table}
It should be noted that the values of ${\Tilde \chi}_{(1)}$ and ${\Tilde \chi}_{(2)}$ values are specific to our field definitions. However, we conjecture that Eq. (\ref{eq6.4}) will hold if the supermultiplet field definitions are in valise form.
The realization that the two $\chi$-values appear as in the table above calculated based on the valise and unfolded adinkras also aligns with the use of the permutahedron
as a means of defining 4D, ${\cal N} = 1$ supermultiplets.

The field theories from which these are derived are all very different. The
CS system contains spinors and scalars, the VS system contains spinors and vectors, while the TS system contains scalars, 2-forms, and spinors. Thus, the Lorentz representation of the 4D, $\cal N$ = 1 system is distinctly encoded in the values different values of the graph-theoretic quantities $\chi$ and ${\Tilde\chi}$.  This is the essence of the concept \cite{ENUF} of ``SUSY holography'' realized on the unfolded adinkras.

In an example drawn from above, the CS and TS systems are known to be super-Hodge dual to one another.  This is apparently the reason relating the two sets of
$\chi$-values and $\Tilde \chi$-values.  If we regard these values as defining a space, then super Hodge duality is a ``parity reflection'' in this space. 

There are other interesting observations to make in the adjacency matrices for the unfolded 
versus folded cases that can be seen by using the results in Eqs. (\ref{eq211}) - (\ref{eq225}) for the CS
network system and in results in Eqs. (\ref{eq245}) -(\ref{eq259}) for the VS network system.  

In particular, substituting those results in Eq. (\ref{eq5.28B}) and for {\em {fixed}} values of ``I-index'' equal the ``J-index'' case
leads for the CS network to the equations
\begin{eqnarray} 
\d{}_{i}{}^{k} ~+~[(2)_b]{}_{i}{}^{k}
&=& 2\,
[{\bm {\rL}}{}_{1}^{(1)}]_{{i}}{}^{\hat j}
[{\bm {\rR}}{}_{1}^{(1)}]_{{\hat j}}{}^{k} ~~~,~~~
\d{}_{i}{}^{k} ~+~[(2)_b]{}_{i}{}^{k}
~=~ 2\,
[{\bm {\rL}}{}_{2}^{(1)}]_{{i}}{}^{\hat j}
[{\bm {\rR}}{}_{2}^{(1)}]_{{\hat j}}{}^{k} ~~~, \nonumber\\
\d{}_{i}{}^{k} ~+~[(2)_b]{}_{i}{}^{k}
&=& 2\,
[{\bm {\rL}}{}_{3}^{(1)}]_{{i}}{}^{\hat j}
[{\bm {\rR}}{}_{3}^{(1)}]_{{\hat j}}{}^{k}
~~~,~~~
\d{}_{i}{}^{k} ~+~[(2)_b]{}_{i}{}^{k}
~=~ 2\,
[{\bm {\rL}}{}_{4}^{(1)}]_{{i}}{}^{\hat j}
[{\bm {\rR}}{}_{4}^{(1)}]_{{\hat j}}{}^{k} ~~~,
 \label{CLSN1a}
\end{eqnarray}
and
\begin{eqnarray}
\d{}_{\hat i}{}^{\hat k} ~+~[(6)_b]{}_{\hat i}{}^{\hat k}
&=& 2\,
[{\bm {\rR}}{}_{1}^{(1)}]_{{\hat i}}{}^{j}
[{\bm {\rL}}{}_{1}^{(1)}]_{{j}}{}^{\hat k} ~~~,~~~
\d{}_{\hat i}{}^{\hat k} ~-~[(6)_b]{}_{\hat i}{}^{\hat k}
~=~ 2\,
[{\bm {\rR}}{}_{2}^{(1)}]_{{\hat i}}{}^{j}
[{\bm {\rL}}{}_{2}^{(1)}]_{{j}}{}^{\hat k} ~~~, \nonumber\\
\d{}_{\hat i}{}^{\hat k} ~-~[(6)_b]{}_{\hat i}{}^{\hat k}
&=& 2\,
[{\bm {\rR}}{}_{3}^{(1)}]_{{\hat i}}{}^{j}
[{\bm {\rL}}{}_{3}^{(1)}]_{{j}}{}^{\hat k}
~~~,~~~
\d{}_{\hat i}{}^{\hat k} ~+~[(6)_b]{}_{\hat i}{}^{\hat k}
~=~ 2\,
[{\bm {\rR}}{}_{4}^{(1)}]_{{\hat i}}{}^{j}
[{\bm {\rL}}{}_{4}^{(1)}]_{{j}}{}^{\hat k} ~~~.
\label{CLSN1b}
\end{eqnarray}

For the VS network the analogous equations are
\begin{eqnarray} 
\d{}_{i}{}^{k} ~+~ [(8)_b]{}_{i}{}^{k}
&=& 2\,
[{\bm {\rL}}{}_{1}^{(1)}]_{{i}}{}^{\hat j}
[{\bm {\rR}}{}_{1}^{(1)}]_{{\hat j}}{}^{k} ~~~,~~~
\d{}_{i}{}^{k} ~+~ [(8)_b]{}_{i}{}^{k}
~=~ 2\,
[{\bm {\rL}}{}_{2}^{(1)}]_{{i}}{}^{\hat j}
[{\bm {\rR}}{}_{2}^{(1)}]_{{\hat j}}{}^{k} ~~~, \nonumber\\
\d{}_{i}{}^{k} ~+~ [(8)_b]{}_{i}{}^{k}
&=& 2\,
[{\bm {\rL}}{}_{3}^{(1)}]_{{i}}{}^{\hat j}
[{\bm {\rR}}{}_{3}^{(1)}]_{{\hat j}}{}^{k}
~~~,~~~
\d{}_{i}{}^{k} ~+~ [(8)_b]{}_{i}{}^{k}
~=~ 2\,
[{\bm {\rL}}{}_{4}^{(1)}]_{{i}}{}^{\hat j}
[{\bm {\rR}}{}_{4}^{(1)}]_{{\hat j}}{}^{k} ~~~,
 \label{CLSN1c}
\end{eqnarray}
and
\begin{eqnarray}
\d{}_{i}{}^{k} ~+~ [(4)_b]{}_{i}{}^{k}
&=& 2\,
[{\bm {\rR}}{}_{1}^{(1)}]_{{\hat i}}{}^{j}
[{\bm {\rL}}{}_{1}^{(1)}]_{{j}}{}^{\hat k} ~~~,~~~
\d{}_{i}{}^{k} ~+~ [(8)_b]{}_{i}{}^{k}
~=~ 2\,
[{\bm {\rR}}{}_{2}^{(1)}]_{{\hat i}}{}^{j}
[{\bm {\rL}}{}_{2}^{(1)}]_{{j}}{}^{\hat k} ~~~, \nonumber\\
\d{}_{i}{}^{k} ~+~ [(1)_b]{}_{i}{}^{k}
&=& 2\,
[{\bm {\rR}}{}_{3}^{(1)}]_{{\hat i}}{}^{j}
[{\bm {\rL}}{}_{3}^{(1)}]_{{j}}{}^{\hat k}
~~~,~~~
\d{}_{i}{}^{k} ~+~ [(2)_b]{}_{i}{}^{k}
~=~ 2\,
[{\bm {\rR}}{}_{4}^{(1)}]_{{\hat i}}{}^{j}
[{\bm {\rL}}{}_{4}^{(1)}]_{{j}}{}^{\hat k} ~~~.
\label{CLSN1d}
\end{eqnarray}
where the ``Boolean Factor''  matrices $ [(1)_2]$, $ [(2)_2]$, $ [(4)_b]$, $ [(6)_b]$, and $ [(8)_b]$
have been defined in the work of
\cite{permutadnk}.

No analogous equations appear for the TS network as its adinkra is a 2-level valise adinkra even in the 4D equations.

For the CLS system, the behaviors of ${\Tilde \chi}_{\rm {(1)}}$ and  ${\Tilde \chi}_{(2)}$ are vastly more complicated.

\be
\small
{\Tilde \chi}_{(1)}(CLS:red) 
 ~=~ {\Tilde \chi}_{(1)}(CLS:yellow) ~=~ {\Tilde \chi}_{(1)}(CLS:blue) ~=~ {\Tilde \chi}_{(1)}(CLS:green) ~=~ +\frac{9}{2^{10}}
 ~~~.
 \label{C1}
\ee

\be
\small
{\Tilde \chi}_{(2)}(CLS:red) 
 ~=~ {\Tilde \chi}_{(2)}(CLS:yellow) ~=~ {\Tilde \chi}_{(2)}(CLS:blue) ~=~ {\Tilde \chi}_{(2)}(CLS:green) ~=~ +\frac{9}{2^{34}}
 ~~~.
 \label{C1}
\ee

\subsection{Nodal Vorticity Definition}
Vorticity is a property that arises among valise adinkras. Since valise adinkras can
be utilized to construct non-valise adinkras, this property extends accordingly.
We define vorticity in terms of an order of colors that we call a ``rainbow" \cite{dorangeometrization}, and each adinkra has two types of rainbows: one for bosonic nodes, and another for fermionic nodes. There is no unique definition of a rainbow, so it suffices
to choose one definition and use it universally thereafter.  

One such definition of the
rainbow for the bosonic nodes is set by picking the first bosonic node on the left, and noting the order in which the colors appear when looking at edges in a clockwise order. Similarly, the rainbow for fermionic nodes is set by picking the leftmost fermionic node, and noting the order in which the colors appear in a counterclockwise order.

For example, in Fig. \ref{fig001} the rainbow for the bosonic field is defined by the colors of edges emanating from $A$, which when read in a clockwise order yields RYBG (red, yellow, blue, and green). For the rainbow for the fermionic field, we look at $\psi_1$ and see the edges coming out of it in counterclockwise order, and see the rainbow is RGYB. Based on these bosonic and fermionic rainbows, we determine the vorticity as either clockwise or counterclockwise for each node. For example in Fig. \ref{fig001}, the node $A$ by definition has clockwise vorticity. However, the node $B$ has counterclockwise vorticity, as one needs to read off the colors of the edges for $B$ in a counterclockwise manner in order to produce the bosonic rainbow of RYBG set by the node $A$. Similarly, we see that the $F$ node has counterclockwise vorticity, and the $G$ node has clockwise vorticity. Meanwhile, the node $\psi_1$ by definition has counterclockwise vorticity, as does $\psi_2$, while $\psi_3$ and $\psi_4$ have clockwise vorticity.

\subsection{Vorticity in Valise Adinkra}

Based on Figs. \ref{fig001}-\ref{fig004} and ${\bm\rL}$ and ${\bm\rR}$ matrix information of VS${}_1$, VS${}_2$, and VS${}_3$ from \cite{permutadnk}, we calculate the vorticities of each node for each supermultiplet.

\vskip3pt
\begin{table}[H]
\begin{minipage}{.5\linewidth}
\centering
\footnotesize
\begin{tabular}{|c||c|c|c|c|}
\hline 
 {$\Phi_i $} &  1 & 2 & 3 & 4 \\ \hline
 \hline
{\rm {CS}}(B) & CW & CCW &  CCW & CW \\  \hline
{\rm {VS}}(B) & CW & CW  &  CCW & CCW \\  \hline  
{\rm {TS}}(B) & CW &  CW & CCW & CCW\\  \hline
{$\rm{VS_1}$}(B) & CW & CCW & CW & CCW \\  \hline
{$\rm{VS_2}$}(B) & CW & CCW & CCW & CW\\  \hline  
{$\rm{VS_3}$}(B) & CW & CCW & CW & CCW\\  \hline
\end{tabular}
\end{minipage}%
\begin{minipage}{.5\linewidth}
\centering
\footnotesize
\begin{tabular}{|c||c|c|c|c|}
\hline 
 {$\Psi_i$} &  1 & 2 & 3 & 4 \\ \hline
 \hline
{\rm {CS}}(F) &  CCW &  CCW & CW & CW \\  \hline 
{\rm {VS}}(F) &  CCW &  CCW & CW  &  CW \\  \hline 
{\rm {TS}}(F) &  CCW & CW & CW & CCW \\  \hline
{$\rm{VS_1}$}(F) & CCW & CW & CCW &  CW \\  \hline 
{$\rm{VS_2}$}(F) & CCW & CW & CW & CCW\\  \hline  {$\rm{VS_3}$}(F) & CCW & CW &  CCW  &CW\\  \hline
\end{tabular}
\end{minipage} 
\caption{Vorticity for CS, VS, TS and $\rm{VS_1}-\rm{VS_3}$ valise Adinkras. The left-hand side table is for bosonic nodes, and the right-hand side is for fermionic nodes. Here CW is clockwise and CCW is the counterclockwise direction of the rainbow.}
\label{tab2}
\end{table}

Furthermore, based on Figs. \ref{fig4}-\ref{fig6} we can calculate the vorticity of CLS valise adinkra.

\vskip3pt
\begin{table}[H]
\begin{center}
\footnotesize
\begin{tabular}{|c||c|c|c|c|c|c|c|c|c|c|c|c|c|}
\hline 
 {$\Phi_i,\Psi_i$} & 1&2&3&4&5&6&7&8&9&10&11&12 \\ \hline
 \hline
{\rm {CLS}}(B) &  CW &  CCW & CW & CCW & CW&CCW&CW&CCW&CW&CCW&CW&CCW\\  \hline
{\rm {CLS}}(F) & CCW &  CW& CCW & CW&CCW&CW&CCW&CW&CCW&CW&CCW&CW\\  \hline
\end{tabular}
\end{center}
\caption{Vorticity for CLS valise Adinkra, (B) is for bosonic, and (F) is for fermionic. Here CW is clockwise and CCW is the counterclockwise direction of the rainbow.}
\label{tab3}
\end{table}
As a result, one can see that the direction of different vorticity appears with the same number in each row of Tables \ref{tab2}-\ref{tab3}.  Stated 
another way, the net nodal vorticity at every level is zero.  Based on this observation and the examination of some other examples, we conjecture that this result is valid for all adinkras.

\section{Conclusion and Outlook}

In this paper, we have defined and constructed unfolded adinkras for the CS, VS, TS, 
and CLS networks. One can show there is a periodicity on the adinkra connection after 
the fifth level. Compared to the folded adinkra construction process, we can also 
verify that the modified garden algebra holds for the unfolded field definitions.

We have also given a first installment on the issue of considering the significance of
the local nodal vorticity in adinkras.  Expanding on our understanding of this property
will certainly extend into the future.

Clearly, more research is required to obtain a deeper understanding of the mathematical structures that govern the
algebraic relationships between folded and unfolded $\bm \rL$ and $\bm \rR$ matrices. One possible 
avenue for this is further study using the matrix information to find a polytopic representation for CLS fields. More generally, the intriguing emergence of a type-level number\footnote{The type-level number is the superscript number on the L and R matrix at Eq. (\ref{eq54}), Eq. (\ref{eq55})} associated with the D-operators
suggests a role for infinite dimensional algebras such filter Clifford algebras.  This is because when completely folded valise adinkra networks are used, ordinary Clifford algebras appear but when unfolded valise adinkra networks appear, then the emergence of the level number
is natural to such constructions.

\vspace{.05in}
\begin{center}
\parbox{4in}{{\it ``
Do not go where the path may lead, go instead where $~~~$ there is no path and leave a trail.
'' \\ ${~}$ 
 ${~}$ 
\\ ${~}$ }\,\,-\,\, \, Ralph Waldo Emerson   
$~~~~~~~~~$}
 \parbox{4in}{
 $~~$}  
 \end{center}

\noindent
{\bf {Acknowledgements}}\\[.1in] \indent
The research of  A.\ C., S.\ J.\ G., Y.\ L., E.\ T.\ L., T. O.\ R., and J.\ R.\
was supported during a portion of the time it was carried out by the endowment of the Ford Foundation Professorship of Physics at Brown 
University.  All also gratefully acknowledge the support of the Brown Theoretical Physics Center.   In addition, the research of S.\ J.\ G. is supported by the
Clark Leadership Chair in Science endowment at the University of Maryland - College Park.  Finally, 
Y.\ L., E.\ L., T. O.\ R., and J.\ R.\ acknowledge their participation in the Summer Student Theoretical Physics Research Session (SSTPRS) during the summer of 2022.

\newpage
\noindent
\section*{\bf {Appendix A: M-S Invariance}}

This appendix is devoted to discussing a symmetry that is present in several cases when a supermultiplet possesses
propagating spin-1/2 fields and a pair of auxiliary spin-1/2 fields.  Let us consider only all of the fermionic terms in the action 
seen Eq. (\ref{ACT-CLS}) of the form
$$
\mathcal{L}_{CLS}^{ferm.}(\zeta_a , \, \rho_a , \,  \beta_b) = i
\fracm{1}{2}(\gamma^\mu)^{ab} (\zeta_a \partial_\mu \zeta_b)+ iC^{ab}(\rho_a\beta_b)
\eqno(A.1)
$$

We can define new fermionic fields via the definitions
$$ 
\zeta_a ~=~ {\Hat \zeta}_a ~+~ \a_0 \, \Hat\rho_a  
{~~~~~~~~~~~~~~~~~~~~~~~}
\eqno(A.2)
$$
$$
\rho_a ~=~ {\Hat \rho}_a {~~~~~~~~~~~~~~~~~~~~~~~~~~~~~~~~~~~}
 \eqno(A.3)
$$
$$ 
\beta_a ~=~ {\Hat \beta}_a ~-~ \a_0 \,  (\gamma^\mu)_{a}{}^{b} \, \partial_\mu \,  (  \Hat\zeta_b + \fracm{1}{2} \a_0 \,  \Hat\rho_b)
\,
\eqno(A.4)
$$
and under this changed of variable, we find
$$
\mathcal{L}_{CLS}^{ferm.}(\zeta_a , \, \rho_a , \,  \beta_b) = 
\mathcal{L}_{CLS}^{ferm.}({\Hat \zeta}_a , \, {\Hat \rho}_a , \,  {\Hat \beta}_b)
~-~i\,  \fracm{1}{2} \alpha_0  \, \partial_\mu [ \, (\gamma^\mu)^{ab} \, \Hat\rho_a \Hat\zeta_b ) \,]  ~~~.
 \eqno(A.5) 
$$ \noindent

Thus, either set of field variables $({ \zeta}_a , \, { \rho}_a , \,  { \beta}_b)$ or
$({\Hat \zeta}_a , \, {\Hat \rho}_a , \,  {\Hat \beta}_b)$ can be used in the action.  The
D-algebra for the two different sets of fermionic
fields are distinct.  These are derived from the application of the D${}_a$-operator to both sides of the equations written in (A.2) - (A.4) above.

This type of invariance,``M-S symmetry,'' was first noted in the work of \cite{2} to our knowledge where its presence was noted in non-minimal supergravity and matter gravitino systems.
Though there is currently no proof, it appears
likely that this symmetry can appear in higher
spin supermultiplet systems also.

\newpage
\noindent
\section*{\bf {Appendix B: Infinitely Unfolded Adinkras }}

To make it clear, in this appendix, we discuss the possibility of the unfolding process
being carried out an infinite number of times.

\begin{figure}[H]
    \centering    \includegraphics[width=17cm]{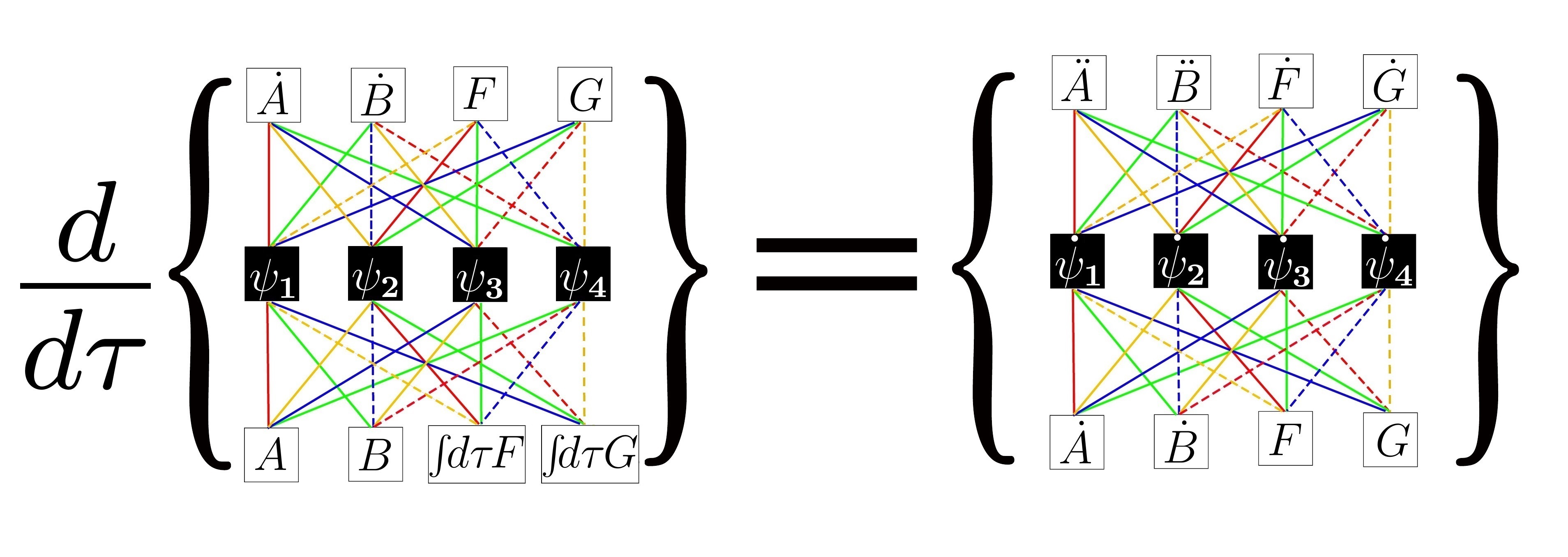}
    \caption{Applying time derivatives on the unfolded Adinkra of 4D $\mathcal{N}=1$ Chiral supermultiplet to make higher layers}
    \label{fig19}
\end{figure}

In Fig.\ \ref{fig19} to the left, we apply a time derivative to the unfolded two-level
adinkra for the CS network.  The result of this differentiation is shown to the right of
the figure.

This calculation shows how to construct the infinitely unfolded adinkras by using the time derivatives on 4D $\mathcal{N}=1$ Chiral supermultiplet. The lowest level of the adinkra on
the right is the same as the highest level of the adinkra on the left.  Therefore, the
lowest level of the rightmost adinkra can be superimposed on the top level of the leftmost
adinkra to construct a new image that possesses five levels.  Applying this process once
more will lead to a new image with seven levels.  Recursion of the process an infinite
number of times leads to infinitely unfolded adinkras.

\newpage
\noindent
\section*{\bf {Appendix C: Computer Simulation and graphical representation}}

In this chapter, we will investigate about the $\chi_{\rm o}$ values for ${\bm\rL}$- and ${\bm\rR}$-matrix on unfolded CLS Adinkra.
The calculation with unfolded ${\bm\rL}$-and ${\bm\rR}$-matrix is noted with a superscript, (unfold) and other general ${\bm\rL}$-and ${\bm\rR}$-matrix calculation is written without superscript notation.

\subsection*{\bf{C.1~~Mathematica code}}

We define the unfolded ${\bm\rL}$- and ${\bm\rR}$-matrix as a sum of all type-level ${\bm\rL}$- and ${\bm\rR}$-matrix respectively.

\begin{algorithm}
	\caption{$\chi_o$ calculation and L and R matrix transpose relation for CLS unfolded Adinkra} 
	\begin{algorithmic}[1]
    \State Update L and R matrix (Update Eq. (\ref{eq4-10})-(\ref{eq4-29}) on Mathematica code)
		\For {$iteration(i)=1,2,3,4$}
		  \For {$type\;level\;number(j)=1,2$}
				\State Sum up all $\rm{L_i^{(j)}}$ matrix
			\EndFor
   		\For {$type\;level\;number(j)=1,2,3$}
				\State Sum up all $\rm{R_i^{(j)}}$ matrix
			\EndFor
		\EndFor

  \Return (1) $\chi_o={\rm Tr(L_1R_2L_3R_4)}/4$   

  \State By using isEqual function, Compare $\rm{L_i}$ and Transposed $\rm{R_i}$

  \Return (2) True when it is same, and return False when it is different
  \end{algorithmic} 
\end{algorithm}

\begin{lcverbatim}
L1 = L[1, 1] + L[1, 2];
R1 = R[1, 1] + R[1, 2] + R[1, 3];
L2 = L[2, 1] + L[2, 2];
R2 = R[2, 1] + R[2, 2] + R[2, 3];
L3 = L[3, 1] + L[3, 2];
R3 = R[3, 1] + R[3, 2] + R[3, 3];
L4 = L[4, 1] + L[4, 2];
R4 = R[4, 1] + R[4, 2] + R[4, 3];

(Example)
Tr[L1 . R2 . L3 . R4]/4

(Example)
isEqual = L1 === Transpose[R1];
isEqual
\end{lcverbatim}

For example, when we calculate $\rm{\bm \rL}_1{\bm \rR}_2{\bm \rL}_3{\bm \rR}_4^{(unfold)}$ one can get the result as,

\begin{equation*}
\left(
\begin{array}{cccccccccccc}
 -3 & 0 & 0 & 0 & 2 & 0 & 0 & 0 & 0 & 0 & 0 & 0 \\
 0 & -3 & 0 & 0 & 0 & 0 & 0 & 0 & 2 & 0 & 0 & 0 \\
 0 & 0 & 1 & 0 & 0 & 0 & 0 & 0 & 0 & 0 & 0 & 0 \\
 0 & 0 & 0 & 1 & 0 & 0 & 0 & 0 & 0 & 0 & 0 & 0 \\
 -4 & 0 & 0 & 0 & 3 & 0 & 0 & 0 & 0 & 0 & 0 & 0 \\
 0 & 0 & 0 & 0 & 0 & -1 & 0 & 0 & 0 & 0 & 0 & 0 \\
 0 & 0 & 0 & 0 & 0 & 0 & -1 & 0 & 0 & 0 & 0 & 0 \\
 0 & 0 & 0 & 0 & 0 & 0 & 0 & -1 & 0 & 0 & 0 & 0 \\
 0 & -4 & 0 & 0 & 0 & 0 & 0 & 0 & 3 & 0 & 0 & 0 \\
 0 & 0 & 0 & 0 & 0 & 0 & 0 & 0 & 0 & -1 & 0 & 0 \\
 0 & 0 & 0 & 0 & 0 & 0 & 0 & 0 & 0 & 0 & -1 & 0 \\
 0 & 0 & 0 & 0 & 0 & 0 & 0 & 0 & 0 & 0 & 0 & -1 \\
\end{array}
\right)
\eqno(C.1)
\end{equation*}

and if we calculate $\rm{{\bm \rR}_1{\bm \rL}_2{\bm \rR}_3{\bm \rL}_4^{(unfold)}}$, we can get output as

\begin{equation*}
\left(
\begin{array}{cccccccccccc}
 1 & 0 & 0 & 0 & 0 & 0 & 0 & 0 & 0 & 0 & 0 & 0 \\
 0 & 1 & 0 & 0 & 0 & 0 & 0 & 0 & 0 & 0 & 0 & 0 \\
 0 & 0 & 1 & 0 & 0 & 0 & 0 & 0 & 0 & 0 & 0 & 0 \\
 0 & 0 & 0 & 1 & 0 & 0 & 0 & 0 & 0 & 0 & 0 & 0 \\
 0 & 0 & 0 & 0 & 0 & 0 & 0 & 0 & 0 & -2 & 0 & 0 \\
 0 & 0 & 0 & 0 & 0 & 0 & 0 & 0 & 2 & 0 & 0 & 0 \\
 0 & 0 & 0 & 0 & 0 & 0 & 0 & 0 & 0 & 0 & 0 & 2 \\
 0 & 0 & 0 & 0 & 0 & 0 & 0 & 0 & 0 & 0 & -2 & 0 \\
 0 & 0 & 0 & 0 & 0 & 0.5 & 0 & 0 & 0 & 0 & 0 & 0 \\
 0 & 0 & 0 & 0 & -0.5 & 0 & 0 & 0 & 0 & 0 & 0 & 0 \\
 0 & 0 & 0 & 0 & 0 & 0 & 0 & -0.5 & 0 & 0 & 0 & 0 \\
 0 & 0 & 0 & 0 & 0 & 0 & 0.5 & 0 & 0 & 0 & 0 & 0 \\
\end{array}
\right)
\eqno(C.2)
\end{equation*}

Based on Eq. (\ref{eq544}) if we choose I, J, K, L in order of 1,2,3,4, then all Kronecker delta terms go zero and we can get $\chi_{\rm o}$ directly. And due to the property of trace in the left-hand side of the equation, if we change the order of I, J, K, L in cyclic order the sign of $\chi_{\rm o}$ values does not change.

$${\rm Tr}\left({\bm \rL}_1 {\bm \rR}_2 {\bm \rL}_3 {\bm \rR}_4 \right) 
={\rm Tr}\left({\bm \rL}_1 {\bm \rR}_2 {\bm \rL}_3 {\bm \rR}_4 \right)\epsilon^{1234}
={\rm Tr}\left({\bm \rL}_4 {\bm \rR}_3 {\bm \rL}_2 {\bm \rR}_1 \right)\epsilon^{4321}
={\rm Tr}\left({\bm \rR}_1 {\bm \rL}_4 {\bm \rR}_3 {\bm \rL}_2 \right)\epsilon^{4321}$$
\begin{equation*}
={\rm Tr}\left({\bm \rR}_1 {\bm \rL}_2 {\bm \rR}_3 {\bm \rL}_4 \right)\epsilon^{2341}
=-{\rm Tr}\left({\bm \rR}_1 {\bm \rL}_2 {\bm \rR}_3 {\bm \rL}_4 \right)
= 4\chi_{\rm{o}}
{~~~~~}
\eqno(C.3)
\end{equation*}

As a result, we can get the $\chi_{\rm o}$ value for unfolded CLS as -1 by using (C.1)-(C.3), and for the L- and R-matrix transpose relation, we got all ``False'' outputs from the {\it{``isEqual"}} function which means they are not equal for unfolded Adinkra ${\bm\rL}$- and ${\bm\rR}$-matrix.

\subsection*{\bf{C.2~~$\chi_{\rm o}$ calculation; Order of colors}}

When we express the $\chi_{\rm o}$ calculation process as a graphical representation, first we can assign a color to $\rm D_1, D_2, D_3, D_4$ as Red, Yellow, Blue, and Green for example, and then we can apply those color orders to the closed path on the Adinkra.
For example, consider the adinkra found in Figs. \ref{fig4}-\ref{fig6}. Starting at \(\Phi_1\) and traversing the red, yellow, blue, and green edges, one encounters two dashed lines and two non-dashed lines, yielding +1. 
This graphical process can be expressed in the equation as (C.4)

$$
\rm D_4D_3D_2D_1\Phi_i=\rm
D_4D_3D_2\left(i({\bm \rL}_1)_{i}{}^{\hat k}\Psi_{\hat k}\right)=
D_4D_3\left(i({\bm \rL}_1{\bm \rR}_2)_{i}{}^{\hat j}\frac{d}{d\tau}\Phi_{j}\right)
{~~~~~~~~~~~~~~~~~~~~~~~}{}
$$
\begin{equation*}
=\rm D_4\left(-({\bm \rL}_1{\bm \rR}_2{\bm \rL}_3)_{i}{}^{\hat l}\frac{d}{d\tau}\Psi_{\hat l}\right)=
-({\bm \rL}_1{\bm \rR}_2{\bm \rL}_3{\bm \rR}_4)_{i}{}^{m}\frac{d^2}{d\tau^2}\Phi_{m}
\eqno(C.4)
\end{equation*}

Also, starting at the same node with reverse color order (green, blue, yellow, and red) one encounters two dashed lines and two non-dashed lines, yielding +1.
This graphical process can be expressed in the equation as (C.5). And this result also fits with (C.3) where the trace of ${\bm \rL}_1{\bm \rR}_2{\bm \rL}_3{\bm \rR}_4$ and ${\bm \rL}_4{\bm \rR}_3{\bm \rL}_2{\bm \rR}_1$ gives the same result.

\begin{equation*}
\rm D_1D_2D_3D_4\Phi_i=-({\bm \rL}_4{\bm \rR}_3{\bm \rL}_2{\bm \rR}_1)_{i}{}^{m}\frac{d^2}{d\tau^2}\Phi_{m}
\eqno(C.5)
\end{equation*}

If we repeat this process for all $\Phi_1-\Phi_{12}$, we can get +1 for $\Phi_1-\Phi_4$, and get -1 for $\Phi_5-\Phi_{12}$
Thus if we sum all these values and divide by four, we can get -1 as a final $\chi_{\rm o}$ value.

When we start from a fermionic node, for example, \(\Psi_1\) with the same color order(Red, Yellow, Blue, and Green), then we can get one dashed line and three non-dashed lines, yielding -1. 
This graphical process can be expressed in the equation as (C.6).

\begin{equation*}
\rm D_4D_3D_2D_1\Psi_{\hat i}=
-({\bm \rR}_1{\bm\rL}_2{\bm \rR}_3{\bm \rL}_4)_{\hat i}{}^{\hat m}\frac{d^2}{d\tau^2}\Psi_{\hat m}
\eqno(C.6)
\end{equation*}

If we repeat this process for all $\Psi_1-\Psi_{12}$, we can get -1 for $\Psi_1-\Psi_4$, and get +1 for $\Psi_5-\Psi_{12}$.
When we sum all these values and divide by four, we can get +1 which gives extra minus signs to trace value based on (C.3).
Thus the net result of this yields
$\chi_{\rm {o}}(CLS) ~=~ - 1$

\newpage
\noindent
\section*{\bf{Appendix D: Adinkra Transformations}}

According to \cite{3,10} there are two adinkra transformations: left-handed and right-handed transformations.\footnote{Adinkra transformations ``cis'' and ``trans'' adinkras \cite{10} have now been renamed to ``right-handed'' and ``left-handed'' respectively.}
These transformations maintain the same 0-brane information from lower level to higher dimensional level in valise adinkras while moving the nodes around left and right and redefining nodes on the same level.

\begin{figure}[H]
    \centering    \includegraphics[width=14cm]{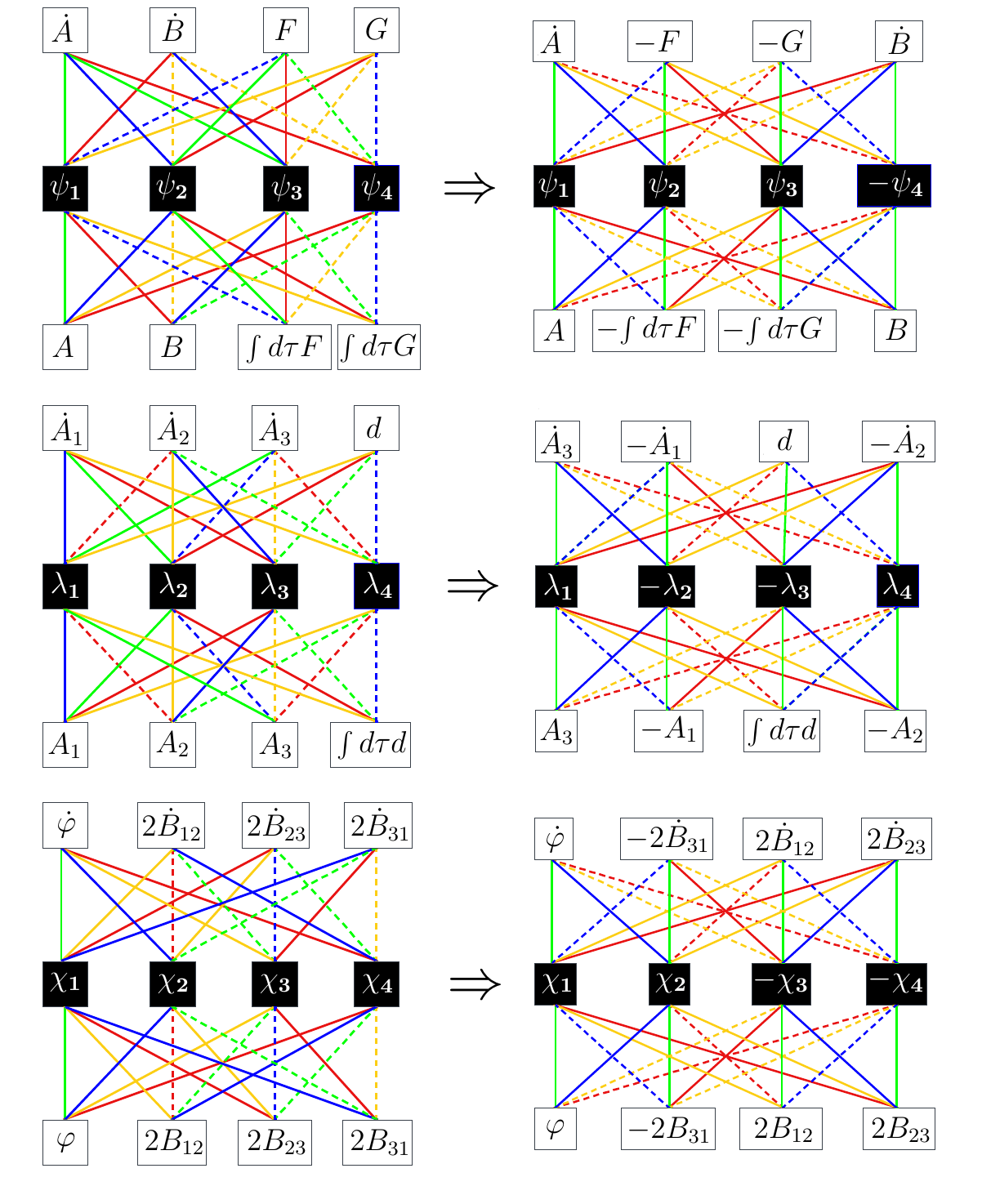}
    \caption{From the upper to lower figure, the left-handed, right-handed, and right-handed adinkra transformations of 4D $\mathcal{N}=1$ CS, VS, TS (Figs. \ref{fig1a}, \ref{fig2}, \ref{fig3}) are shown respectively.\protect\cite{10}}
    \label{fig20}
\end{figure}

\newpage
Note that the images in Fig. \ref{fig20} allow one to read off the ${\cal X}$-matrices and ${\cal Y}$-matrices that were defined by the equations (69) and (70) indicated in \cite{3}. This will imply that each pair has the same value for $\chi_{\rm o}$.

\vskip3pt
\begin{table}[H]
\begin{center}
\footnotesize
\begin{tabular}{|c||c|c|c|}
\hline 
 {} &  ~~~~\, $ \chi_{\rm {o}}$ &  ${\Tilde \chi}_{(1)}$ ~ &  ${\Tilde \chi}_{(2)}$ ~  \\ \hline
 \hline
{\rm {left-handed CS}} &  ~\,~  1 &  ~~~~\,~ 1 ~   &  ~~~~\,~ 1 ~ \\  \hline
{\rm {right-handed VS}}  &  ~\,~ - 1 &  ~~~~\,~ 1 ~  &  ~~~~\,~ 1 ~ \\  \hline  
{\rm {right-handed TS}}  &  ~\,~ - 1 &  ~~~~\,~ 1 ~ &  ~~~~\,~ 1 ~  \\  \hline
\end{tabular}
\end{center}
\caption{$\chi$-values \& ${\Tilde\chi}$-Values for  
left-handed CS, right-handed VS, and right-handed TS Networks}
\end{table}

For example, in the case of left-handed CS, the result is the same as $\rm{VS_2}$, $\rm{VS_3}$, and in the case of right-handed VS and right-handed TS, the result is identical to $\rm{VS_1}$.

\end{document}